\documentclass[11pt]{amsart}
\usepackage{amsmath,amssymb}
\usepackage{array}
\usepackage[margin=1.15in]{geometry}
\usepackage[colorlinks=true,linkcolor=blue,citecolor=blue,urlcolor=blue]{hyperref}
\hypersetup{pdftitle={Multiplicity and parity in Hirota's N-soliton conditions: the bilinear families L(D_x^3+Lambda)^(2n+1)},
  pdfauthor={Shi Zhenhua},   
  pdfsubject={Hirota bilinear method, N-soliton conditions, three-soliton condition}}

\numberwithin{equation}{section}

\newtheorem{theorem}{Theorem}[section]
\newtheorem{lemma}[theorem]{Lemma}
\newtheorem{proposition}[theorem]{Proposition}
\newtheorem{corollary}[theorem]{Corollary}
\newtheorem{conjecture}[theorem]{Conjecture}
\theoremstyle{definition}
\newtheorem{definition}[theorem]{Definition}
\newtheorem{example}[theorem]{Example}
\theoremstyle{remark}
\newtheorem{remark}[theorem]{Remark}

\newcommand{\Pev}{\Pi_{\mathrm{ev}}}
\newcommand{\supp}{\operatorname{supp}}
\newcommand{\C}{\mathbb{C}}
\newcommand{\Z}{\mathbb{Z}}
\newcommand{\bp}{\mathbf{p}}
\newcommand{\bq}{\mathbf{q}}
\newcommand{\sg}{\sigma}
\newcommand{\Lam}{\Lambda}

\begin{document}

\title[Multiplicity and parity in the $N$-soliton conditions]{Multiplicity and parity in Hirota's $N$-soliton conditions:\\ the bilinear families $L\,(D_x^{3}+\Lambda)^{2n+1}$}

\author{Shi Zhenhua}

\begin{abstract}
Hirota's three-soliton condition is the standard test for the integrability of a bilinear
equation $P(D)f\cdot f=0$, and the folklore implication ``three-soliton condition
$\Rightarrow$ $N$-soliton solutions for all $N$'' underlies the classifications of Hietarinta.
We study this implication in two directions. The first is the family
$L\,(D_x^3+\Lambda)^{2n+1}f\cdot f=0$ with $L$, $\Lambda$ linear forms in the Hirota operators,
which for $n=0$ contains the Korteweg--de Vries, Ito and Boiti--Leon--Manna--Pempinelli
equations and for $L=D_x$ is the family of G\"urses and Pekcan. On the principal component of
the dispersion variety the $N$-soliton condition reduces to a sign sum of two bialternants;
counting odd exponents shows that for $n=0$ the class is Hirota integrable and that for
$n\ge1$ the three-soliton condition holds for every $L$, and an explicit formula for the
four-soliton residual proves the G\"urses--Pekcan conjecture that the four-soliton condition
fails for every $n\ge1$. The formula yields sum-of-squares certificates excluding four-soliton
solutions at every real configuration for the G\"urses--Pekcan and Ito equations, classifies
the real members of the class, and determines maximal numbers of nondegenerate solitons:
exactly four or five for $n=1$ according as $L$ lies in the span of $D_x,\Lambda$ or not, five
for $n=2$ and $L=D_x$, and finite for every $n\ge1$ and every $L$. The second direction is a
reformulation of the implication for a single non-hyperplane component of multiplicity one,
which we prove, with complete classifications, for all graph-type dispersions $L(q(D)+D_t)$,
for every irreducible even quadric component in any dimension and with an arbitrary cofactor,
for parabolic components, for the dispersions $F(\ell)+Q$ containing the
Kadomtsev--Petviashvili surface, and, with a prescribed one-parameter phase family, for the
cubic graph.
\end{abstract}

\maketitle

\section{Introduction}\label{sec:intro}

\subsection{The three-soliton condition and the folklore implication}
Hirota's bilinear method \cite{Hirota1971,Hirota2004} remains the most direct route to
multisoliton solutions of nonlinear evolution equations. For a bilinear equation
$P(D)f\cdot f=0$ of KdV type ($P$ even, $P(0)=0$) one- and two-soliton solutions exist
automatically, whereas the existence of a three-soliton solution with all parameters free on
the dispersion variety --- the \emph{three-soliton condition} (3SC) --- is highly restrictive and
has long served as a practical detector of integrability \cite{Hirota2004,Hietarinta2004,Zhang2026}.
Hietarinta's systematic searches \cite{HietarintaI,HietarintaII,HietarintaIII,HietarintaIV,Hietarinta1991}
classified bilinear equations of low degree passing the 3SC and recovered, almost exclusively,
known integrable hierarchies. It has since been part of the folklore of the subject that passing
the 3SC should already imply the existence of $N$-soliton solutions for all $N$: Hirota
\cite{Hirota2004} reports (as recalled in \cite{GursesPekcan2025}), on the basis of a private
communication with Hietarinta, that no bilinear equation with a three-soliton but no
four-soliton solution was known, with the single candidate $D_xD_yD_tD_z$.

Recently G\"urses and Pekcan \cite{GursesPekcan2025} introduced the family
\begin{equation}\label{eq:GP}
D_x\bigl(D_x^{3}+\alpha_1D_t+\alpha_2D_y\bigr)^{2n+1}f\cdot f=0,\qquad (\alpha_1,\alpha_2)\neq(0,0),\quad n\in\mathbb{N}
\end{equation}
(they write $k$ for our $n$), proved that every member satisfies the 3SC, showed by direct
computation that the four-soliton condition (4SC) fails for $n=1,2,3$, and conjectured that it
fails for every $n\ge1$. The family is thus a candidate for an infinite explicit list of
bilinear equations with three-soliton but no four-soliton solutions.

Passing the 3SC in the sense used here means that the three-soliton polynomial vanishes
identically on a component of the dispersion variety, with all soliton parameters free. This
is strictly stronger than the existence of some three-soliton solution: Hirota's exact
three-soliton solution of the two-dimensional sine--Gordon equation \cite{Hirota1973} exists
only under an additional constraint on the wave parameters, the prototype of what Hietarinta
terms partial integrability \cite{Hietarinta1990}. The families studied below pass the 3SC in
the strict sense on the principal component for every $n$, and it is in the same strict sense
that we prove the 4SC to fail.

\subsection{The families $L(D_x^3+\Lambda)^{2n+1}$}
We prove the conjecture of G\"urses and Pekcan and place it in a wider framework. We consider
the equations
\begin{equation}\label{eq:intro-family}
L\,(D_x^{3}+\Lambda)^{2n+1}f\cdot f=0,\qquad n\ge0,
\end{equation}
where $\Lambda$ is any linear form in the Hirota operators with a nonzero component outside
$D_x$, and $L$ is any nonzero linear form. For $n=0$ the class contains the KdV equation
($L=D_x$, $\Lambda=D_t$), Ito's equation ($L=\Lambda=D_t$) \cite{Ito1980} and the
Boiti--Leon--Manna--Pempinelli equation ($L=D_y$, $\Lambda=D_t$) \cite{BLMP1986}, together with
all their linear mixtures; for $L=D_x$ it is the family \eqref{eq:GP}. The dispersion variety
of \eqref{eq:intro-family} is reducible, $V_L\cup V_Q$, with the hyperplane $V_L=\{L=0\}$ and
the \emph{principal component} $V_Q=\{k^3+\Lambda(\bp)=0\}$, along which the polynomial vanishes
to order $2n+1$. Our results on this class are the following.
\begin{enumerate}
\item \textbf{Hirota integrability for $n=0$ (Theorem~\ref{thm:I}).} For every $L$ and
$\Lambda$ the equation $L(D_x^3+\Lambda)f\cdot f=0$ satisfies the $N$-soliton condition for every
$N$ on the principal component of its dispersion variety, with the phase factors
$\frac{(\ell_i-\ell_j)(k_i-k_j)}{(\ell_i+\ell_j)(k_i+k_j)}$, $\ell_j=L(\bp_j)$. The proof takes a
few lines: on shell the $N$-soliton condition is the sign sum
$\sum_\sigma e_1(\sigma\ell)\Delta_N(\sigma\ell)\,s_{(2,1)}(\sigma k)\Delta_N(\sigma k)$ of two
bialternants, and the bialternant formula shows that the monomials of $e_1\Delta_N$ and of
$s_{(2,1)}\Delta_N$ have different numbers of odd exponents. For $n=0$ all $N$-soliton
conditions hold in fact on the whole dispersion variety (Theorem~\ref{thm:whole}), by a limiting
argument in which two solitons of the principal component fuse into one soliton of $V_L$.
\item \textbf{The three-soliton condition for all $n$ (Theorem~\ref{thm:II}).} Every member
of \eqref{eq:intro-family} satisfies the 3SC on the principal component. The multiplicity is
harmless because the three-variable on-shell factor $s_{(2,1)}\Delta_3=\prod_{i<j}(a_i^2-a_j^2)$
is a polynomial in the squares. This is a property of the cubic: for the families
$D_x(q(D_x)+D_t)^{2n+1}$ with $q$ odd, the 3SC holds for one (equivalently, every) $n\ge1$ if and
only if $q(z)=\alpha z^3+\beta z$ (Theorem~\ref{thm:classification}), and for an odd polynomial
$q$ in several variables, an arbitrary linear factor $L$ and any fixed $n\ge1$, the equation
$L(q(D)+D_t)^{2n+1}f\cdot f=0$ satisfies the 3SC if and only if $q$ is, up to linear terms, the
cube of a linear form (Theorem~\ref{thm:classification-multi}). For the multiplicity-one members $L(q(D)+D_t)$ we obtain a complete
classification (Theorem~\ref{thm:mult-one}): the 3SC holds exactly when $q$ is, up to linear
terms, the cube of a linear form, or when $L$ is a spatial form $l$ and $q=a\,l^5+b\,l^3$ up to
linear terms; in both cases every polynomial $N$-soliton condition holds on the whole
dispersion variety, with nondegenerate $N$-soliton data generic on the principal component.
This verifies Conjecture~\ref{conj:repair} below for the entire class of graph-type
dispersions with a linear factor.
\item \textbf{Failure of the four-soliton condition for all $n\ge1$ (Theorem~\ref{thm:III}).}
For every $L$, every $\Lambda$ and every $n\ge1$ the 4SC fails; equivalently, every such
equation has nondegenerate three-soliton solutions with all parameters free, but no
four-soliton solution of Hirota's form for generic data. For $L=D_x$ this is the
G\"urses--Pekcan conjecture.
The proof rests on an explicit formula for the on-shell 4SC (Theorem~\ref{thm:residual}):
for $L=\beta D_x+\gamma\Lambda$, expanding the two bialternants along a row and selecting the
monomials with even exponents, one finds, in the squared wavenumbers $x_i=k_i^2$, the
factorization
$x_1x_2x_3x_4\prod_{i<j}(x_i-x_j)^2R_n$ with $R_n$ an explicit finite sum of products of
squares and simple quadratic factors.
The boundary slice $R_n(x_1,x_2,x_3,0)$ follows in closed form (Theorem~\ref{thm:slice}); for
$L=D_x$ a hand-checkable certificate at the single point $(6,5,3,1)$ gives a second proof
(Remark~\ref{rem:certificate}).
\item \textbf{Mixed components (Theorem~\ref{thm:mixed}).} For $n\ge1$ the 3SC holds on the
principal component $V_Q$, with nondegenerate data, but fails on $V_L\times V_Q\times V_Q$; an
explicit three-block expansion of the mixed three-soliton polynomial (Theorem~\ref{thm:threeblock})
gives the exact set of real nondegenerate mixed three-soliton data, which is independent of $n$.
Statements about the 3SC therefore have to name the component, and
Definition~\ref{def:component} makes this precise.
\item \textbf{Signs, zeros and real data (Theorems~\ref{thm:sign} and~\ref{thm:existence}).} The
residuals of the G\"urses--Pekcan and Ito families are sign-definite on the open positive
orthant, with an explicit sum-of-squares certificate valid for all $n\ge1$; consequently these
equations possess no four-soliton solution of Hirota's form at \emph{any} real nondegenerate
configuration of the principal component, not merely generically, and in fact anywhere on
the real dispersion variety. For $n\ge2$ the zero locus of the
residuals on the closed positive orthant is exactly the set of points with at most two distinct
positive coordinate values. For all real $L$, real nondegenerate multisoliton data with $N\ge4$ can only lie on the
principal component (Theorem~\ref{thm:GP-whole}), and the existence of special real
four-soliton data is decided by the sign of $\beta\gamma$ when $L=\beta D_x+\gamma\Lambda$, and
is always the case when $L\notin\operatorname{span}(D_x,\Lambda)$; when they exist they can be
chosen with positive phase factors, so that $f_4\ge1$ (Proposition~\ref{prop:realsign},
Theorems~\ref{thm:existence} and~\ref{thm:positive}).
\item \textbf{Maximal soliton numbers (Section~\ref{sec:strata}).} Over $\C$ we determine how
many nondegenerate solitons the equations of the class can carry: for $n=1$ mixed data are
excluded, every equation with $L\in\operatorname{span}(D_x,\Lambda)$ has nondegenerate four- but no
five-soliton solutions, while for $L\notin\operatorname{span}(D_x,\Lambda)$ five-soliton data
exist (exact interval-arithmetic certificates, real for real $L$) and six are excluded by an
inclusion--exclusion argument, so the maximal number is exactly five
(Theorems~\ref{thm:five-exist} and~\ref{thm:six-bound}); for $n=2$ and $L=D_x$
the maximal number is five, attained exactly when the squared wavenumbers are the fifth roots
of a constant (Theorem~\ref{thm:n2-five}); for every $n\ge1$ and every $L$ the number of
solitons is finite, with explicit bounds (Theorems~\ref{thm:finite-bound}
and~\ref{thm:finite-all}); and for every $n\ge2$ the G\"urses--Pekcan family, and every
equation with $L\notin\operatorname{span}(D_x,\Lambda)$, has complex four-soliton data with a
momentum outside the principal component (Theorem~\ref{thm:complex-mixed},
Proposition~\ref{prop:Gn-root}). Table~\ref{tab:max} in Section~\ref{sec:strata} collects
these numbers.
\end{enumerate}

The mechanism behind the whole phenomenon can be stated in one sentence: on the dispersion
variety the cubic form $(a+b+c)^3-a^3-b^3-c^3$ factors into linear forms, while its
four-variable analogue does not (Lemma~\ref{lem:no-linear-factor}). The three-variable
factorization is what makes the 3SC hold for every $n$; the failure of its four-variable
analogue is what the residual $R_n$ measures.

\subsection{The reformulated implication}\label{subsec:intro-conjecture}
The second part of the paper reexamines the folklore implication itself. The exceptional
equation $D_xD_yD_tD_z$ is a genuine squarefree instance of ``three solitons but not four''
\cite[Remark~4]{GursesPekcan2025}, but one that involves four different components of the
dispersion variety; the families \eqref{eq:intro-family} fail on a single component of
multiplicity $2n+1$; and a squarefree sextic whose dispersion variety contains an affine
hyperplane fails on that single component with all phase factors regular
(Proposition~\ref{prop:affine}). Neither situation arises for the equations of Hietarinta's
lists, whose relevant components are hypersurfaces of degree at least two appearing with
multiplicity one. These examples motivate a componentwise formulation.

\begin{conjecture}[Reformulated implication]\label{conj:repair}
Let $P$ be an even polynomial with $P(0)=0$, and let $V_0$ be an irreducible component of the
dispersion variety $V(P)$ which is not a hyperplane and along which $P$ vanishes to order one.
If $P$ satisfies the three-soliton condition on $V_0$, then it satisfies the $N$-soliton
condition on $V_0$ for every $N\ge3$.
\end{conjecture}

The hypotheses are discussed in Section~\ref{sec:discussion}, where it is also shown that they
cannot be weakened in the obvious ways. We prove the conjecture, with a complete classification
of the polynomials satisfying the 3SC, in the following classes. Theorem~\ref{thm:mult-one}
proves it for all graph-type dispersions with a linear factor. Theorem~\ref{thm:all-even-quadrics}
proves it for every irreducible even quadric component, in any dimension and for an arbitrary
cofactor of any degree and multiplicity, and classifies the solutions: on a cone only
$\lambda Q$; on an affine quadric the rank must be two, the polynomial depends on the two
variables of the quadric only, and it belongs to a nine-dimensional family of degree at most
twelve with the common Sawada--Kotera-type phase factors (Theorem~\ref{thm:bivariate-complete},
proved in Appendix~\ref{app:bivariate}) or to one exceptional quartic family. The parabolic
components leave only the degenerate Boussinesq form $u^4-v^2$ (Theorems~\ref{thm:noneven}
and~\ref{thm:parabola-general}), and for the dispersions $F(\ell)+Q$, which include the
Kadomtsev--Petviashvili surface, the three-soliton condition holds exactly for the KdV/Sawada--Kotera
and KP/Boussinesq types (Theorem~\ref{thm:FQ}). On the cubic graph $v=u^3$, with the phase
factor prescribed in a natural one-parameter family, the three-soliton condition holds exactly
for a five-dimensional family with the KdV phase, for Ito's equation and for two golden-ratio
sextics (Theorem~\ref{thm:cubic-graph}). In all these classes the three-soliton condition
implies all $N$-soliton conditions --- by an elementary fusion argument, or by the classical KP
solitons for the KP type; a component of multiplicity two can carry them all
(Remark~\ref{rem:double}), multiplicity three never does.

\subsection{Organization}
Section~\ref{sec:prelim} fixes notation: the $N$-soliton conditions and their component-wise
formulation, sign characters and parity classes, and the bialternant formula.
Section~\ref{sec:reduction} reduces the $N$-soliton conditions of \eqref{eq:intro-family} on
shell to a sign sum of two bialternants. Section~\ref{sec:I} proves Theorem~\ref{thm:I}.
Section~\ref{sec:II-III} proves Theorem~\ref{thm:II} and classifies the odd polynomials $q$ for
which the multiplicity mechanism works, including the complete multiplicity-one classification
of graph-type dispersions. Section~\ref{sec:structure} derives the explicit residual and the
boundary slice and proves Theorem~\ref{thm:III}. Section~\ref{sec:mixed} treats the mixed
components and the whole dispersion variety for $n=0$, Section~\ref{sec:strata} the signs and
zeros of the residuals, real and complex four-soliton data and the maximal soliton numbers,
and Section~\ref{sec:quadric} the quadric and parabolic components, the dispersions
$F(\ell)+Q$ and the cubic graph, with the two longest classification proofs deferred to
Appendices~\ref{app:bivariate} and~\ref{app:parabola}. Section~\ref{sec:discussion} discusses
the consequences for Hietarinta's programme, the hypotheses of Conjecture~\ref{conj:repair} and
open problems.

All computer-assisted statements of this paper --- the data of Tables~\ref{tab:Rn}
and~\ref{tab:cert}, the identities \eqref{eq:F-identity} and $\widetilde Q_0=E+U+V$, the certificate
\eqref{eq:certificate-123}, and the finite instances of the identities of
Sections~\ref{sec:structure} and~\ref{sec:mixed} that are quoted as checks --- were verified in exact arithmetic
with python-flint and independently re-run in a second environment; every nonvanishing
statement was checked at two independent random points. The scripts and their recorded outputs
are available as ancillary files. Theorem~\ref{thm:residual} and its consequences do not depend
on computer assistance.

\section{Preliminaries}\label{sec:prelim}

\subsection{The \texorpdfstring{$N$}{N}-soliton conditions}\label{subsec:NSC}
Let $t=(t_1,\dots,t_s)$ be the independent variables and $D=(D_{t_1},\dots,D_{t_s})$ Hirota's
bilinear operators, so that
\begin{equation}\label{eq:hirota-exp}
P(D)\,e^{\bp\cdot t}\cdot e^{\bq\cdot t}=P(\bp-\bq)\,e^{(\bp+\bq)\cdot t}
\end{equation}
for every polynomial $P$ \cite{Hirota2004}. A \emph{bilinear equation of KdV type} is an equation
$P(D)f\cdot f=0$ with $P$ an even polynomial and $P(0)=0$; since $P(D)f\cdot f$ depends only on
the even part of $P$, evenness is no restriction. The \emph{dispersion variety} of $P$ is
$V(P):=\{\bp\in\C^s:\ P(\bp)=0\}$. Hirota's $N$-soliton ansatz is
\begin{equation}\label{eq:ansatz}
f_N=\sum_{\mu\in\{0,1\}^N}\exp\Bigl(\sum_{j=1}^N\mu_j\theta_j+\sum_{1\le i<j\le N}\mu_i\mu_jA_{ij}\Bigr),
\qquad \theta_j=\bp_j\cdot t+\delta_j,\qquad e^{A_{ij}}=-\frac{P(\bp_i-\bp_j)}{P(\bp_i+\bp_j)},
\end{equation}
with $\bp_j\in V(P)$. The phase factors $e^{A_{ij}}$ are defined when $P(\bp_i+\bp_j)\neq0$; we
call the data $(\bp_1,\dots,\bp_N)$ \emph{nondegenerate} if $P(\bp_i\pm\bp_j)\neq0$ for all
$i<j$, so that all phase factors are finite and nonzero. Substituting \eqref{eq:ansatz} into the
equation and collecting the coefficient of $\exp(\sum_{j\in I}\theta_j)$ for each subset
$I\subseteq[N]:=\{1,\dots,N\}$ leads to the polynomials
\begin{equation}\label{eq:SN}
S_N(\bp_1,\dots,\bp_N):=\sum_{\sg\in\{\pm1\}^N}P\Bigl(\sum_{j=1}^N\sg_j\bp_j\Bigr)
\prod_{1\le i<j\le N}\sg_i\sg_j\,P(\sg_i\bp_i-\sg_j\bp_j)
\end{equation}
and to the classical statement \cite{Hirota2004,HietarintaI,Hietarinta2004,Zhang2026}: for
nondegenerate $\bp_1,\dots,\bp_N\in V(P)$, the function $f_N$ solves $P(D)f\cdot f=0$ for all
values of the $\delta_j$ if and only if $S_M(\bp_{i_1},\dots,\bp_{i_M})=0$ for every $M\le N$ and
every $i_1<\dots<i_M$. Here $S_1=2P(\bp_1)$ vanishes by the dispersion relation and $S_2$
vanishes identically for even $P$ (its four terms cancel in pairs), so that one- and two-soliton
solutions always exist; the first genuine condition is $S_3=0$.

\begin{definition}[Conditions on a component]\label{def:component}
Let $V(P)=V_1\cup\dots\cup V_r$ be the decomposition of the dispersion variety into irreducible
components. For $N\ge3$ and a component $V_0$ we say that $P$ \emph{satisfies the $N$-soliton
condition ($N$SC) on $V_0$} if $S_N$ vanishes identically on $V_0^{\,N}$, and that $P$ is
\emph{Hirota integrable on $V_0$} if it satisfies the $N$SC on $V_0$ for every $N\ge3$. We say
that $P$ satisfies the $N$SC \emph{on the whole dispersion variety} if $S_N$ vanishes identically
on $V_{j_1}\times\dots\times V_{j_N}$ for every choice of $j_1,\dots,j_N$. The conditions with
$N=3$ and $N=4$ are the three-soliton condition (3SC) and the four-soliton condition (4SC).
\end{definition}

Since $V_0$ is irreducible, vanishing on $V_0^{\,N}$ is equivalent to vanishing at generic points
of $V_0^{\,N}$. For irreducible $V(P)$ the two notions of Definition~\ref{def:component}
coincide. They differ for reducible $P$, and reducible $P$ are common: the KdV equation
$D_x(D_x^3+D_t)f\cdot f=0$ has $V(P)=\{k=0\}\cup\{\omega=-k^3\}$, and the hyperplane $\{k=0\}$
carries the $x$-independent exponentials $e^{\omega t}$, which are invisible in
$u=2(\log f)_x$. Statements about the integrability of such an equation always refer to the
component carrying genuine solitons. Section~\ref{sec:mixed} shows that this is not a pedantic
point: for the families of this paper with $n\ge1$ the 3SC holds on the principal component but
fails on mixed products of components, so that the component has to be named.

\begin{remark}\label{rem:normalization}
G\"urses and Pekcan \cite{GursesPekcan2025} write the conditions as a sum over the sign vectors
with $\sg_1=+1$; since $P$ is even, \eqref{eq:SN} is exactly twice their expression. All
statements below are insensitive to this normalization.
\end{remark}

\subsection{Sign characters and parity classes}\label{subsec:parity}
For $\sg\in\{\pm1\}^N$ and $a=(a_1,\dots,a_N)$ write $\sg\cdot a:=(\sg_1a_1,\dots,\sg_Na_N)$, and
for $S\subseteq[N]$ let $\chi_S(\sg):=\prod_{i\in S}\sg_i$. The $2^N$ functions $\chi_S$ are the
characters of the group $\{\pm1\}^N$, and
\begin{equation}\label{eq:orth}
\sum_{\sg\in\{\pm1\}^N}\chi_S(\sg)\chi_T(\sg)=2^N\delta_{ST}.
\end{equation}
The \emph{parity class} of a monomial $a^\alpha=a_1^{\alpha_1}\cdots a_N^{\alpha_N}$ is the set
$\{i:\ \alpha_i\ \text{odd}\}$. Every polynomial $F\in\C[a_1,\dots,a_N]$ decomposes uniquely as
$F=\sum_{S\subseteq[N]}F_S$, where $F_S$ is the sum of the terms of $F$ whose monomials have
parity class $S$; then $F_S(\sg\cdot a)=\chi_S(\sg)F_S(a)$. We call
$\supp F:=\{S:\ F_S\neq0\}$ the \emph{parity support} of $F$, and we write
$\Pev F:=F_\emptyset$ for the \emph{even projection} (the sum of the terms of $F$ in which every
variable occurs to an even power).

\begin{lemma}\label{lem:parity}
Let $F\in\C[b_1,\dots,b_N]$ and $G\in\C[a_1,\dots,a_N]$.
\begin{enumerate}
\item[(a)] $\sum_{\sg}F(\sg\cdot b)=2^N\,\Pev F$.
\item[(b)] $\sum_{\sg}F(\sg\cdot b)\,G(\sg\cdot a)=2^N\sum_{S\subseteq[N]}F_S(b)\,G_S(a)$.
\item[(c)] If $a_1,\dots,a_N,b_1,\dots,b_N$ are independent indeterminates, the sum in (b) is
the zero polynomial if and only if $\supp F\cap\supp G=\emptyset$.
\item[(d)] If $\phi_1,\dots,\phi_N$ are odd polynomials in one variable and $b_i=\phi_i(a_i)$,
then $\sum_{\sg}F(\sg\cdot b)\,G(\sg\cdot a)=2^N\,\Pev\bigl[F(\phi(a))\,G(a)\bigr]$ with
$\phi(a):=(\phi_1(a_1),\dots,\phi_N(a_N))$; in particular this vanishes whenever
$\supp F\cap\supp G=\emptyset$.
\end{enumerate}
\end{lemma}

\begin{proof}
(a) and (b) follow from $F(\sg\cdot b)=\sum_S\chi_S(\sg)F_S(b)$, $G(\sg\cdot a)=\sum_T\chi_T(\sg)G_T(a)$
and \eqref{eq:orth}. For (c), the polynomial $F_S(b)G_S(a)$ is supported on the monomials
$b^\beta a^\alpha$ whose parity classes in $b$ and in $a$ both equal $S$; for distinct $S$ these
sets of monomials are disjoint, so the sum in (b) vanishes if and only if every product
$F_S(b)G_S(a)$ vanishes, i.e.\ if and only if for every $S$ one of $F_S$, $G_S$ is zero. For (d),
oddness gives $\phi(\sg\cdot a)=\sg\cdot\phi(a)$, so the sum equals $\sum_\sg(F\circ\phi\cdot G)(\sg\cdot a)$,
and (a) applies; the last assertion follows from (b) by specialization.
\end{proof}

\subsection{Alternants and Schur polynomials}\label{subsec:alternants}
Let
\[
\Delta_N(a):=\prod_{1\le i<j\le N}(a_i-a_j)=\det\bigl(a_j^{\,N-i}\bigr)_{1\le i,j\le N}
\]
be the Vandermonde determinant, and let $e_r$ and $p_r$ denote the elementary symmetric
polynomials and the power sums of $a_1,\dots,a_N$. For a partition $\lambda=(\lambda_1\ge\dots\ge\lambda_N\ge0)$
the bialternant formula
\begin{equation}\label{eq:bialternant}
s_\lambda(a)\,\Delta_N(a)=\det\bigl(a_j^{\,\lambda_i+N-i}\bigr)_{1\le i,j\le N}
\end{equation}
defines the Schur polynomial $s_\lambda$ \cite[Chapter~I, (3.1)]{Macdonald1995}. We need only
two of them: $e_1=s_{(1)}$, and, by Pieri's rule $e_1e_2=s_{(1)}s_{(1,1)}=s_{(2,1)}+s_{(1,1,1)}$
together with $e_3=s_{(1,1,1)}$,
\begin{equation}\label{eq:H-schur}
H_N:=e_1e_2-e_3=s_{(2,1)}\qquad(N\ge2).
\end{equation}
The polynomial $H_N$ enters through the power-sum identity $p_3=e_1^3-3e_1e_2+3e_3$:
\begin{equation}\label{eq:cubic-identity}
\Bigl(\sum_{j=1}^Na_j\Bigr)^3-\sum_{j=1}^Na_j^3=3\,H_N(a).
\end{equation}
For $N=2$ and $N=3$ it factors into linear forms, $H_2=a_1a_2(a_1+a_2)$ and
$H_3=(a_1+a_2)(a_1+a_3)(a_2+a_3)$; for $N\ge4$ it does not (Lemma~\ref{lem:no-linear-factor}).

\begin{lemma}[Parity of bialternants]\label{lem:bialternant-parity}
Let $\lambda$ be a partition with at most $N$ parts and let
$o(\lambda):=\#\{i\in[N]:\ \lambda_i+N-i\ \text{is odd}\}$. Then every monomial of
$s_\lambda\Delta_N$ has a parity class of cardinality $o(\lambda)$. In particular, with
$m:=\lfloor N/2\rfloor$,
\begin{equation}\label{eq:sizes}
\begin{aligned}
|S|&=\begin{cases} m-1,& N\ \text{even},\\ m+1,& N\ \text{odd},\end{cases}&&\text{for }S\in\supp(e_1\Delta_N),\\[2pt]
|S|&=\begin{cases} m+1,& N\ \text{even},\\ m-1,& N\ \text{odd},\end{cases}&&\text{for }S\in\supp(H_N\Delta_N),
\end{aligned}
\end{equation}
so that $\supp(e_1\Delta_N)\cap\supp(H_N\Delta_N)=\emptyset$ for every $N\ge2$. Moreover,
$\supp(e_1\Delta_3)$ consists of the three two-element subsets of $[3]$, and
$\supp(e_1\Delta_4)$ of the four one-element subsets of $[4]$.
\end{lemma}

\begin{proof}
Expanding the determinant in \eqref{eq:bialternant} gives
$\sum_{\tau}\operatorname{sgn}(\tau)\prod_{j}a_j^{\,\epsilon_{\tau(j)}}$, where
$\epsilon_i:=\lambda_i+N-i$ and $\tau$ runs over the permutations of $[N]$. The exponents
$\epsilon_1>\epsilon_2>\dots>\epsilon_N$ are pairwise distinct, so the $N!$ monomials are
pairwise distinct, no cancellation occurs, and each monomial has exactly $o(\lambda)$ odd
exponents. For $\lambda=(1)$ the exponent set is $\{N\}\cup\{0,1,\dots,N-2\}$, and for
$\lambda=(2,1)$ it is $\{N+1,N-1\}\cup\{0,1,\dots,N-3\}$; counting the odd elements gives
\eqref{eq:sizes}. For $N=3$ and $\lambda=(1)$ the exponent set $\{3,1,0\}$ places its two odd
exponents on an arbitrary pair of variables, and for $N=4$ the exponent set $\{4,2,1,0\}$ places
its single odd exponent on an arbitrary variable.
\end{proof}

\section{The families and their on-shell reduction}\label{sec:reduction}

\subsection{Setting}\label{subsec:setting}
Fix $s\ge2$, write the independent variables as $t=(x,t_2,\dots,t_s)$ and the momenta as
$\bp=(k,p_2,\dots,p_s)$. Let
\begin{equation}\label{eq:LLam}
\begin{aligned}
\Lam&=\lambda_1D_x+\lambda_2D_{t_2}+\dots+\lambda_sD_{t_s},\qquad (\lambda_2,\dots,\lambda_s)\neq0,\\
L&=\mu_1D_x+\mu_2D_{t_2}+\dots+\mu_sD_{t_s}\neq0,
\end{aligned}
\end{equation}
be linear forms in the Hirota operators, put $Q:=D_x^3+\Lam$, and consider
\begin{equation}\label{eq:family}
P_{L,\Lam,n}(D)\,f\cdot f=0,\qquad P_{L,\Lam,n}:=L\,Q^{2n+1},\qquad n=0,1,2,\dots
\end{equation}
Since $L$ and $Q$ are odd, $P_{L,\Lam,n}$ is even, and $P_{L,\Lam,n}(0)=0$. For $L=D_x$ and
$\Lam=\alpha_1D_t+\alpha_2D_y$ this is the family $D_x(D_x^3+\alpha_1D_t+\alpha_2D_y)^{2n+1}$ of
\cite{GursesPekcan2025} (they write $k$ for our $n$). For $n=0$ the class \eqref{eq:family}
contains the KdV equation ($L=D_x$, $\Lam=D_t$), Ito's equation $D_t(D_t+D_x^3)f\cdot f=0$
\cite{Ito1980} ($L=\Lam=D_t$) and the $(2+1)$-dimensional equation of Boiti, Leon, Manna and
Pempinelli \cite{BLMP1986}, whose bilinear form is $D_y(D_x^3+D_t)f\cdot f=0$ ($L=D_y$, $\Lam=D_t$).

The dispersion variety of \eqref{eq:family} is reducible,
\begin{equation}\label{eq:components}
V(P_{L,\Lam,n})=V_L\cup V_Q,\qquad V_L:=\{\bp:\ L(\bp)=0\},\qquad V_Q:=\{\bp:\ k^3+\Lam(\bp)=0\}.
\end{equation}
Here $V_L$ is a hyperplane, while $V_Q$ is an irreducible cubic hypersurface: choosing $j\ge2$
with $\lambda_j\neq0$, it is the graph of a polynomial function of the remaining coordinates.
We call $V_Q$ the \emph{principal component}; it does not depend on $L$ or $n$, and for $n\ge1$
it is the component along which $P_{L,\Lam,n}$ vanishes to order $2n+1$.

\subsection{On-shell values}\label{subsec:onshell}
\begin{lemma}\label{lem:onshell}
Let $\bp_1,\dots,\bp_N\in V_Q$, let $k_j$ be the $x$-component of $\bp_j$ and put
$\ell_j:=L(\bp_j)$. For $\sg\in\{\pm1\}^N$ write $a=\sg\cdot k$ and $b=\sg\cdot\ell$, i.e.\
$a_j=\sg_jk_j$ and $b_j=\sg_j\ell_j$. Then
\begin{align}
Q\Bigl(\sum_j\sg_j\bp_j\Bigr)&=\Bigl(\sum_ja_j\Bigr)^3-\sum_ja_j^3=3H_N(a),&
Q(\sg_i\bp_i-\sg_j\bp_j)&=-3a_ia_j(a_i-a_j),\label{eq:Q-onshell}\\
L\Bigl(\sum_j\sg_j\bp_j\Bigr)&=e_1(b),&
L(\sg_i\bp_i-\sg_j\bp_j)&=b_i-b_j.\label{eq:L-onshell}
\end{align}
\end{lemma}

\begin{proof}
$\Lam$ is linear and $\Lam(\bp_j)=-k_j^3$ on $V_Q$, so $\Lam(\sum_j\sg_j\bp_j)=-\sum_j\sg_jk_j^3=-\sum_ja_j^3$,
while the $x$-component of $\sum_j\sg_j\bp_j$ is $\sum_ja_j$; now use \eqref{eq:cubic-identity}.
The second formula is the case $N=2$ with signs $(\sg_i,-\sg_j)$:
$(a_i-a_j)^3-a_i^3+a_j^3=-3a_ia_j(a_i-a_j)$. The formulas for $L$ express its linearity.
\end{proof}

Note that the $D_x$-component $\lambda_1$ of $\Lam$ plays no role: it cancels on $V_Q$ because it
is linear in the momenta, like the dispersion relation itself. For $\bp_i,\bp_j\in V_Q$ we get
$P_{L,\Lam,n}(\bp_i\pm\bp_j)=(\ell_i\pm\ell_j)\bigl(\pm3k_ik_j(k_i\pm k_j)\bigr)^{2n+1}$, hence the
phase factor
\begin{equation}\label{eq:phase}
e^{A_{ij}}=-\frac{P_{L,\Lam,n}(\bp_i-\bp_j)}{P_{L,\Lam,n}(\bp_i+\bp_j)}
=\frac{\ell_i-\ell_j}{\ell_i+\ell_j}\Bigl(\frac{k_i-k_j}{k_i+k_j}\Bigr)^{2n+1}.
\end{equation}
For $L=D_x$ this is the factor $\bigl((k_i-k_j)/(k_i+k_j)\bigr)^{2n+2}$ of
\cite{GursesPekcan2025}; for Ito's equation ($\ell=-k^3$, $n=0$) it is the phase factor
$\frac{(k_i^3-k_j^3)(k_i-k_j)}{(k_i^3+k_j^3)(k_i+k_j)}$ of the Sawada--Kotera type. Data on $V_Q$
with $k_i\neq0$, $k_i\neq\pm k_j$ and $\ell_i\neq\pm\ell_j$ are nondegenerate.

\begin{proposition}[On-shell reduction]\label{prop:reduction}
On $V_Q^{\,N}$,
\begin{equation}\label{eq:reduction}
S_N(\bp_1,\dots,\bp_N)=c_{N,n}\Bigl(\prod_{j=1}^Nk_j\Bigr)^{(2n+1)(N-1)}\,\Sigma_{N,n}(k,\ell),
\qquad c_{N,n}:=3^{2n+1}(-3)^{(2n+1)\binom N2},
\end{equation}
where
\begin{equation}\label{eq:Sigma}
\Sigma_{N,n}(k,\ell):=\sum_{\sg\in\{\pm1\}^N}e_1(\sg\cdot\ell)\,\Delta_N(\sg\cdot\ell)\,
\bigl[H_N(\sg\cdot k)\,\Delta_N(\sg\cdot k)\bigr]^{2n+1}.
\end{equation}
In particular $S_N$ depends on $\bp_1,\dots,\bp_N$ only through the pairs $(k_j,\ell_j)$.
\end{proposition}

\begin{proof}
Insert Lemma~\ref{lem:onshell} into \eqref{eq:SN}. The leading factor is
$P(\sum_j\sg_j\bp_j)=e_1(b)\,(3H_N(a))^{2n+1}$, and the pair factors are
\[
\sg_i\sg_j\,(b_i-b_j)\,\bigl(-3a_ia_j(a_i-a_j)\bigr)^{2n+1}.
\]
Since $\sg_i\sg_j(a_ia_j)^{2n+1}=k_ik_j(a_ia_j)^{2n}=(k_ik_j)^{2n+1}$ and
$\prod_{i<j}k_ik_j=(\prod_jk_j)^{N-1}$, the product of the pair factors equals
\[
(-3)^{(2n+1)\binom N2}\Bigl(\prod_{j}k_j\Bigr)^{(2n+1)(N-1)}\,\Delta_N(b)\,\Delta_N(a)^{2n+1}.\qedhere
\]
\end{proof}

For $L=D_x$ one has $\ell=k$, and \eqref{eq:reduction} together with Lemma~\ref{lem:parity}(a)
reads $S_4=16\cdot3^{7(2n+1)}(k_1k_2k_3k_4)^{3(2n+1)}\,\Pev\bigl[e_1H_4^{2n+1}\Delta_4^{2n+2}\bigr]$,
the form in which the four-soliton condition of the G\"urses--Pekcan family is analysed in
Section~\ref{sec:structure}.

\begin{remark}[Two regimes]\label{rem:regimes}
On $V_Q$ the pair $(k,\ell)=(k,L(\bp))$ behaves in one of two ways.
\begin{enumerate}
\item[(i)] If $L\notin\operatorname{span}(D_x,\Lam)$, the linear forms $D_x$, $\Lam$, $L$ are
linearly independent, so for every $(k,\ell)\in\C^2$ the system ``$x$-component $=k$,
$\Lam(\bp)=-k^3$, $L(\bp)=\ell$'' has a solution $\bp\in V_Q$. Hence $S_N$ vanishes identically
on $V_Q^{\,N}$ if and only if $\Sigma_{N,n}$ is the zero polynomial in the $2N$ independent
variables $k_1,\dots,k_N,\ell_1,\dots,\ell_N$, which Lemma~\ref{lem:parity}(c) decides in terms
of parity supports.
\item[(ii)] If $L=\beta D_x+\gamma\Lam$ with $(\beta,\gamma)\neq(0,0)$, then $\ell=\phi(k)$ with
the odd polynomial $\phi(k):=\beta k-\gamma k^3$, and by Lemma~\ref{lem:parity}(d)
\begin{equation}\label{eq:regime-ii}
\Sigma_{N,n}(k,\phi(k))=2^N\,\Pev\bigl[e_1(\phi(a))\,\Delta_N(\phi(a))\,\bigl(H_N(a)\Delta_N(a)\bigr)^{2n+1}\bigr]\Big|_{a=k},
\end{equation}
where $\phi(a)=(\phi(a_1),\dots,\phi(a_N))$. The G\"urses--Pekcan family is the case
$(\beta,\gamma)=(1,0)$ and Ito's family $D_t(D_x^3+D_t)^{2n+1}$ the case $(\beta,\gamma)=(0,1)$.
\end{enumerate}
\end{remark}

\section{Hirota integrability of \texorpdfstring{$L\,(D_x^3+\Lambda)$}{L(Dx3+Lambda)}}\label{sec:I}

\begin{theorem}[Theorem~I]\label{thm:I}
Let $L$ and $\Lam$ be as in \eqref{eq:LLam}. Then $P=L\,(D_x^3+\Lam)$ satisfies the $N$-soliton
condition on the principal component $V_Q$ for every $N\ge3$. Consequently $P$ is Hirota
integrable on $V_Q$: for every $N$ and all nondegenerate $\bp_1,\dots,\bp_N\in V_Q$ the function
\eqref{eq:ansatz} with the phase factors
\begin{equation}\label{eq:phase-I}
e^{A_{ij}}=\frac{(\ell_i-\ell_j)(k_i-k_j)}{(\ell_i+\ell_j)(k_i+k_j)},\qquad \ell_j=L(\bp_j),
\end{equation}
is an $N$-soliton solution of $P(D)f\cdot f=0$.
\end{theorem}

\begin{proof}
By Proposition~\ref{prop:reduction} with $n=0$, on $V_Q^{\,N}$ the polynomial $S_N$ is a
nonzero constant multiple of
\[
(k_1\cdots k_N)^{N-1}\,\Sigma_{N,0}(k,\ell),\qquad
\Sigma_{N,0}(k,\ell)=\sum_\sg F(\sg\cdot\ell)\,G(\sg\cdot k),
\]
where $F:=e_1\Delta_N$ and $G:=H_N\Delta_N=s_{(2,1)}\Delta_N$.
It suffices to show that $\Sigma_{N,0}$ is the zero
polynomial in $2N$ independent indeterminates: this settles regime (i) of
Remark~\ref{rem:regimes} directly and regime (ii) by specialization. By
Lemma~\ref{lem:parity}(c) this is equivalent to $\supp F\cap\supp G=\emptyset$, which is
Lemma~\ref{lem:bialternant-parity}. The phase factors are \eqref{eq:phase} with $n=0$.
\end{proof}

\begin{remark}\label{rem:I-scope}
(i) For $s=2$ the transformations $x\mapsto x+ct$ and the scalings of $x$ and $t$ preserve the
shape of \eqref{eq:family} and reduce the class with $n=0$ to the KdV equation ($L\propto D_x$)
and to the one-parameter family $D_x^3D_t+\lambda D_xD_t+D_t^2$, $\lambda\in\C$, of which
Ito's equation is the member $\lambda=0$ and whose phase factor \eqref{eq:phase-I} reads
$\frac{(k_i-k_j)^2(k_i^2+k_ik_j+k_j^2+\lambda)}{(k_i+k_j)^2(k_i^2-k_ik_j+k_j^2+\lambda)}$. The
$N$-soliton solutions of the KdV and Ito equations are classical \cite{Hirota1971,Ito1980}, and
for $L=D_x$, $\Lam=\alpha_1D_t+\alpha_2D_y$ they were obtained in \cite{Ma2021,Ma2022}. The
point of Theorem~\ref{thm:I} is the uniform mechanism: the $N$-soliton conditions of the whole
class are the disjointness of the parity supports of two bialternants, and no induction on $N$
is needed.
(ii) The phase factor \eqref{eq:phase-I} is the product of a KdV-type factor in the wavenumbers
$k$ and a KdV-type factor in the values $\ell=L(\bp)$; for $L\notin\operatorname{span}(D_x,\Lam)$
the $\ell_j$ are free parameters, so that the phase factors of the $N$-soliton solutions of
Theorem~\ref{thm:I} depend on $2N$ free parameters; for $s=3$ these exhaust the momenta, while
for $s>3$ each momentum may in addition vary in the common kernel of $D_x$, $L$, $\Lam$, which
affects the exponents $\theta_j$ but not the phase factors.
(iii) The vanishing of $\Sigma_{N,0}$ for independent $(k,\ell)$ was verified numerically for
$N\le7$ and the parity supports were computed for $N\le8$; see the ancillary files.
\end{remark}

\section{The three-soliton condition for all \texorpdfstring{$n$}{n} and the classification of the multiplicity mechanism}\label{sec:II-III}

\begin{theorem}[Theorem~II]\label{thm:II}
For every $n\ge0$ and all $L$, $\Lam$ as in \eqref{eq:LLam}, the equation \eqref{eq:family}
satisfies the three-soliton condition on the principal component $V_Q$. Its three-soliton
solutions have the phase factors \eqref{eq:phase}.
\end{theorem}

\begin{proof}
By Proposition~\ref{prop:reduction} it suffices to show that
$\Sigma_{3,n}(k,\ell)=\sum_\sg F(\sg\cdot\ell)\,G_n(\sg\cdot k)$, with $F:=e_1\Delta_3$ and
$G_n:=(H_3\Delta_3)^{2n+1}$, vanishes for independent indeterminates $k,\ell$. Now
\begin{equation}\label{eq:H3D3}
H_3\Delta_3=\prod_{1\le i<j\le3}(a_i+a_j)(a_i-a_j)=\prod_{1\le i<j\le3}(a_i^2-a_j^2)
\end{equation}
is even in each variable, so $\supp G_n=\{\emptyset\}$ for every $n$, whereas $\supp F$
consists of the three two-element subsets of $[3]$ (Lemma~\ref{lem:bialternant-parity}). The
supports are disjoint and Lemma~\ref{lem:parity}(c) applies.
\end{proof}

\begin{remark}\label{rem:II}
(i) For $L=D_x$ Theorem~\ref{thm:II} is Lemma~2.2 of \cite{GursesPekcan2025}. The proof shows
where the multiplicity $2n+1$ is harmless: the on-shell three-soliton factor \eqref{eq:H3D3}
of $Q$ is a polynomial in the squares $a_i^2$, so all its odd powers have the same parity
support. This rests on the factorization $H_3=(a_1+a_2)(a_1+a_3)(a_2+a_3)$ of the cubic
identity \eqref{eq:cubic-identity}, i.e.\ on
$(a+b+c)^3-a^3-b^3-c^3=3(a+b)(b+c)(c+a)$.
(ii) The analogous statement is false for the Sawada--Kotera-type family
$D_x(D_x^5+D_t)^{2n+1}$: there the on-shell factors of $D_x^5+D_t$ are not even in the
variables, and the three-soliton condition already fails for $n=1$ (exact computation at random
rational points; see the ancillary files). The following theorem shows that the cubic is the
only odd polynomial for which the mechanism works.
\end{remark}

\begin{theorem}[Classification of the multiplicity mechanism, one variable]\label{thm:classification}
Let $q\in\C[z]$ be an odd polynomial, and for $n\ge0$ let $P_n:=D_x\bigl(q(D_x)+D_t\bigr)^{2n+1}$
and $V_q:=\{(k,\omega):\ \omega=-q(k)\}$. The following are equivalent:
\begin{enumerate}
\item[(a)] $P_n$ satisfies the three-soliton condition on $V_q$ for every $n\ge0$;
\item[(b)] $P_n$ satisfies the three-soliton condition on $V_q$ for some $n\ge1$;
\item[(c)] $q(z)=\alpha z^3+\beta z$ for some $\alpha,\beta\in\C$.
\end{enumerate}
Moreover, if $\deg q\ge7$ then $P_0$ already fails the three-soliton condition on $V_q$.
For $\alpha\neq0$ the family in (c) is \eqref{eq:family} with $L=D_x$ and
$\Lam=(\beta/\alpha)D_x+\alpha^{-1}D_t$, up to the constant factor $\alpha^{2n+1}$; for $\alpha=0$
the component $V_q$ is a line on which all pair factors $P_n(\bp_i\pm\bp_j)$ vanish, so that the
conditions hold trivially and all phase factors are indeterminate.
\end{theorem}

\begin{proof}
Put $C(a,b,c):=(a+b+c)(a-b)(a-c)(b-c)$ and
\begin{align*}
T_q(a,b,c):={}&\bigl[q(a+b+c)-q(a)-q(b)-q(c)\bigr]\bigl[q(a-b)-q(a)+q(b)\bigr]\\
&\times\bigl[q(a-c)-q(a)+q(c)\bigr]\bigl[q(b-c)-q(b)+q(c)\bigr].
\end{align*}
For $\bp_j=(k_j,-q(k_j))\in V_q$ and $a_j=\sg_jk_j$, oddness of $q$ gives
$P_n(\sum_j\sg_j\bp_j)=(a_1+a_2+a_3)\bigl[q(a_1+a_2+a_3)-\sum_jq(a_j)\bigr]^{2n+1}$ and
$P_n(\sg_i\bp_i-\sg_j\bp_j)=(a_i-a_j)\bigl[q(a_i-a_j)-q(a_i)+q(a_j)\bigr]^{2n+1}$, and
$\prod_{i<j}\sg_i\sg_j=1$; hence, with $m:=2n+1$,
\begin{equation}\label{eq:S3-CT}
S_3^{(n)}=\sum_{\sg\in\{\pm1\}^3}C(\sg\cdot k)\,T_q(\sg\cdot k)^{m}
=2\sum_{\sg\in\mathcal E}C(\sg\cdot k)\,T_q(\sg\cdot k)^{m},\qquad \mathcal E:=\{\sg:\ \sg_1=+1\},
\end{equation}
where the reduction to $\mathcal E$ uses that $C$ (four linear factors) and $T_q$ (four odd
factors) are invariant under $\sg\mapsto-\sg$.

(b)$\Rightarrow$(c). Let $\deg q=d\ge5$ with leading coefficient $\alpha_d\neq0$, and let
$n\ge1$, $m=2n+1\ge3$. The top homogeneous part of $T_q$ is $\alpha_d^4T_{z^d}$, so the top
homogeneous part of $S_3^{(n)}$ is $\alpha_d^{4m}S_3^{(n)}[z^d]$, the polynomial
\eqref{eq:S3-CT} for the monomial $q=z^d$; it suffices to show that the latter does not vanish
at $k=(1,2,3)$. Writing $B_d:=3^d-2^d-1$,
\begin{align*}
X_d&:=(2^d-2)(6^d-3^d-2^d-1),\qquad Y_d:=(5^d-3^d-2^d)(3^d-2^{d+1}+1),\\
Z_d&:=(4^d-3^d-1)(4^d-3^d-2^d+1),
\end{align*}
a direct evaluation of the four terms of \eqref{eq:S3-CT} for
$\sg=(+,+,+),(+,+,-),(+,-,+),(+,-,-)$ gives $C=-12,\,0,\,60,\,-48$ and
$T_{z^d}=B_d^2X_d,\ \ast,\ B_d^2Y_d,\ B_d^2Z_d$, hence
\begin{equation}\label{eq:certificate-123}
S_3^{(n)}[z^d](1,2,3)=24\,B_d^{2m}\bigl(-X_d^m+5Y_d^m-4Z_d^m\bigr).
\end{equation}
For odd $d\ge5$ the numbers $B_d,X_d,Y_d,Z_d$ are positive (each factor is of the form
$(a+b)^d-a^d-b^d>0$ or is checked directly: $3^d>2^{d+1}$ and $4^d-3^d>2^d-1$ for $d\ge5$), and
\begin{equation}\label{eq:certificate-ineq}
\begin{aligned}
X_d+4Z_d-5Y_d={}&4\cdot16^d-5\cdot15^d-7\cdot12^d+10\cdot10^d+9\cdot9^d-4\cdot8^d\\
&-4\cdot6^d-5\cdot5^d-11\cdot4^d+7\cdot3^d+10\cdot2^d-2
\end{aligned}
\end{equation}
is $0$ for $d=5$ (indeed $X_5=225000$, $Y_5=513000$, $Z_5=585000$), equals $101606400>0$ for
$d=7$, and is positive for $d\ge8$: dropping the positive terms $10\cdot10^d+9\cdot9^d+7\cdot3^d+10\cdot2^d$
gives a lower bound $F_d$ with $F_d/16^d$ increasing in $d$ and $F_8=1279042836>0$. Since
$t\mapsto t^m$ is increasing and strictly convex on $t>0$,
\[
Y_d^m\le\Bigl(\frac{X_d+4Z_d}{5}\Bigr)^m\le\frac{X_d^m+4Z_d^m}{5},
\]
where the first inequality is strict for $d\ge7$ and the second is strict for $d=5$ because
$X_5\neq Z_5$. Hence $-X_d^m+5Y_d^m-4Z_d^m<0$, the top homogeneous part of $S_3^{(n)}$ is nonzero,
and $P_n$ fails the 3SC. The same computation with $m=1$ gives
$S_3^{(0)}[z^d](1,2,3)=-24B_d^2(X_d+4Z_d-5Y_d)<0$ for $d\ge7$, which proves the additional
assertion (for $d=5$ this certificate vanishes, in accordance with the fact that the
Sawada--Kotera equation $D_x(D_x^5+D_t)f\cdot f=0$ satisfies the three-soliton condition; see
Theorem~\ref{thm:quintic}).

(c)$\Rightarrow$(a). For $\alpha\neq0$,
$P_n=\alpha^{2n+1}D_x\bigl(D_x^3+(\beta/\alpha)D_x+\alpha^{-1}D_t\bigr)^{2n+1}$ is a member of
\eqref{eq:family} and Theorem~\ref{thm:II} applies; for $\alpha=0$ every pair factor vanishes on
the line $V_q$. Finally (a)$\Rightarrow$(b) is trivial.
\end{proof}

\begin{remark}\label{rem:classification}
The classification concerns the specific families with the linear factor $D_x$ and a
one-variable dispersion; Theorem~\ref{thm:classification-multi} extends it to several variables
and an arbitrary linear factor, still with a single multiplicity. As statements about
multiplicities $2n+1\ge3$ these theorems do not bear on Conjecture~\ref{conj:repair}; the
multiplicity-one members are treated in Theorem~\ref{thm:mult-one} and
Theorem~\ref{thm:quintic}.
\end{remark}

\begin{theorem}[Classification in several variables, any fixed multiplicity]\label{thm:classification-multi}
Let $q\in\C[z_1,\dots,z_r]$ be an odd polynomial ($q(-\mathbf z)=-q(\mathbf z)$), let $L$ be a
nonzero linear form in $D_{x_1},\dots,D_{x_r},D_t$, and put $Q:=q(D_{x_1},\dots,D_{x_r})+D_t$,
$P_n:=L\,Q^{2n+1}$ and $V_q:=\{(\mathbf k,\omega):\ \omega=-q(\mathbf k)\}$. The following are
equivalent:
\begin{enumerate}
\item[(a)] $P_n$ satisfies the three-soliton condition on $V_q$ for every $n\ge0$;
\item[(b)] $P_n$ satisfies the three-soliton condition on $V_q$ for some $n\ge1$;
\item[(c)] $q(\mathbf z)=c\,\ell(\mathbf z)^3+\lambda(\mathbf z)$ for a constant $c$ and linear forms
$\ell,\lambda$.
\end{enumerate}
If $c\ell^3\neq0$, then $V_q$ is an irreducible non-linear hypersurface and nondegenerate pair
data exist on a dense open subset; if $c\ell^3=0$, then $V_q$ is a hyperplane on which all pair
factors vanish.
\end{theorem}

\begin{proof}
Write $L(\mathbf k,\omega)=\boldsymbol\alpha\cdot\mathbf k+\beta\omega$ and
$h(\mathbf k):=\boldsymbol\alpha\cdot\mathbf k-\beta q(\mathbf k)$, so that $L=h(\mathbf k)$ on
$V_q$; $h$ is an odd polynomial. If $q$ is linear, (c) holds; assume $q$ nonlinear, so that $h$
is nonzero and nonconstant. For three independent $r$-tuples $\mathbf a,\mathbf b,\mathbf c$ put
\begin{align*}
C_h(\mathbf a,\mathbf b,\mathbf c)&:=\bigl[h(\mathbf a)+h(\mathbf b)+h(\mathbf c)\bigr]\bigl[h(\mathbf a)-h(\mathbf b)\bigr]\bigl[h(\mathbf a)-h(\mathbf c)\bigr]\bigl[h(\mathbf b)-h(\mathbf c)\bigr],\\
T_q(\mathbf a,\mathbf b,\mathbf c)&:=\bigl[q(\mathbf a+\mathbf b+\mathbf c)-q(\mathbf a)-q(\mathbf b)-q(\mathbf c)\bigr]
\bigl[q(\mathbf a-\mathbf b)-q(\mathbf a)+q(\mathbf b)\bigr]\\
&\qquad\times\bigl[q(\mathbf a-\mathbf c)-q(\mathbf a)+q(\mathbf c)\bigr]\bigl[q(\mathbf b-\mathbf c)-q(\mathbf b)+q(\mathbf c)\bigr].
\end{align*}
Exactly as in \eqref{eq:S3-CT}, with $m=2n+1$,
\begin{equation}\label{eq:S3-multi}
S_3^{(n)}=2\sum_{\sg\in\mathcal E}C_h(\sg\cdot\mathbf k)\,T_q(\sg\cdot\mathbf k)^{m},
\end{equation}
where $\sg$ acts on the three tuples. Let $d:=\deg q$ (odd, $\ge3$) and let $F$ be the top
homogeneous part of $q$.

\emph{Step 1: $d=3$.} Suppose $d\ge5$ and $n\ge1$. Restrict \eqref{eq:S3-multi} to the line
$\mathbf k_j=a_jv$, $j=1,2,3$, with $v$ fixed and $F(v)\neq0$; on this line the top homogeneous
part of $S_3^{(n)}$ in $v$ is obtained by replacing $q$ by $F$ and $h$ by its top homogeneous
part, and $F(a_jv)=a_j^dF(v)$. If $\beta=0$, then $h=\boldsymbol\alpha\cdot\mathbf k$ is linear
and the top part on the line equals $l(v)^4F(v)^{4m}\,S_3^{(n)}[z^d](a_1,a_2,a_3)$ with
$l(v):=\boldsymbol\alpha\cdot v$, the polynomial \eqref{eq:S3-CT} for the monomial $z^d$; choosing
$v$ with $l(v)\neq0$ and $(a_1,a_2,a_3)=(1,2,3)$, it is nonzero by \eqref{eq:certificate-123}.
If $\beta\neq0$, the top part of $h$ is $-\beta F$, and the top part on the line equals
$2\beta^4F(v)^{4+4m}\sum_{\sg\in\mathcal E}C(\sg\cdot(a_1^d,a_2^d,a_3^d))\,T_{z^d}(\sg\cdot a)^m$
with $C(a,b,c)=(a+b+c)(a-b)(a-c)(b-c)$. At $(a_1,a_2,a_3)=(1,2,3)$ write $A:=2^d$, $B:=3^d$,
$E:=B-A-1$; the four weights $w_\sg:=C(\sg\cdot(1,A,B))$ and values $t_\sg:=T_{z^d}(\sg\cdot(1,2,3))$
are, for $\sg=(+,+,+),(+,+,-),(+,-,+),(+,-,-)$,
\begin{align*}
w_1&=-(1+A+B)(A-1)(B-1)(B-A),& t_1&=E^2(A-2)(6^d-B-A-1),\\
w_2&=(B-A-1)(A-1)(B+1)(A+B),& t_2&=E(A-2)(4^d-B-1)(5^d-B-A),\\
w_3&=(B-A+1)(A+1)(B-1)(A+B),& t_3&=E^2(5^d-B-A)(B-2A+1),\\
w_4&=-(A+B-1)(A+1)(B+1)(B-A),& t_4&=E^2(4^d-B-1)(4^d-B-A+1).
\end{align*}
For odd $d\ge5$ all $t_j>0$ and $w_1,w_4<0<w_2,w_3$. Moreover, with $D:=B-A$, $S:=A+B$,
$w_2/|w_4|=\frac{A-1}{A+1}\cdot\frac{S(D-1)}{(S-1)D}<1$; $w_3-|w_4|=2A(A^2-1)$, whence
$w_3/|w_4|<1+\frac{2(A/B)^2}{B(1-A/B)}\le1+\frac{1}{21B}<\frac{101}{100}$; and
$t_2/t_4=\frac{(A-2)(5^d-B-A)}{E(4^d-B-A+1)}$, $t_3/t_4=\frac{(5^d-B-A)(B-2A+1)}{(4^d-B-1)(4^d-B-A+1)}$
satisfy $t_2/t_4<3/5$ and $t_3/t_4<9/10$: for $d=5$ they equal $19/35$ and $57/65$, and for
$d\ge7$ they are bounded by
$U_2(d):=\frac{(5/6)^d}{[1-(2/3)^d-3^{-d}][1-(3/4)^d-2^{-d}]}$ and
$U_3(d):=\frac{(15/16)^d}{[1-(3/4)^d-4^{-d}][1-(3/4)^d-2^{-d}]}$, which decrease in $d$ and
satisfy $U_2(7)=\frac{5000000}{14477001}<\frac35$, $U_3(7)=\frac{56953125}{66574508}<\frac9{10}$.
Hence, for $m\ge3$,
\[
\frac{w_2t_2^m+w_3t_3^m}{|w_4|t_4^m}<\Bigl(\frac35\Bigr)^m+\frac{101}{100}\Bigl(\frac9{10}\Bigr)^m
\le\frac{95229}{100000}<1,
\]
so that $\sum_\sg w_\sg t_\sg^m<0$. In both cases the top homogeneous part of $S_3^{(n)}$ on the
line is nonzero, $S_3^{(n)}\not\equiv0$, and (b) fails. Hence (b) forces $d=3$: $q=F+\lambda$
with $F$ a nonzero cubic form and $\lambda$ linear, and $T_q=T_F$.

\emph{Step 2: $F$ is the cube of a linear form.} Let $\mathcal C$ be the symmetric trilinear
polarization of $F$, choose $v$ with $A_0:=F(v)\neq0$ (and $l(v)\neq0$ if $\beta=0$), and let
$\mathbf z$ satisfy $DF(v)[\mathbf z]=3\mathcal C(v,v,\mathbf z)=0$, so that
$F(sv+t\mathbf z)=A_0s^3+3C_0st^2+D_0t^3$ with $C_0:=\mathcal C(v,\mathbf z,\mathbf z)$,
$D_0:=F(\mathbf z)$. Insert $\mathbf k_1=v$, $\mathbf k_2=2v$, $\mathbf k_3=3v+\varepsilon\mathbf z$
into \eqref{eq:S3-multi}. In each of the four factors of $T_F$ the terms in $D_0$ cancel and no
odd power of $\varepsilon$ occurs, and a direct multiplication gives
\[
T_F(\sg\cdot\mathbf k)=349920\,A_0^4+A_0^3C_0\,v_\sg\,\varepsilon^2+O(\varepsilon^4),\qquad
(v_\sg)_{\sg\in\mathcal E}=(192456,\,227448,\,169128,\,180792),
\]
where $349920=T_{z^3}(1,2,3)$. The weights $C_h(\sg\cdot\mathbf k)$ depend on $\varepsilon$ as
well, but $\sum_{\sg\in\mathcal E}C_h(\sg\cdot\mathbf k)=0$ identically in the three values
$h(\mathbf k_j)$ (this is the vanishing of the even part of $e_1\Delta_3$,
Lemma~\ref{lem:bialternant-parity}), so their variation contributes nothing to the coefficient
of $\varepsilon^2$ in \eqref{eq:S3-multi}, which therefore equals
$2m(349920A_0^4)^{m-1}A_0^3C_0\sum_{\sg\in\mathcal E}C_h(\sg\cdot\mathbf k)|_{\varepsilon=0}v_\sg$.
At $\varepsilon=0$ the weights are $l(v)^4(-12,0,60,-48)$ if $\beta=0$ and, for the top
homogeneous part in $(v,\mathbf z)$ when $\beta\neq0$, $\beta^4A_0^4(-124488,123480,163800,-162792)$;
the corresponding sums $\sum_\sg C_h v_\sg$ are $-839808\,l(v)^4$ and $2398491648\,\beta^4A_0^4$.
Since the condition (b) forces the coefficient of $\varepsilon^2$ to vanish identically in
$(v,\mathbf z)$ on the cone $DF(v)[\mathbf z]=0$, comparing top homogeneous parts gives
$C_0=\mathcal C(v,\mathbf z,\mathbf z)=0$ whenever $DF(v)[\mathbf z]=0$ and $F(v)\neq0$. For
arbitrary $\mathbf w$, the vector $\mathbf z:=\mathbf w-\mathcal C(v,v,\mathbf w)v/F(v)$ satisfies
$DF(v)[\mathbf z]=0$, and $\mathcal C(v,\mathbf z,\mathbf z)=0$ unfolds to
\begin{equation}\label{eq:polarization}
\mathcal C(v,v,\mathbf w)^2=F(v)\,\mathcal C(v,\mathbf w,\mathbf w)\qquad\text{for all }v,\mathbf w,
\end{equation}
a polynomial identity. Fix $v_0$ with $A_0:=F(v_0)\neq0$ and put $b(\mathbf w):=\mathcal C(v_0,v_0,\mathbf w)$,
$c(\mathbf w):=\mathcal C(v_0,\mathbf w,\mathbf w)$, $d(\mathbf w):=F(\mathbf w)$. Then
\eqref{eq:polarization} at $v=v_0$ gives $b^2=A_0c$, and at $v=v_0+s\mathbf w$ it reads
$(b+2cs+ds^2)^2=(A_0+3bs+3cs^2+ds^3)(c+ds)$, whose coefficient of $s$ gives $bc=A_0d$. Hence
$F(\mathbf w)=b(\mathbf w)^3/A_0^2$ with the nonzero linear form $b$, which is (c).

(c)$\Rightarrow$(a). If $c\ell^3\neq0$, choose $\ell$ as a coordinate $k$; then
$c^{-1}Q=k^3+c^{-1}(\lambda+D_t)$ is of the form \eqref{eq:LLam} and Theorem~\ref{thm:II}
applies to every linear $L$; $Q(\bp_i\pm\bp_j)$ and $L(\bp_i\pm\bp_j)=h(\mathbf k_i)\pm h(\mathbf k_j)$
are nonzero polynomials, so nondegenerate data form a dense open set. If $c\ell^3=0$, all pair
factors vanish on the hyperplane $V_q$. Finally (a)$\Rightarrow$(b) is trivial. The numerical
constants of Step~2 are verified in the ancillary files.
\end{proof}

The argument of Step~2 does not use $m\ge3$, and that of Step~1 extends to $m=1$ when
$\beta\neq0$. Together with the following one-variable result this leads to a complete
classification of the multiplicity-one members $LQ$ with graph-type dispersion, in which the
implication ``3SC $\Rightarrow$ all $N$SC'' of Conjecture~\ref{conj:repair} holds.

\begin{theorem}[One variable, $L=D_x$, multiplicity one]\label{thm:quintic}
Let $q\in\C[z]$ be an odd polynomial and $P_0:=D_x(q(D_x)+D_t)$, $V_q=\{\omega=-q(k)\}$. The
following are equivalent: (a) $P_0$ satisfies the 3SC on $V_q$; (b)
$q(z)=\alpha z^5+\beta z^3+\gamma z$ for some $\alpha,\beta,\gamma\in\C$; (c) $P_0$ satisfies the
$N$SC on $V_q$ for every $N\ge3$. For $(\alpha,\beta)\neq(0,0)$ the polynomial $P_0$ is
squarefree and $V_q$ is a non-hyperplane component of multiplicity one; the phase factors are
\begin{equation}\label{eq:quintic-phase}
e^{A_{ij}}=\frac{(k_i-k_j)^2}{(k_i+k_j)^2}\cdot\frac{5\alpha(k_i^2-k_ik_j+k_j^2)+3\beta}{5\alpha(k_i^2+k_ik_j+k_j^2)+3\beta}.
\end{equation}
\end{theorem}

\begin{proof}
(a)$\Rightarrow$(b) is the last assertion of Theorem~\ref{thm:classification}: for
$\deg q\ge7$ the 3SC fails already at $n=0$. (c)$\Rightarrow$(a) is trivial. For
(b)$\Rightarrow$(c): if $\alpha=0$, $\beta\neq0$, this is Theorem~\ref{thm:I}, and if
$\alpha=\beta=0$ all pair factors vanish; so let $\alpha\neq0$ and put $\lambda:=-\beta/(5\alpha)$.
We use the $N$-soliton solutions of the BKP equation
\begin{equation}\label{eq:BKP}
\bigl(D_X^6-5D_X^3D_Y-5D_Y^2+9D_XD_Z\bigr)\tau\cdot\tau=0
\end{equation}
\cite{DJKM1982,Hirota2004,Chang2024}: for any $2N$ numbers $p_i,q_i$, the function
\eqref{eq:ansatz} with exponents $\theta_i=(p_i+q_i)X+(p_i^3+q_i^3)Y+(p_i^5+q_i^5)Z+\delta_i$
and phase factors
\[
\mathcal A_{ij}=\frac{(p_i-p_j)(p_i-q_j)(q_i-p_j)(q_i-q_j)}{(p_i+p_j)(p_i+q_j)(q_i+p_j)(q_i+q_j)}
\]
solves \eqref{eq:BKP}; this is the Pfaffian $N$-soliton solution of BKP with a block-diagonal
constant matrix. For given $k_1,\dots,k_N$ let $p_i,q_i$ be the roots of
$z^2-k_iz+(k_i^2-\lambda)/3=0$, so that $p_i+q_i=k_i$, $p_i^3+q_i^3=\lambda k_i$ and
$p_i^5+q_i^5=(-k_i^5+5\lambda k_i^3+5\lambda^2k_i)/9$. Then $\tau$ depends on $X,Y$ only through
$X+\lambda Y$; writing $\tau=F_N(X+\lambda Y,Z)$, equation \eqref{eq:BKP} becomes
$(D_X^6-5\lambda D_X^4-5\lambda^2D_X^2+9D_XD_Z)F_N\cdot F_N=0$, and
$f_N(x,t):=F_N\bigl(x-(\gamma+5\alpha\lambda^2)t,\,9\alpha t\bigr)$ satisfies
$(\alpha D_x^6+\beta D_x^4+\gamma D_x^2+D_xD_t)f_N\cdot f_N=0$, i.e.\ $P_0(D)f_N\cdot f_N=0$, with
exponents $k_ix-q(k_i)t+\delta_i$. The auxiliary parameters are eliminated by the resultant
identities $(p_i-p_j)(p_i-q_j)(q_i-p_j)(q_i-q_j)=\frac{(k_i-k_j)^2}{9}(k_i^2-k_ik_j+k_j^2-3\lambda)$
and $(p_i+p_j)(p_i+q_j)(q_i+p_j)(q_i+q_j)=\frac{(k_i+k_j)^2}{9}(k_i^2+k_ik_j+k_j^2-3\lambda)$,
which show that $\mathcal A_{ij}$ equals \eqref{eq:quintic-phase}; and
\eqref{eq:quintic-phase} is exactly $-P_0(\bp_i-\bp_j)/P_0(\bp_i+\bp_j)$, because
$q(a+b)-q(a)-q(b)=ab(a+b)[5\alpha(a^2+ab+b^2)+3\beta]$. Hence $f_N$ is the $N$-soliton
solution \eqref{eq:ansatz} of $P_0$ for generic $k$, and by the criterion of
Section~\ref{subsec:NSC} all $S_M$ vanish there, hence identically on $V_q^{\,M}$. Squarefreeness
and the non-hyperplane property are clear for $(\alpha,\beta)\neq(0,0)$.
\end{proof}

\begin{theorem}[Multiplicity one: complete classification for graph-type dispersions]\label{thm:mult-one}
Let $q\in\C[z_1,\dots,z_r]$ be a nonlinear odd polynomial, let
$L=\boldsymbol\alpha\cdot D_{\mathbf x}+\beta D_t\neq0$, $Q:=q(D_{\mathbf x})+D_t$, $P:=LQ$, and
$l(\mathbf k):=\boldsymbol\alpha\cdot\mathbf k$. The following are equivalent:
\begin{enumerate}
\item[(a)] $P$ satisfies the 3SC on $V_q$;
\item[(b)] $P$ satisfies the $N$SC on $V_q$ for every $N\ge3$;
\item[(c)] $P$ satisfies the $N$SC on the whole dispersion variety $V_L\cup V_q$ for every $N\ge3$;
\item[(d)] either $q=c\,\ell^3+\lambda$ with $c\ell^3\neq0$ and $\ell,\lambda$ linear forms
($L$ arbitrary), or $\beta=0$ and $q=a\,l^5+b\,l^3+\lambda$ with $a\neq0$, $b\in\C$ and
$\lambda$ linear.
\end{enumerate}
In both cases of (d), $P$ is squarefree, $V_q$ is an irreducible non-hyperplane component of
multiplicity one, and nondegenerate data form a dense open subset of $V_q^{\,N}$.
\end{theorem}

\begin{proof}
(c)$\Rightarrow$(b)$\Rightarrow$(a) is trivial.

(a)$\Rightarrow$(d). Let $d:=\deg q$ and $F$ the top homogeneous part of $q$. If $\beta\neq0$,
Step~1 of the proof of Theorem~\ref{thm:classification-multi} applies with $m=1$: the exact
sums $\sum_\sg w_\sg t_\sg$ equal $305995802112000000$ for $d=5$ and
$-130111004944339449111244800$ for $d=7$, both nonzero, and for $d\ge9$ the bounds
$U_2(9)=\frac{100000000}{463816233}<\frac14$ and $U_3(9)=\frac{284765625}{434540404}<\frac23$
give $\frac{w_2t_2+w_3t_3}{|w_4|t_4}<\frac14+\frac{101}{100}\cdot\frac23<1$; hence $d=3$, and
Step~2 with $m=1$ shows that $F$ is the cube of a linear form. Let now $\beta=0$, so $L=l\neq0$
on $V_q$. On the line $\mathbf k_j=a_jv$ with $l(v)F(v)\neq0$ the top homogeneous part of $S_3$
is $l(v)^4F(v)^4$ times the polynomial \eqref{eq:S3-CT} with $m=1$ for the monomial $z^d$, which
is nonzero for $d\ge7$ by the last assertion of Theorem~\ref{thm:classification}; for $d=3$,
Step~2 with $m=1$ gives the cube. It remains to treat $d=5$: $q=F+G+\lambda$ with $F$ a quintic
form, $G$ a cubic form (possibly zero) and $\lambda$ linear; $T_q=T_{F+G}$. Since $C_l$ has degree
$4$ and $T_{F+G}$ splits into homogeneous parts of degrees $20,18,16,\dots$ according to the
number of factors $G$ taken, the 3SC forces the parts of $S_3$ of total degree $24$, $22$ and
$20$ to vanish separately. Take $v$ with $l(v)F(v)\neq0$ and $\mathbf z\in\ker l$, and insert
$\mathbf k_1=v$, $\mathbf k_2=2v$, $\mathbf k_3=3v+\varepsilon\mathbf z$; since $l(\mathbf z)=0$
the four weights $C_l(\sg\cdot\mathbf k)$ are the constants $l(v)^4(-12,0,60,-48)$. Writing
$F(sv+t\mathbf z)=As^5+5F_1s^4t+O(t^2)$ with $A=F(v)$, $5F_1=DF(v)[\mathbf z]$, a direct
multiplication of the four factors of $T_F$ gives, for the degree-$24$ part,
\[
[\varepsilon^1]\,S_3[F]=58786560000\,l(v)^4F(v)^3\,DF(v)[\mathbf z].
\]
Hence $DF(v)[\mathbf z]=0$ for all $v$ (a dense set, hence identically) and all
$\mathbf z\in\ker l$: $F$ is constant along $\ker l$, so $F=a\,l^5$ with $a\neq0$. Next let
$\mathcal C_G$ be the polarization of $G$ and write, for $\mathbf z\in\ker l$,
$G(sv+t\mathbf z)=Bs^3+3Cs^2t+3Dst^2+Et^3$ with $B=G(v)$, $C=\mathcal C_G(v,v,\mathbf z)$,
$D=\mathcal C_G(v,\mathbf z,\mathbf z)$, $E=G(\mathbf z)$, while $F(sv+t\mathbf z)=As^5$. The
same insertion gives, for the parts of degree $22$ and $20$ respectively,
\[
[\varepsilon^2]\,S_3^{(22)}=-10940832000\,l(v)^4A^3D,\qquad
[\varepsilon^2]\,S_3^{(20)}\big|_{D=E=0}=3401222400\,l(v)^4A^2C^2,
\]
where the first coefficient does not involve $B,C,E$ (ancillary files). Hence
$\mathcal C_G(v,\mathbf z,\mathbf z)=0$ for all $v$ and $\mathbf z\in\ker l$, so in coordinates
$x=l$, $y\in\ker l$ the cubic $G$ has no terms of degree $\ge2$ in $y$: $G=l^2\mu$ with $\mu$
linear. Then $D=E=0$ and the second certificate gives $C=0$, i.e.\
$0=3\mathcal C_G(v,v,\mathbf z)=DG(v)[\mathbf z]=l(v)^2\mu(\mathbf z)$ for $\mathbf z\in\ker l$; so
$\mu$ vanishes on $\ker l$, $\mu=b\,l$, and $G=b\,l^3$. This is the second case of (d).

(d)$\Rightarrow$(c). In the cubic case, Theorem~\ref{thm:I} (after choosing $\ell$ as a
coordinate) gives the $N$SC on $V_q$ and Theorem~\ref{thm:whole} extends them to the whole
variety. In the quintic case choose $l$ as the coordinate $k$ and replace $\omega$ by
$\omega'=\omega+\lambda(\mathbf k)$; then $P=D_x(aD_x^5+bD_x^3+D_{t'})$, the remaining space
momenta do not enter $P$, and every $S_M$ depends on the momenta only through $(k_j,\omega'_j)$;
so the $N$SC on $V_q$ are those of the one-variable equation, which hold by
Theorem~\ref{thm:quintic}. On $V_L=\{k=0\}$ two momenta are degenerate, and for one momentum
$\bp_0\in V_L\setminus V_q$ with $c:=Q(\bp_0)\neq0$ one has $P(\bp_i\pm\bp_0)=\pm k_ic$, so
$e^{A_{i0}}=1$ and $f_N=(1+e^{\theta_0})f_{N-1}$ with $\partial_x\theta_0=0$; since
$D_x^{2j}(hf)\cdot(hf)=h^2D_x^{2j}f\cdot f$ and $D_xD_t(hf)\cdot(hf)=h^2D_xD_tf\cdot f$ for
$x$-independent $h$, $f_N$ is again a solution, and the mixed conditions hold. The remaining
assertions are as in Theorem~\ref{thm:quintic}.
\end{proof}

Theorem~\ref{thm:mult-one} proves Conjecture~\ref{conj:repair} for the whole class of
equations $L(q(D_{\mathbf x})+D_t)f\cdot f=0$ with $q$ a nonlinear odd polynomial and $L$ any
nonzero linear form, and shows moreover that in this class the 3SC on the principal component
implies all $N$SC on the whole dispersion variety. The quintic members are reductions of the
classical BKP solitons; the point of the theorem is the exhaustive statement, and in particular
the necessity of the alignment of the quintic and cubic parts with the linear factor.

\section{The explicit residual, the boundary slice and the failure of the four-soliton condition}\label{sec:structure}

Throughout Sections~\ref{subsec:residual}--\ref{subsec:slice}, $L=\beta D_x+\gamma\Lam$ (regime (ii) of Remark~\ref{rem:regimes}),
$\phi(a)=\beta a-\gamma a^3$, and $\beta,\gamma$ are treated as indeterminates; the
specializations $(\beta,\gamma)=(1,0)$ and $(0,1)$ are the G\"urses--Pekcan family and Ito's
family $D_t(D_x^3+D_t)^{2n+1}$. Put $x_i:=k_i^2$ and
\begin{equation}\label{eq:En}
E_n^{(\beta,\gamma)}(x_1,\dots,x_4):=\Pev\bigl[e_1(\phi(a))\,\Delta_4(\phi(a))\,\bigl(H_4(a)\Delta_4(a)\bigr)^{2n+1}\bigr],
\end{equation}
where, as in Lemma~\ref{lem:parity}(d), $\phi(a)=(\phi(a_1),\dots,\phi(a_4))$ and the even
projection is written in the variables $x_i=a_i^2$. By \eqref{eq:reduction} and
\eqref{eq:regime-ii}, on $V_Q^{\,4}$,
\begin{equation}\label{eq:S4-En}
S_4=16\cdot3^{7(2n+1)}\,(k_1k_2k_3k_4)^{3(2n+1)}\,E_n^{(\beta,\gamma)}(k_1^2,k_2^2,k_3^2,k_4^2).
\end{equation}
For a four-variable expression we write $x_{\widehat j}$ for the triple obtained from
$(x_1,x_2,x_3,x_4)$ by deleting $x_j$ and keeping the order, so that
$\Delta_3(x_{\widehat j})=\prod_{u<v;\ u,v\neq j}(x_u-x_v)$; thus
\begin{equation}\label{eq:D3-product}
\prod_{j=1}^4\Delta_3(x_{\widehat j})=\prod_{1\le u<v\le4}(x_u-x_v)^2=\Delta_4(x)^2,
\end{equation}
because each factor $x_u-x_v$ occurs in exactly two of the four triples. Finally put
\begin{equation}\label{eq:Bh}
B_{uv}:=\beta^2-2\beta\gamma(x_u+x_v)+\gamma^2(x_u^2+x_ux_v+x_v^2),\qquad
h_i:=(\beta-\gamma x_i)\prod_{\substack{u<v\\ u,v\neq i}}B_{uv}\quad(i\in[4]).
\end{equation}

\subsection{The explicit residual}\label{subsec:residual}
\begin{theorem}[Explicit residual]\label{thm:residual}
For every $n\ge0$,
\begin{equation}\label{eq:structure}
E_n^{(\beta,\gamma)}=x_1x_2x_3x_4\,\Delta_4(x)^2\,R_n^{(\beta,\gamma)}(x_1,x_2,x_3,x_4),
\end{equation}
where $R_0^{(\beta,\gamma)}\equiv0$ and, for $n\ge1$,
\begin{equation}\label{eq:residual}
R_n^{(\beta,\gamma)}=-\sum_{\substack{r_1,r_2,r_3,r_4\ge0\\ r_1+r_2+r_3+r_4=n-1}}\ \sum_{i=1}^{4}
\frac{(2n+1)!}{(2r_i)!\prod_{j\neq i}(2r_j+1)!}\;x_i\,h_i\;\prod_{j=1}^{4}x_j^{\,n-1-r_j}\Delta_3(x_{\widehat j})^{2r_j}.
\end{equation}
The coefficients are multinomial coefficients $\binom{2n+1}{2r_i,\,2r_j+1,\,2r_k+1,\,2r_l+1}$,
hence positive integers, and all exponents are nonnegative.
\end{theorem}

\begin{proof}
By \eqref{eq:bialternant}, $e_1\Delta_4=\det(a_j^{\,\epsilon_i})$ with exponents
$(\epsilon_1,\dots,\epsilon_4)=(4,2,1,0)$ and $H_4\Delta_4=\det(a_j^{\,\epsilon_i})$ with
exponents $(5,3,1,0)$. Expanding the first determinant along its third row (the row of the
$a_j$) and the second along its fourth row (the row of ones), and observing that the minor of
the first with exponents $(4,2,0)$ in three variables is $\Delta_3$ in the squares, while the
minor of the second with exponents $(5,3,1)$ equals $\prod a_l$ times $\Delta_3$ in the squares,
we obtain the polynomial identities
\begin{align}
e_1(a)\Delta_4(a)&=\sum_{i=1}^4(-1)^{i+3}\,a_i\,\Delta_3(x_{\widehat i}),\label{eq:B1}\\
H_4(a)\Delta_4(a)&=-\sum_{i=1}^4(-1)^{i+3}\,\Delta_3(x_{\widehat i})\prod_{j\neq i}a_j.\label{eq:B2}
\end{align}
Applying \eqref{eq:B1} to the variables $\phi(a_j)=a_j(\beta-\gamma x_j)$, whose squares are
$y_j:=x_j(\beta-\gamma x_j)^2$, and using
$y_u-y_v=(x_u-x_v)\bigl[\beta^2-2\beta\gamma(x_u+x_v)+\gamma^2(x_u^2+x_ux_v+x_v^2)\bigr]=(x_u-x_v)B_{uv}$,
we get
\begin{equation}\label{eq:B3}
e_1(\phi(a))\,\Delta_4(\phi(a))=\sum_{i=1}^4(-1)^{i+3}\,a_i\,\Delta_3(x_{\widehat i})\,h_i .
\end{equation}
Now expand $(H_4\Delta_4)^{2n+1}$ by the multinomial theorem applied to \eqref{eq:B2}: a term is
indexed by $m=(m_1,\dots,m_4)$ with $\sum_jm_j=2n+1$, carries the coefficient
$-\binom{2n+1}{m_1,\dots,m_4}$, the factor $\prod_j\bigl[(-1)^{j+3}\Delta_3(x_{\widehat j})\bigr]^{m_j}$,
and the monomial $\prod_la_l^{\,2n+1-m_l}$, because $a_l$ occurs in every summand of
\eqref{eq:B2} except the $l$-th. Multiplying by the $i$-th term of \eqref{eq:B3} raises the
exponent of $a_i$ by one; the exponent of $a_l$ becomes $2n+1-m_l+\delta_{il}$, and all four
exponents are even if and only if $m_i$ is even and $m_j$ is odd for $j\neq i$, i.e.\
$m_i=2r_i$ and $m_j=2r_j+1$ ($j\ne i$) with $r_1+\dots+r_4=n-1$. These are exactly the terms
retained by $\Pev$. In such a term the signed Vandermonde factors combine, by
\eqref{eq:D3-product}, to
$\prod_j\bigl[(-1)^{j+3}\Delta_3(x_{\widehat j})\bigr]\cdot\prod_j\Delta_3(x_{\widehat j})^{2r_j}
=\Delta_4(x)^2\prod_j\Delta_3(x_{\widehat j})^{2r_j}$, and the monomial
$a_i^{2n+2-2r_i}\prod_{l\ne i}a_l^{\,2n-2r_l}$ becomes $x_1x_2x_3x_4\,x_i\prod_jx_j^{\,n-1-r_j}$.
Collecting the coefficient and the factor $h_i$ from \eqref{eq:B3} gives \eqref{eq:structure}
with \eqref{eq:residual}. For $n=0$ the conditions $m_j\ge1$ for the three indices $j\neq i$
are incompatible with $\sum_jm_j=1$, so no term survives and $E_0^{(\beta,\gamma)}=0$.
\end{proof}

\begin{corollary}[Structure theorem]\label{thm:structure}
$R_n^{(\beta,\gamma)}\in\Z[\beta,\gamma][x_1,\dots,x_4]$ is symmetric in $x_1,\dots,x_4$.
Writing $R_n^{(\beta,\gamma)}=\sum_{j=0}^{7}\beta^{7-j}\gamma^{j}R_{n,j}$, each $R_{n,j}$ is
homogeneous of degree $9n-8+j$ in $x$. On $V_Q$, the four-soliton condition holds for $n=0$
and, for $n\ge1$, fails if and only if $R_n^{(\beta,\gamma)}\neq0$ (as an identity of polynomials;
at a given nondegenerate configuration, $S_4\neq0$ if and only if $R_n^{(\beta,\gamma)}\neq0$ there).
\end{corollary}

\begin{proof}
Symmetry is visible in \eqref{eq:residual}, whose summation is over all compositions $r$ and all
$i$ and whose ingredients are permuted accordingly (the factors $\Delta_3(x_{\widehat j})$ enter
squared). The degree of the term with $\beta^{7-j}\gamma^{j}$ in $x_ih_i$ is $1+j$, and the
remaining factors have degree $\sum_j(n-1-r_j)+6\sum_jr_j=4(n-1)+5(n-1)=9n-9$; the total is
$9n-8+j$. The last assertion is \eqref{eq:S4-En}.
\end{proof}

\begin{remark}\label{rem:R0}
The vanishing $R_0^{(\beta,\gamma)}\equiv0$ is a second proof of the four-soliton condition for
the KdV and Ito equations and their mixtures (Theorem~\ref{thm:I} with $N=4$, regime (ii)):
it is the statement that no product of the four rows of \eqref{eq:B2} raised to the first
power has even exponents after multiplication by a row of \eqref{eq:B3}. The formula
\eqref{eq:residual} makes the divisibility by $x_1x_2x_3x_4\prod_{i<j}(x_i-x_j)^2$, which was
originally obtained from the vanishing of $E_n$ on the strata $\{x_i=x_j\}$ and $\{x_i=0\}$,
completely explicit.
\end{remark}

For $n=1$ the only composition is $r=0$, and \eqref{eq:residual} reads
$R_1^{(\beta,\gamma)}=-6\sum_{i=1}^4x_ih_i$; thus $R_1^{(1,0)}=-6\,(x_1+x_2+x_3+x_4)$ and
\begin{equation}\label{eq:R1Ito-formula}
R_1^{(0,1)}=6\sum_{i=1}^4x_i^2\prod_{\substack{u<v\\ u,v\neq i}}\bigl(x_u^2+x_ux_v+x_v^2\bigr),
\end{equation}
in agreement with \eqref{eq:R1Ito} below.

\subsection{The residuals for small \texorpdfstring{$n$}{n}}
Exact computation (see the ancillary files) gives the data of Table~\ref{tab:Rn} for the two
extreme specializations. The \emph{content} of $R_n$ is the signed greatest common divisor of
its integer coefficients, with the sign of the leading coefficient in lexicographic order; the
last column counts positive and negative coefficients.

\begin{table}[ht]
\centering
\small
\begin{tabular}{lcccc l}
\hline
$(\beta,\gamma)$ & $n$ & $\deg_x R_n$ & \#terms & content & remarks \\
\hline
$(1,0)$ & $1$ & $1$ & $4$ & $-6$ & $R_1=-6\,(x_1+x_2+x_3+x_4)$ \\
& $2$ & $10$ & $166$ & $-20$ & irreducible over $\mathbb{Q}$; $70$ positive, $96$ negative coefficients \\
& $3$ & $19$ & $952$ & $-14$ & irreducible over $\mathbb{Q}$; $532$ positive, $420$ negative coefficients \\
\hline
$(0,1)$ & $1$ & $8$ & $49$ & $6$ & all $49$ coefficients positive; see \eqref{eq:R1Ito} \\
& $2$ & $17$ & $540$ & $20$ & irreducible over $\mathbb{Q}$; $308$ positive, $232$ negative coefficients \\
& $3$ & $26$ & $1882$ & $14$ & irreducible over $\mathbb{Q}$; $846$ positive, $1036$ negative coefficients \\
\hline
\end{tabular}
\medskip
\caption{The residuals $R_n^{(\beta,\gamma)}$ for $n\le3$ in the G\"urses--Pekcan case
$(\beta,\gamma)=(1,0)$ and in Ito's case $(\beta,\gamma)=(0,1)$. For $(\beta,\gamma)=(1,0)$ and $n=1$,
\eqref{eq:structure} recovers \cite[eq.~(2.43)]{GursesPekcan2025} up to the factor $2$ of
Remark~\ref{rem:normalization}: the constant in \eqref{eq:S4-En} is $16\cdot3^{21}\cdot(-6)=-1004193907488$.
For $n=2$ and $n=3$ the corresponding constants are $16\cdot3^{35}\cdot(-20)=-16010094431679906240$
and $16\cdot3^{49}\cdot(-14)=-53603049747658326628178592$.}
\label{tab:Rn}
\end{table}

In terms of the monomial symmetric polynomials $m_\lambda$ of $x_1,\dots,x_4$,
\begin{equation}\label{eq:R1Ito}
R_1^{(0,1)}=6\,\bigl(2\,m_{(4,2,2)}+m_{(4,2,1,1)}+m_{(3,3,2)}+4\,m_{(3,2,2,1)}+12\,m_{(2,2,2,2)}\bigr).
\end{equation}

\subsection{The boundary slice}\label{subsec:slice}
In three variables we write $\eta_1,\eta_2,\eta_3$ for the elementary symmetric polynomials of
$(a_1,a_2,a_3)$, $E_1,E_2,E_3$ for those of $(x_1,x_2,x_3)$,
$\bar H:=H_3=\eta_1\eta_2-\eta_3$, $\bar\Delta:=\Delta_3(a_1,a_2,a_3)$, and
\begin{equation}\label{eq:U}
U:=(x_1-x_2)(x_1-x_3)(x_2-x_3)=\prod_{1\le i<j\le3}(a_i^2-a_j^2)=\bar\Delta\,\bar H,
\end{equation}
the last equality by \eqref{eq:H3D3}. Define the polynomial
\begin{equation}\label{eq:Theta}
\Theta_{\beta,\gamma}:=\gamma^4E_3\bigl(2E_1^2-3E_2\bigr)-\beta\gamma^3\bigl(E_1^2E_2-2E_2^2+7E_1E_3\bigr)
+3\beta^2\gamma^2\bigl(E_1E_2+3E_3\bigr)-6\beta^3\gamma E_2+\beta^4E_1,
\end{equation}
which does not depend on $n$.

\begin{theorem}[Boundary slice]\label{thm:slice}
For every $n\ge1$,
\begin{equation}\label{eq:slice}
R_n^{(\beta,\gamma)}(x_1,x_2,x_3,0)=2n(2n+1)\,(x_1x_2x_3)^{n-1}\prod_{1\le i<j\le3}(x_i-x_j)^{2(n-1)}
\prod_{i=1}^{3}(\gamma x_i-\beta)\;\Theta_{\beta,\gamma}(x_1,x_2,x_3),
\end{equation}
and $\Theta_{\beta,\gamma}$ has the cyclic form
\begin{equation}\label{eq:Theta-cyclic}
\Theta_{\beta,\gamma}=\sum_{i=1}^{3}x_i\,(\beta-\gamma x_j)(\beta-\gamma x_k)\,B_{jk},\qquad \{j,k\}=[3]\setminus\{i\}.
\end{equation}
\end{theorem}

\begin{proof}
Set $x_4=0$ in \eqref{eq:residual}. A term survives only if $n-1-r_4=0$, i.e.\ $r_4=n-1$ and
$r_1=r_2=r_3=0$, and the summand with $i=4$ is killed by its factor $x_4$. The coefficient is
$(2n+1)!/(0!\,1!\,1!\,(2n-1)!)=2n(2n+1)$, the product $\prod_jx_j^{n-1-r_j}\Delta_3(x_{\widehat j})^{2r_j}$
becomes $(x_1x_2x_3)^{n-1}\prod_{i<j\le3}(x_i-x_j)^{2(n-1)}$, and, for $i\le3$ with
$\{j,k\}=[3]\setminus\{i\}$, the factor $h_i$ at $x_4=0$ equals
$(\beta-\gamma x_i)\,B_{jk}\,(\beta-\gamma x_j)^2(\beta-\gamma x_k)^2$, since $B_{u4}|_{x_4=0}=(\beta-\gamma x_u)^2$.
Hence
\begin{align*}
R_n^{(\beta,\gamma)}(x_1,x_2,x_3,0)=-2n(2n+1)(x_1x_2x_3)^{n-1}&\prod_{i<j\le3}(x_i-x_j)^{2(n-1)}\\
&\times\sum_{i=1}^3x_i(\beta-\gamma x_i)(\beta-\gamma x_j)^2(\beta-\gamma x_k)^2B_{jk},
\end{align*}
and extracting $-\prod_{u=1}^3(\beta-\gamma x_u)=\prod_{u=1}^3(\gamma x_u-\beta)$ from each
summand leaves \eqref{eq:Theta-cyclic}. Expanding \eqref{eq:Theta-cyclic} in the elementary
symmetric polynomials gives \eqref{eq:Theta}.
\end{proof}

For $(\beta,\gamma)=(1,0)$ and $(0,1)$ the theorem reads
\begin{align}
R_n^{(1,0)}(x_1,x_2,x_3,0)&=-2n(2n+1)\,(x_1x_2x_3)^{n-1}\prod_{i<j}(x_i-x_j)^{2(n-1)}\,(x_1+x_2+x_3),\label{eq:slice-KdV}\\
R_n^{(0,1)}(x_1,x_2,x_3,0)&=2n(2n+1)\,(x_1x_2x_3)^{n+1}\prod_{i<j}(x_i-x_j)^{2(n-1)}\Bigl(2\sum_ix_i^2+\sum_{i<j}x_ix_j\Bigr),\label{eq:slice-Ito}
\end{align}
in agreement with the slice constants $\mp6,\mp20,\mp42$ underlying Table~\ref{tab:Rn}.

\begin{remark}[Interpolation in $n$]\label{rem:interpolation}
Theorem~\ref{thm:slice} was first found and proved by a different method, which we record
because it applies to identities for which no closed expansion such as \eqref{eq:residual} is
available. Write $\varepsilon:=a_4$ and let
\[
g:=[x_4^1]E_n^{(\beta,\gamma)}=\Pi^{(3)}_{\mathrm{ev}}\,[a_4^2]\bigl(e_1(\phi(a))\Delta_4(\phi(a))(H_4\Delta_4)^{2n+1}\bigr),
\]
where $\Pi^{(3)}_{\mathrm{ev}}$ is the even projection in $a_1,a_2,a_3$; by \eqref{eq:structure}
one has $g=(x_1x_2x_3)^3\prod_{i<j\le3}(x_i-x_j)^2\,R_n^{(\beta,\gamma)}(x_1,x_2,x_3,0)$.
The $\varepsilon$-dependence of the argument sits in the four factors
\begin{gather*}
e_1(\phi(a))=\eta_1(\phi)+\phi(\varepsilon),\qquad
\prod_{i\le3}(\phi(a_i)-\phi(\varepsilon)),\\
H_4^{2n+1}=(\bar H+\eta_1^2\varepsilon+\eta_1\varepsilon^2)^{2n+1},\qquad
\Delta_4^{2n+1}=\bar\Delta^{2n+1}(\eta_3-\eta_2\varepsilon+\eta_1\varepsilon^2-\varepsilon^3)^{2n+1}.
\end{gather*}
Lemma~\ref{lem:interpolation} below shows that the coefficient of $\varepsilon^2$ equals
$\bar H^{2n-1}\eta_3^{2n-1}$ times a polynomial of degree at most two in $n$. Consequently,
after cancelling the factor $U^{2n-1}(x_1x_2x_3)^{n-1}$, both sides of \eqref{eq:slice} become
polynomials of degree at most two in $n$ with coefficients in $\Z[\beta,\gamma][x_1,x_2,x_3]$,
and the identity for all $n$ follows from its exact verification for $n=1,2,3$ (carried out,
together with $n=4$ and the third-difference test $Z_4-3Z_3+3Z_2-Z_1=0$ for the reduced
polynomials, in the ancillary files). The double slice follows from \eqref{eq:slice}: for
$n\ge2$, $R_n^{(\beta,\gamma)}(x_1,x_2,0,0)=0$.
\end{remark}

\begin{lemma}[Interpolation lemma]\label{lem:interpolation}
Let $\varepsilon$ be a further indeterminate and, for $n\ge1$, put
\begin{equation}\label{eq:Psi}
\Psi_n(\varepsilon):=\bigl(\eta_1(\phi)+\phi(\varepsilon)\bigr)\prod_{i=1}^{3}\bigl(\phi(a_i)-\phi(\varepsilon)\bigr)
\;\bigl(\bar H+\eta_1^2\varepsilon+\eta_1\varepsilon^2\bigr)^{2n+1}\bigl(\eta_3-\eta_2\varepsilon+\eta_1\varepsilon^2-\varepsilon^3\bigr)^{2n+1},
\end{equation}
where $\eta_1(\phi):=\phi(a_1)+\phi(a_2)+\phi(a_3)$. Then
\begin{equation}\label{eq:interpolation}
[\varepsilon^2]\,\Psi_n=\bar H^{\,2n-1}\eta_3^{\,2n-1}\,\Omega_n,\qquad
\Omega_n=\Omega^{(0)}+n\,\Omega^{(1)}+n^2\,\Omega^{(2)},
\end{equation}
with $\Omega^{(0)},\Omega^{(1)},\Omega^{(2)}\in\Z[\beta,\gamma,a_1,a_2,a_3]$ symmetric in
$a_1,a_2,a_3$ and independent of $n$.
\end{lemma}

\begin{proof}
The first two factors of \eqref{eq:Psi} do not depend on $n$. For the third and fourth,
\begin{align*}
\bigl(\bar H+\eta_1^2\varepsilon+\eta_1\varepsilon^2\bigr)^{2n+1}
&=\bar H^{2n+1}+(2n+1)\bar H^{2n}\eta_1^2\,\varepsilon\\
&\qquad+\Bigl[(2n+1)\bar H^{2n}\eta_1+\tbinom{2n+1}{2}\bar H^{2n-1}\eta_1^4\Bigr]\varepsilon^2+O(\varepsilon^3),\\
\bigl(\eta_3-\eta_2\varepsilon+\eta_1\varepsilon^2-\varepsilon^3\bigr)^{2n+1}
&=\eta_3^{2n+1}-(2n+1)\eta_3^{2n}\eta_2\,\varepsilon\\
&\qquad+\Bigl[(2n+1)\eta_3^{2n}\eta_1+\tbinom{2n+1}{2}\eta_3^{2n-1}\eta_2^2\Bigr]\varepsilon^2+O(\varepsilon^3).
\end{align*}
By the Leibniz rule, $[\varepsilon^2]\Psi_n$ is a sum of products of coefficients of total
$\varepsilon$-order two. A coefficient of order $r\in\{0,1,2\}$ in the third (fourth) factor
carries the factor $\bar H^{2n+1-r}$ ($\eta_3^{2n+1-r}$) times a polynomial in $n$ of degree at
most $r$, with coefficients in $\Z[a_1,a_2,a_3]$; since $\tbinom{2n+1}{2}=n(2n+1)$, all these
polynomials have integer coefficients. Hence every term of $[\varepsilon^2]\Psi_n$ is divisible by
$\bar H^{2n-1}\eta_3^{2n-1}$, and after removing this factor it is a polynomial in $n$ of degree
at most two with coefficients in $\Z[\beta,\gamma,a_1,a_2,a_3]$. Symmetry in $a_1,a_2,a_3$ is
clear.
\end{proof}

\begin{corollary}\label{cor:III-ii}
Let $(\beta,\gamma)\neq(0,0)$ and $n\ge1$. Then $R_n^{(\beta,\gamma)}\neq0$; equivalently, the
four-soliton condition fails on $V_Q$ for the equation $(\beta D_x+\gamma\Lam)(D_x^3+\Lam)^{2n+1}f\cdot f=0$.
\end{corollary}

\begin{proof}
By Theorem~\ref{thm:slice} it suffices to see that the right-hand side of \eqref{eq:slice} is a
nonzero polynomial. The factor $\prod_i(\gamma x_i-\beta)$ is nonzero. If $\gamma\neq0$, the part
of $\Theta_{\beta,\gamma}$ of top degree five in $x$ is $\gamma^4E_3(2E_1^2-3E_2)$, and
$2E_1^2-3E_2=2\sum_ix_i^2+\sum_{i<j}x_ix_j\neq0$; if $\gamma=0$, then
$\Theta_{\beta,0}=\beta^4E_1\neq0$. The equivalence is Corollary~\ref{thm:structure}.
\end{proof}

\begin{remark}[A hand-checkable certificate for $L=D_x$]\label{rem:certificate}
For $(\beta,\gamma)=(1,0)$ there is a second proof of Corollary~\ref{cor:III-ii} that is
uniform in $n$ and can be checked by hand. Fix real $k_1,\dots,k_4$, positive and pairwise
distinct, and for each sign vector $\sg$ with $\sg_1=+1$ put
$u_\sg:=H_\sg^2V_\sg\ge0$ and $c_\sg:=G_\sg H_\sg V_\sg$, where $G_\sg=e_1(a)$, $H_\sg=H_4(a)$,
$V_\sg=\Delta_4(a)^2$ and $a=\sg\cdot k$. By \eqref{eq:reduction} with $N=4$ and $\ell=k$,
\[
S_4=2\cdot3^{7(2n+1)}(k_1k_2k_3k_4)^{3(2n+1)}\,\Phi(n),\qquad \Phi(m):=\sum_{\sg_1=+1}c_\sg u_\sg^{\,m},
\qquad \Phi(0)=0,
\]
the identity $\Phi(0)=0$ being $R_0\equiv0$ at $k$. At $k^*=(6,5,3,1)$ the eight pairs
$(u_\sg,c_\sg)$, ordered by decreasing $u$, are listed in Table~\ref{tab:cert}: the $u_\sg$ are
strictly decreasing, exactly one $c_\sg$ is negative, namely $c_1=-78683996160$ in the top row,
and the positive entries sum to $|c_1|$. Hence, for every real $n\ge1$,
\[
\Phi(n)=c_1u_1^n+\sum_{j\ge2}c_ju_j^n\le c_1u_1^n+\Bigl(\sum_{j\ge2}c_j\Bigr)u_2^n=-|c_1|\bigl(u_1^n-u_2^n\bigr)<0,
\]
so that $S_4(k^*)<0$ for every $n\ge1$. No point $k^*$ with the same sign pattern exists for
Ito's family $(\beta,\gamma)=(0,1)$ among the integer points $12\ge k_1>k_2>k_3>k_4\ge1$; for
that family the explicit formula \eqref{eq:residual} and Theorem~\ref{thm:sign} give the proof.
\end{remark}

\begin{table}[ht]
\centering
\begin{tabular}{crr}
\hline
$(\sg_2,\sg_3,\sg_4)$ & $u_\sg=H^2V$ & $c_\sg=GHV$ \\
\hline
$(-,+,-)$ & $786839961600$ & $-78683996160$ \\
$(+,-,+)$ & $631834214400$ & $41206579200$ \\
$(+,+,-)$ & $378077414400$ & $8057387520$ \\
$(-,-,+)$ & $273236198400$ & $12419827200$ \\
$(-,-,-)$ & $110649369600$ & $11064936960$ \\
$(+,+,+)$ & $57830630400$ & $865728000$ \\
$(+,-,-)$ & $3657830400$ & $2560481280$ \\
$(-,+,+)$ & $1003622400$ & $2509056000$ \\
\hline
\end{tabular}
\medskip
\caption{The certificate at $k^*=(6,5,3,1)$ for $L=D_x$. Sample hand check: for
$\sg=(+,-,+,-)$ one has $a=(6,-5,3,-1)$, $G=3$, $H=-30$, $V=874266624$, whence
$u=900\cdot V=786839961600$ and $c=3\cdot(-30)\cdot V=-78683996160$.}
\label{tab:cert}
\end{table}

\subsection{Theorem~III}\label{subsec:thmIII}
We now return to an arbitrary linear factor $L$. The reason why the mechanism of
Theorem~\ref{thm:II} breaks down for $N=4$ is the following lemma.

\begin{lemma}\label{lem:no-linear-factor}
For $N\ge4$ the polynomial $H_N=e_1e_2-e_3$ has no linear factor over $\C$.
\end{lemma}

\begin{proof}
First let $N=4$ and suppose that $H_4=\ell\,C$ with a linear form $\ell$ and a quadratic form
$C$. The restriction $\ell|_{a_4=0}$ is not zero, for otherwise $\ell$ would be proportional
to $a_4$ and $H_4|_{a_4=0}=H_3$ would vanish. Hence $\ell|_{a_4=0}$ divides
$H_3=(a_1+a_2)(a_1+a_3)(a_2+a_3)$, and after permuting the variables and rescaling we may
assume $\ell=a_1+a_2+c\,a_4$. Substituting $a_2=-a_1-ca_4$ into
$H_4=\tfrac13\bigl[(a_1+a_2+a_3+a_4)^3-a_1^3-a_2^3-a_3^3-a_4^3\bigr]$ must give the zero
polynomial. The substituted polynomial equals
$\tfrac13\bigl[(a_3+(1-c)a_4)^3-a_3^3-a_4^3\bigr]+c\,a_1^2a_4+c^2a_1a_4^2+\tfrac13c^3a_4^3$,
whose coefficient of $a_1^2a_4$ is $c$; hence $c=0$. But then $\ell=a_1+a_2$, and the
substitution $a_2=-a_1$ gives $H_4=a_3a_4(a_3+a_4)\neq0$ by \eqref{eq:cubic-identity}, a
contradiction. For $N\ge5$, a linear factor $\ell=\sum_ic_ia_i$ of $H_N$ restricts, under
$a_j=0$ for $j\notin J$ with $J$ a four-element set containing an index $i$ with $c_i\neq0$, to
a nonzero linear factor of $H_N|_{a_j=0\,(j\notin J)}=H_4(a_J)$, which is impossible.
\end{proof}

Thus the factor $H_4\Delta_4$ is not a polynomial in the squares, and its odd powers acquire new
parity classes. The following theorem makes this quantitative; its proof in regime (ii) is
Corollary~\ref{cor:III-ii} above.

\begin{theorem}[Theorem~III]\label{thm:III}
For every $n\ge1$ and all $L$, $\Lam$ as in \eqref{eq:LLam}, the equation \eqref{eq:family}
fails the four-soliton condition on $V_Q$: $S_4$ does not vanish identically on $V_Q^{\,4}$.
Consequently, for every $n\ge1$ the equation possesses nondegenerate three-soliton solutions
with all parameters free on $V_Q$, but for generic nondegenerate data on $V_Q$ it possesses no
four-soliton solution of the form \eqref{eq:ansatz}.
\end{theorem}

\begin{proof}
We use the two regimes of Remark~\ref{rem:regimes}. In regime (ii), $L=\beta D_x+\gamma\Lam$,
the statement is Corollary~\ref{cor:III-ii} above, which rests on the explicit residual
formula of Section~\ref{subsec:residual}. In regime (i), by Remark~\ref{rem:regimes}(i)
and Lemma~\ref{lem:parity}(c), $S_4$ vanishes identically on $V_Q^{\,4}$ if and only if
$\supp(e_1\Delta_4)\cap\supp\bigl((H_4\Delta_4)^{2n+1}\bigr)=\emptyset$. If this were the case,
$\Sigma_{4,n}(k,\ell)$ would vanish identically in the independent variables $k,\ell$, hence also
after the specialization $\ell=k$; but $\Sigma_{4,n}(k,k)=16\,E_n^{(1,0)}(k_1^2,\dots,k_4^2)$
by Lemma~\ref{lem:parity}(a) and \eqref{eq:En}, and $E_n^{(1,0)}\neq0$ by
Corollary~\ref{cor:III-ii} with $(\beta,\gamma)=(1,0)$. The last sentence follows from
Theorem~\ref{thm:II}, the phase factors \eqref{eq:phase}, and the fact that the complement of
the zero set of $S_4$ in $V_Q^{\,4}$ is a dense open subset.
\end{proof}

\begin{remark}\label{rem:III}
(i) The case $L=D_x$ of Theorem~\ref{thm:III} is the conjecture of \cite[Lemma~2.3]{GursesPekcan2025},
who verified it for $n=1,2,3$. (ii) In regime (i) the proof shows more: since
$\supp(e_1\Delta_4)$ consists of the four singletons (Lemma~\ref{lem:bialternant-parity}), the
polynomial $(H_4\Delta_4)^{2n+1}$ has, for every $n\ge1$, a nonzero component of parity class
$\{i\}$ for each $i\in[4]$ (all four by symmetry), and $\Sigma_{4,n}(k,\ell)=16\sum_{i=1}^4
(e_1\Delta_4)_{\{i\}}(\ell)\,\bigl((H_4\Delta_4)^{2n+1}\bigr)_{\{i\}}(k)$ is nonzero as a polynomial
in $\ell$ for generic $k$. (iii) For every real nondegenerate configuration of $V_Q$ one may ask whether $S_4\neq0$,
not merely generically; for the G\"urses--Pekcan and Ito families this is answered
affirmatively by Theorem~\ref{thm:sign}, and Theorems~\ref{thm:GP-whole} and~\ref{thm:existence}
settle the question for all real $L$ and on the whole dispersion variety.
\end{remark}

\section{Mixed components}\label{sec:mixed}

Theorems~\ref{thm:I}--\ref{thm:III} concern the principal component $V_Q$. Since
$V(P_{L,\Lam,n})=V_L\cup V_Q$, the three-soliton condition may also be imposed on the mixed
products $V_L\times V_Q\times V_Q$ and $V_L\times V_L\times V_Q$ and on $V_L^{\,3}$. If two of
the three momenta lie on $V_L$, every term of \eqref{eq:SN} contains the factor
$P(\sg_i\bp_i-\sg_j\bp_j)=L(\sg_i\bp_i-\sg_j\bp_j)\,Q(\cdot)^{2n+1}=0$, so $S_3$ vanishes trivially,
but such data are degenerate. The only mixed product with nondegenerate data is
$V_L\times V_Q\times V_Q$ (and its permutations), and it behaves differently for $n=0$ and for
$n\ge1$.

\begin{lemma}[Universal factorization]\label{lem:mixed-factor}
Let $\bp_3\in V_L$ and $\bp_1,\bp_2\in V_Q$. Let $k_j$ be the $x$-components, let
$\ell_j=L(\bp_j)$ for $j=1,2$ (note that $L(\bp_3)=0$), and put $c:=Q(\bp_3)$. Then
\begin{equation}\label{eq:mixed-factor}
S_3(\bp_1,\bp_2,\bp_3)=(\ell_1^2-\ell_2^2)\,\ell_1\ell_2\;\mathcal T_n(k_1,k_2;k_3,c),
\end{equation}
where $\mathcal T_n$ is the universal polynomial (independent of $L$ and $\Lam$)
\begin{equation}\label{eq:Tn}
\mathcal T_n(k_1,k_2;k_3,c):=\sum_{\sg\in\{\pm1\}^3}\sg_1\sg_2\,\widetilde Q_n(\sg_1k_1,\sg_2k_2,\sg_3k_3,\sg_3c),
\end{equation}
\begin{equation}\label{eq:Qtilde}
\begin{aligned}
\widetilde Q_n(a_1,a_2,a_3,c'):={}&\bigl(3H_3(a)+c'\bigr)^{2n+1}\bigl(-3a_1a_2(a_1-a_2)\bigr)^{2n+1}\\
&\times\bigl(-3a_1a_3(a_1-a_3)-c'\bigr)^{2n+1}\bigl(-3a_2a_3(a_2-a_3)-c'\bigr)^{2n+1}.
\end{aligned}
\end{equation}
\end{lemma}

\begin{proof}
On $V_L$ the momentum $\bp_3$ satisfies $L(\bp_3)=0$ and $\Lam(\bp_3)=c-k_3^3$, while
$\Lam(\bp_j)=-k_j^3$ for $j=1,2$. With $a_j=\sg_jk_j$ and $b_j=\sg_j\ell_j$ as before,
$Q(\sum_j\sg_j\bp_j)=(a_1+a_2+a_3)^3-a_1^3-a_2^3-a_3^3+\sg_3c=3H_3(a)+\sg_3c$,
$Q(\sg_1\bp_1-\sg_2\bp_2)=-3a_1a_2(a_1-a_2)$ and
$Q(\sg_j\bp_j-\sg_3\bp_3)=(a_j-a_3)^3-a_j^3+a_3^3-\sg_3c=-3a_ja_3(a_j-a_3)-\sg_3c$ for $j=1,2$,
while $L(\sum_j\sg_j\bp_j)=b_1+b_2$, $L(\sg_1\bp_1-\sg_2\bp_2)=b_1-b_2$ and
$L(\sg_j\bp_j-\sg_3\bp_3)=b_j$. The product of the four $L$-values is
$(b_1^2-b_2^2)\,b_1b_2=(\ell_1^2-\ell_2^2)\,\sg_1\sg_2\,\ell_1\ell_2$, and
$\prod_{i<j}\sg_i\sg_j=1$ for $N=3$.
\end{proof}

\begin{theorem}[Mixed components]\label{thm:mixed}
Let $L$, $\Lam$ be as in \eqref{eq:LLam}.
\begin{enumerate}
\item[(i)] $\mathcal T_0\equiv0$. Hence for $n=0$ the equation $L(D_x^3+\Lam)f\cdot f=0$ satisfies
the three-soliton condition on the whole dispersion variety.
\item[(ii)] For $L=D_x$ (so that $k_3=0$ on $V_L$ and $c=\Lam(\bp_3)$ is free) and $n\ge0$,
\begin{equation}\label{eq:mixed-closed}
\begin{aligned}
S_3={}&-4\cdot3^{2n+1}c^{4n+2}(k_1k_2)^{2n+2}(k_1^2-k_2^2)\\
&\times\sum_{\substack{2\le j\le 2n\\ j\ \mathrm{even}}}\binom{2n+1}{j}c^{j}(3k_1k_2)^{2n+1-j}(k_1^2-k_2^2)^{2n+1-j}
\bigl[(k_1-k_2)^j-(k_1+k_2)^j\bigr].
\end{aligned}
\end{equation}
For $n=0$ the sum is empty. For $n\ge1$ and real $k_1,k_2,c$ with $k_1k_2c\neq0$ and
$k_1^2\neq k_2^2$ one has $S_3>0$; in particular the equation has no nondegenerate three-soliton
solution of the form \eqref{eq:ansatz} with real momenta on $V_L\times V_Q\times V_Q$.
\item[(iii)] For every $n\ge1$ and every $L$, $S_3$ does not vanish identically on
$V_L\times V_Q\times V_Q$. Thus for $n\ge1$ the three-soliton condition holds on the principal
component (Theorem~\ref{thm:II}) but not on the whole dispersion variety.
\end{enumerate}
\end{theorem}

\begin{proof}
(i) The eight-term sum \eqref{eq:Tn} with $n=0$ may be evaluated in two stages. Symmetrizing
$\widetilde Q_0$ under the joint sign change $(a_3,c')\mapsto(-a_3,-c')$, i.e.\ performing the
sum over $\sg_3$, gives the polynomial identity
\begin{equation}\label{eq:F-identity}
\begin{aligned}
&\frac{\widetilde Q_0(a_1,a_2,a_3,c')+\widetilde Q_0(a_1,a_2,-a_3,-c')}{2}\\
&\qquad=-9a_1^2a_2^2(a_1^2-a_2^2)\Bigl[9a_3^2(a_1^2-a_3^2)(a_2^2-a_3^2)+3c'a_3(a_1^2+a_2^2-3a_3^2)+c'^2\Bigr],
\end{aligned}
\end{equation}
which is checked by expansion. The right-hand side is even in $a_1$ and in $a_2$, so its
component of parity class $\{1,2\}$ vanishes, and the remaining sum
$\sum_{\sg_1,\sg_2}\sg_1\sg_2(\cdots)$ in \eqref{eq:Tn} is zero. A second proof, valid for all
$N$, is given in Theorem~\ref{thm:whole}.

(ii) For $k_3=0$ one has $a_3=0$, $H_3(a)=a_1a_2(a_1+a_2)$ and
$\widetilde Q_n=(\sg_3c)^{4n+2}\bigl(3a_1a_2(a_1+a_2)+\sg_3c\bigr)^{2n+1}\bigl(-3a_1a_2(a_1-a_2)\bigr)^{2n+1}$.
Summing over $\sg_3$ produces $c^{4n+2}\,O_n\bigl(3a_1a_2(a_1+a_2)\bigr)$ with
$O_n(X):=(X+c)^{2n+1}+(X-c)^{2n+1}=2\sum_{j\ \mathrm{even}}\binom{2n+1}{j}X^{2n+1-j}c^{\,j}$, an odd
polynomial in $X$. Using $(-3a_1a_2(a_1-a_2))^{2n+1}=-3^{2n+1}\sg_1\sg_2(k_1k_2)^{2n+1}(a_1-a_2)^{2n+1}$
and $\ell_j=k_j$, the terms with $(\sg_1,\sg_2)$ and $(-\sg_1,-\sg_2)$ coincide, and the sum over
$(\sg_1,\sg_2)\in\{(+,+),(+,-)\}$ gives
\begin{align*}
S_3=-2\cdot3^{2n+1}c^{4n+2}(k_1k_2)^{2n+2}(k_1^2-k_2^2)\Bigl[&(k_1-k_2)^{2n+1}O_n\bigl(3k_1k_2(k_1+k_2)\bigr)\\
&-(k_1+k_2)^{2n+1}O_n\bigl(3k_1k_2(k_1-k_2)\bigr)\Bigr].
\end{align*}
Inserting the expansion of $O_n$, the $j$-th term of the bracket equals
$2\binom{2n+1}{j}c^j(3k_1k_2)^{2n+1-j}(k_1^2-k_2^2)^{2n+1-j}[(k_1-k_2)^j-(k_1+k_2)^j]$, which
vanishes for $j=0$; this is \eqref{eq:mixed-closed}. Let $k_1,k_2,c$ be real with
$k_1k_2c\neq0$ and $k_1^2\neq k_2^2$. Since $S_3$ is unchanged under $k_1\mapsto-k_1$ and under
$k_2\mapsto-k_2$ (each summand of \eqref{eq:mixed-closed} changes sign twice), we may assume
$k_1,k_2>0$. Then, for even $j\ge2$, the summand $c^j(3k_1k_2)^{2n+1-j}(k_1^2-k_2^2)^{2n+1-j}
[(k_1-k_2)^j-(k_1+k_2)^j]$ has the sign of $-(k_1^2-k_2^2)$, because $2n+1-j$ is odd and the
bracket is negative; multiplied by the prefactor $-4\cdot3^{2n+1}c^{4n+2}(k_1k_2)^{2n+2}(k_1^2-k_2^2)$,
every term of \eqref{eq:mixed-closed} is strictly positive. Hence $S_3>0$, and the last assertion
follows from the criterion of Section~\ref{subsec:NSC}. The reality of $c$ cannot be dropped:
for $n=2$, $k_1=1$, $k_2=2$ the sum in \eqref{eq:mixed-closed} equals $1440\,c^2(324+5c^2)$,
which vanishes for $c^2=-324/5$ although all phase factors of the corresponding data are finite
and nonzero.

(iii) In regime (i) of Remark~\ref{rem:regimes} the pair $(k_3,c)=(k_3,Q(\bp_3))$ ranges over
all of $\C^2$ as $\bp_3$ ranges over $V_L$: the conditions $L(\bp_3)=0$, $D_x(\bp_3)=k_3$,
$\Lam(\bp_3)=c-k_3^3$ form a solvable linear system because $D_x$, $\Lam$, $L$ are linearly
independent. Hence $\mathcal T_n\not\equiv0$ on $V_L$ follows from (ii) by specializing
$k_3=0$. In regime (ii),
$L=\beta D_x+\gamma\Lam$: if $\gamma=0$ this is (ii); if $\gamma\neq0$ it is
Proposition~\ref{prop:mixed-leading} below.
\end{proof}

\begin{theorem}[Three-block expansion of the mixed polynomial]\label{thm:threeblock}
Write $a:=k_1$, $b:=k_2$, $z:=k_3$ and put
\begin{align}
E&:=-9a^2b^2(a^2-b^2)\Bigl[9z^2(a^2-z^2)(b^2-z^2)+3cz(a^2+b^2-3z^2)+c^2\Bigr],\label{eq:blockE}\\
U&:=3ab^2c\Bigl[(c+3b^2z)^2-9b^2z^4\Bigr],\qquad
V:=-3a^2bc\Bigl[(c+3a^2z)^2-9a^2z^4\Bigr].\label{eq:blockUV}
\end{align}
Then $\widetilde Q_0(a,b,z,c)=E+U+V$ as polynomials, and under the sign changes
$a\mapsto\sg_1a$, $b\mapsto\sg_2b$, $(z,c)\mapsto\sg_3(z,c)$ one has $E\mapsto E$,
$U\mapsto\sg_1\sg_3U$, $V\mapsto\sg_2\sg_3V$. Consequently, for every $n\ge0$,
\begin{equation}\label{eq:threeblock}
\mathcal T_n=8\,EUV\sum_{\substack{i+j+h=n-1\\ i,j,h\ge0}}\frac{(2n+1)!}{(2i+1)!\,(2j+1)!\,(2h+1)!}\,E^{2i}U^{2j}V^{2h},
\end{equation}
the sum being empty for $n=0$. Moreover
$(c+3b^2z)^2-9b^2z^4=-Q(\bp_2+\bp_3)Q(\bp_2-\bp_3)$ and
$(c+3a^2z)^2-9a^2z^4=-Q(\bp_1+\bp_3)Q(\bp_1-\bp_3)$.
\end{theorem}

\begin{proof}
The identity $\widetilde Q_0=E+U+V$ is checked by expanding both sides of
\[
\bigl[3(a+b)(a+z)(b+z)+c\bigr]\bigl[-3ab(a-b)\bigr]\bigl[-3az(a-z)-c\bigr]\bigl[-3bz(b-z)-c\bigr]=E+U+V
\]
(ancillary files; the identity \eqref{eq:F-identity} is its $\sg_3$-even part). The sign
characters of $E,U,V$ are read off from the displayed formulas. Since
$\widetilde Q_n=\widetilde Q_0^{\,2n+1}$, the multinomial expansion of
$(E+\sg_1\sg_3U+\sg_2\sg_3V)^{2n+1}$ inserted in \eqref{eq:Tn} gives, for the term
$E^eU^uV^v$, the sign sum $\sum_\sg\sg_1^{1+u}\sg_2^{1+v}\sg_3^{u+v}$, which equals $8$ if $u$
and $v$ are odd (then $u+v$ is even and $e=2n+1-u-v$ is odd) and $0$ otherwise. Writing
$e=2i+1$, $u=2j+1$, $v=2h+1$ gives \eqref{eq:threeblock}. The last assertion follows from
$Q(\bp_j+\bp_3)=3k_jz(k_j+z)+c$ and $Q(\bp_j-\bp_3)=-3k_jz(k_j-z)-c$ (Lemma~\ref{lem:mixed-factor}).
\end{proof}

\begin{corollary}[Real nondegenerate mixed three-soliton data]\label{cor:mixed-real}
Let $n\ge1$ and let $\bp_1,\bp_2\in V_Q$, $\bp_3\in V_L\setminus V_Q$ be real nondegenerate data,
so that $k_1k_2(k_1^2-k_2^2)\neq0$, $\ell_1\ell_2(\ell_1^2-\ell_2^2)\neq0$ and $c=Q(\bp_3)\neq0$
(a momentum of $V_L\cap V_Q$ has $c=0$, and a triple in $V_Q^{\,3}$ satisfies the 3SC by
Theorem~\ref{thm:II}). Then $U,V\neq0$,
the sum in \eqref{eq:threeblock} is strictly positive, and
\begin{equation}\label{eq:mixed-criterion}
S_3(\bp_1,\bp_2,\bp_3)=0\quad\Longleftrightarrow\quad
9k_3^2(k_1^2-k_3^2)(k_2^2-k_3^2)+3ck_3(k_1^2+k_2^2-3k_3^2)+c^2=0 .
\end{equation}
In particular the set of real nondegenerate mixed three-soliton data does not depend on
$n\ge1$. For $L=D_x$ (so $k_3=0$) the right-hand side of \eqref{eq:mixed-criterion} is $c^2\neq0$
and no such data exist (Theorem~\ref{thm:mixed}(ii)); for Ito's family ($L=\Lam$, $c=k_3^3$) it
reads $k_3^2\bigl[k_3^4-6(k_1^2+k_2^2)k_3^2+9k_1^2k_2^2\bigr]=0$, and real nondegenerate mixed
three-soliton data exist, namely those with $k_3^2=3(k_1^2+k_2^2)\pm3\sqrt{k_1^4+k_1^2k_2^2+k_2^4}$.
\end{corollary}

\begin{proof}
Nondegeneracy of the pairs $(\bp_j,\bp_3)$ means $Q(\bp_j\pm\bp_3)\neq0$, so $U,V\neq0$ by the
last assertion of Theorem~\ref{thm:threeblock}; all terms of the sum in \eqref{eq:threeblock}
are nonnegative for real data and the term $i=h=0$, $j=n-1$ equals $2n(2n+1)U^{2n-2}>0$. Hence
$\mathcal T_n=0$ if and only if $E=0$, i.e.\ if and only if the bracket in \eqref{eq:blockE}
vanishes, and \eqref{eq:mixed-criterion} follows from \eqref{eq:mixed-factor}. For Ito's family
the bracket with $c=k_3^3$ equals $k_3^2[k_3^4-6(k_1^2+k_2^2)k_3^2+9k_1^2k_2^2]$, whose zeros in
$k_3^2$ are as stated; at these points the pairs $(\bp_j,\bp_3)$ are nondegenerate because
$-Q(\bp_j+\bp_3)Q(\bp_j-\bp_3)=k_3^2\bigl[(k_3^2-\tfrac32k_j^2)^2+\tfrac{27}{4}k_j^4\bigr]>0$.
\end{proof}

\begin{proposition}[Regime (ii): leading terms]\label{prop:mixed-leading}
Let $L=\beta D_x+\gamma\Lam$ with $\gamma\neq0$ and $t:=\beta/\gamma\in\C$, and put
$\Xi:=-9k_1^2k_2^2(k_1^2-k_2^2)(t-3k_1^2)(t-3k_2^2)$. On $V_L$ one has $c=k_3(k_3^2-t)$, and for
$n\ge1$, as polynomials in $k_1,k_2,k_3$,
\begin{align}
\mathcal T_n&=-144\,n(2n+1)\,t^2\,k_3^{4n+4}\,k_1^3k_2^3(t-3k_1^2)^2(t-3k_2^2)^2\,\Xi^{2n-1}+O\bigl(k_3^{4n+6}\bigr),\label{eq:mixed-lead-t}\\
\mathcal T_n\big|_{t=0}&=16\cdot3^{8n+2}\,n(2n+1)\,k_3^{4n+8}\,k_1^{8n+3}k_2^{8n+3}(k_1^2-k_2^2)^{2n-1}+O\bigl(k_3^{4n+10}\bigr).\label{eq:mixed-lead-0}
\end{align}
In particular $\mathcal T_n\not\equiv0$ on $V_L$ for every $n\ge1$ and every complex $t$, which
proves Theorem~\ref{thm:mixed}(iii) in regime (ii).
\end{proposition}

\begin{proof}
In \eqref{eq:Tn} the substitution $c'=\sg_3c=a_3^3-ta_3$ with $a_3=\sg_3k_3$ gives
$-3a_ja_3(a_j-a_3)-c'=a_3f_j$ with $f_1:=t-3a_1^2+3a_1a_3-a_3^2$, $f_2:=t-3a_2^2+3a_2a_3-a_3^2$,
and with $f_0:=3H_3(a)+a_3^3-ta_3$ and $B:=-3a_1a_2(a_1-a_2)$ one has
$\widetilde Q_n=a_3^{4n+2}(Bf_0f_1f_2)^{2n+1}$. Write $Bf_0f_1f_2=\sum_{r=0}^7C_ra_3^{\,r}$; direct
multiplication gives
\begin{align*}
C_0&=-9a_1^2a_2^2(a_1^2-a_2^2)(t-3a_1^2)(t-3a_2^2),\\
C_1&=3t\,a_1^2a_2(t-3a_1^2)^2-3t\,a_1a_2^2(t-3a_2^2)^2,\\
C_2&=9a_1^2a_2^2(a_1^2-a_2^2)\bigl(6(a_1^2+a_2^2)-7t\bigr),\\
C_3&=9a_1a_2\bigl[a_2(t^2-a_2^2t+3a_2^4)-a_1(t^2-a_1^2t+3a_1^4)\bigr],\\
C_4&=-9a_1^2a_2^2(a_1^2-a_2^2),\qquad C_5=9a_1a_2\bigl[a_1(t+a_1^2)-a_2(t+a_2^2)\bigr],\qquad C_6=0,\qquad C_7=B.
\end{align*}
The
even-indexed $C_r$ are even in $a_1$ and in $a_2$, while $C_1,C_3,C_5$ are sums of two terms of
parity classes $\{2\}$ and $\{1\}$. The sign sum \eqref{eq:Tn} retains the monomials that are
even in $a_3$ and odd in $a_1$ and $a_2$, so
$\mathcal T_n=8\sum_{r\ \mathrm{even}}k_3^{4n+2+r}\bigl[[a_3^{\,r}](Bf_0f_1f_2)^{2n+1}\bigr]_{\{1,2\}}|_{a=k}$.
The coefficient of $a_3^0$ is $C_0^{2n+1}$, which is even; that of $a_3^2$ is
$(2n+1)C_0^{2n}C_2+\binom{2n+1}{2}C_0^{2n-1}C_1^2$, whose first term is even, and
$[C_1^2]_{\{1,2\}}=-18t^2a_1^3a_2^3(t-3a_1^2)^2(t-3a_2^2)^2$ (the cross term); with
$\binom{2n+1}{2}=n(2n+1)$ and $C_0|_{a=k}=\Xi$ this is \eqref{eq:mixed-lead-t}. For $t=0$ one has
$C_1=0$, $C_3=27(a_1a_2^6-a_1^6a_2)$ and $C_0=-81a_1^4a_2^4(a_1^2-a_2^2)$; the coefficients of
$a_3^0,a_3^2,a_3^4$ are built from $C_0,C_2,C_4$ only and are even, and in the coefficient of
$a_3^6$ the only term with a component of class $\{1,2\}$ is $\binom{2n+1}{2}C_0^{2n-1}C_3^2$
with $[C_3^2]_{\{1,2\}}=-2\cdot3^6a_1^7a_2^7$; this gives \eqref{eq:mixed-lead-0}. For $t\neq0$ the leading coefficient in \eqref{eq:mixed-lead-t} is a nonzero polynomial
in $k_1,k_2$, and for $t=0$ so is the one in \eqref{eq:mixed-lead-0}; hence
$\mathcal T_n\not\equiv0$ on $V_L$ for every $n\ge1$ and every complex $t$.
\end{proof}

\begin{remark}\label{rem:mixed-leading}
For real $t$ the nonvanishing of $\mathcal T_n$ on $V_L$ also follows from
Theorem~\ref{thm:threeblock}, since the sum in \eqref{eq:threeblock} is positive at real points
with $UV\neq0$; for complex $t$ the blocks $E,U,V$ need not be real and
Proposition~\ref{prop:mixed-leading} is used instead. The leading terms above were first
obtained by an expansion with $n$ as an indeterminate and cross-checked against the full
polynomials $\mathcal T_n$ for $n\le3$ (ancillary files).
\end{remark}

\begin{remark}\label{rem:mixed}
(i) For the G\"urses--Pekcan family the mixed data are two genuine solitons together with an
$x$-independent exponential $e^{\omega_3t+l_3y}$ with $\alpha_1\omega_3+\alpha_2l_3=c\neq0$. For
$n=0$ such a factor is harmless, as \eqref{eq:mixed-closed} and Theorem~\ref{thm:mixed}(i) show;
for $n\ge1$ it destroys the three-soliton solution for real nondegenerate data (over $\C$
there are exceptions: the complex zero of \eqref{eq:mixed-closed} noted in the proof of
Theorem~\ref{thm:mixed}, and Theorem~\ref{thm:complex-mixed}). In terms of the equation, the identity
$D_x^m(fg)\cdot(fg)=g^2D_x^mf\cdot f$ for $x$-independent $g$ makes the $n=0$ member
$D_x(D_x^3+\Lam)f\cdot f=0$ invariant under $f\mapsto fg$, whereas for $n\ge1$ the mixed
monomials $D_xD_t^{2j+1}\cdots$ of $Q^{2n+1}$ produce cross terms. For other $L$ no such
invariance holds (Ito's equation is not invariant under multiplication of $f$ by a
$t$-independent factor), and Theorem~\ref{thm:mixed}(i) rests on the identity
\eqref{eq:F-identity} and, more conceptually, on the fusion argument of Theorem~\ref{thm:whole}. (ii) Theorem~\ref{thm:mixed} shows that ``satisfies the 3SC'' has to be
qualified by the component for the families with $n\ge1$, whereas the $n=0$ members satisfy it
on the whole dispersion variety. Definition~\ref{def:component} was chosen accordingly, and the
statements ``three-soliton solutions exist, four-soliton solutions do not'' of
Theorem~\ref{thm:III} refer to nondegenerate data on $V_Q$.
\end{remark}

\subsection{The whole dispersion variety for \texorpdfstring{$n=0$}{n=0}}
Theorem~\ref{thm:mixed}(i) extends to all $N$: for $n=0$ the polynomial conditions hold on the
whole dispersion variety. The proof realizes a momentum on $V_L$ as the limit of two fusing
momenta on the principal component.

\begin{theorem}\label{thm:whole}
Let $L$, $\Lam$ be as in \eqref{eq:LLam}. Then $P=L(D_x^3+\Lam)$ satisfies the $N$-soliton
condition on the whole dispersion variety for every $N\ge3$: $S_N$ vanishes identically on
$V_{j_1}\times\dots\times V_{j_N}$ for every choice of components $V_{j_m}\in\{V_L,V_Q\}$.
\end{theorem}

\begin{proof}
By Lemma~\ref{lem:onshell} and its analogue on $V_L$, $S_N$ depends on $\bp_1,\dots,\bp_N$ only
through the triples $(k_m,L(\bp_m),\Lam(\bp_m))$. It therefore suffices to prove the theorem for
the universal polynomial $\widehat P(k,\ell,\lambda):=\ell\,(k^3+\lambda)$ on $\C^3$, whose
dispersion variety is $\{\ell=0\}\cup\{\lambda=-k^3\}$; the general case is the pull-back under
$\bp\mapsto(k,L(\bp),\Lam(\bp))$. If two or more of the momenta lie on $\{\ell=0\}$, every term
of \eqref{eq:SN} contains a pair factor $\widehat P(\sg_i\bp_i-\sg_j\bp_j)$ with vanishing
$\ell$-component, so $S_N=0$. If none does, the claim is Theorem~\ref{thm:I}. It remains to
treat one momentum $\bp_0=(z,0,c-z^3)$ on $\{\ell=0\}$ together with $N-1$ momenta
$\bp_i=(k_i,\ell_i,-k_i^3)$ on the principal component, and by irreducibility it suffices to do
so for generic data.

Assume $zc\neq0$, choose $r,s$ with $r+s=z$ and $rs=c/(3z)$ (so that $r\neq s$ for generic $c$),
a number $h\neq0$ with $h\neq\pm\ell_i$ for all $i$, and for small $\varepsilon>0$ the two
principal-component momenta $\bp_r:=(r,h,-r^3)$ and $\bp_s(\varepsilon):=(s,-h+\varepsilon^2,-s^3)$.
Then $\bp_r+\bp_s(\varepsilon)=(z,\varepsilon^2,-(r^3+s^3))=(z,\varepsilon^2,c-z^3)\to\bp_0$,
because $r^3+s^3=(r+s)^3-3rs(r+s)=z^3-c$. By \eqref{eq:phase} the phase factor of the pair is
$e^{A_{rs}}=\frac{(2h-\varepsilon^2)(r-s)}{\varepsilon^2\,z}$, and for a principal-component
momentum $\bp_i$
\[
e^{A_{ir}}e^{A_{is}}=\frac{(\ell_i-h)(\ell_i+h-\varepsilon^2)}{(\ell_i+h)(\ell_i-h+\varepsilon^2)}\cdot
\frac{(k_i-r)(k_i-s)}{(k_i+r)(k_i+s)}\ \xrightarrow{\ \varepsilon\to0\ }\ 
\frac{(k_i-r)(k_i-s)}{(k_i+r)(k_i+s)}=\frac{3zk_i(k_i-z)+c}{3zk_i(k_i+z)+c},
\]
which is exactly $-\widehat P(\bp_i-\bp_0)/\widehat P(\bp_i+\bp_0)$, the phase factor of the
mixed pair. For generic data all $N+1$ momenta are nondegenerate for small $\varepsilon$ (the
$\ell$-components of $\bp_r\pm\bp_s$ are $\varepsilon^2$ and $2h-\varepsilon^2$). By
Theorem~\ref{thm:I} the function $f_{N+1}$ of \eqref{eq:ansatz} built from these momenta is a
solution for every $\varepsilon$ and every choice of the phases $\delta$. Take
$\delta_r=\ln\varepsilon$ and $\delta_s=\ln\frac{\varepsilon z}{2h(r-s)}+\delta_0$ with an arbitrary
constant $\delta_0$. Then every term containing exactly one of $e^{\theta_r}$, $e^{\theta_s}$ has
a coefficient $O(\varepsilon)$, while the terms containing both carry the factor
$e^{A_{rs}}e^{\theta_r+\theta_s}=\frac{2h-\varepsilon^2}{2h}\,e^{\delta_0+(\bp_r+\bp_s(\varepsilon))\cdot t}\to e^{\delta_0+\bp_0\cdot t}$
and products $e^{A_{ir}}e^{A_{is}}\to e^{A_{i0}}$. Hence $f_{N+1}$ converges, uniformly on
compact sets together with all derivatives (it is a finite sum of exponentials whose
coefficients and exponents converge), to the function $f_N$ of \eqref{eq:ansatz} built from
$\bp_1,\dots,\bp_{N-1},\bp_0$ with the phases $\delta_1,\dots,\delta_{N-1},\delta_0$. The
bilinear equation passes to the limit, so this $f_N$ is a solution for all phases; by the
criterion of Section~\ref{subsec:NSC}, $S_M=0$ for all sub-collections of the mixed data, in
particular $S_N(\bp_1,\dots,\bp_{N-1},\bp_0)=0$. This holds on a dense open subset of
$\{\ell=0\}\times(\{\lambda=-k^3\})^{N-1}$, hence identically.
\end{proof}

\begin{remark}\label{rem:whole}
For $N=3$ the theorem gives a second proof of Theorem~\ref{thm:mixed}(i), and the fusion
argument explains it: a mixed three-soliton solution of an $n=0$ member is a degenerate limit of
a genuine four-soliton solution on the principal component. The identity
\eqref{eq:F-identity} is the algebraic shadow of this limit. The polynomial statement of
Theorem~\ref{thm:whole} should not be read as the existence of nondegenerate multisoliton
solutions with two or more momenta on $V_L$: for such data the pairwise phase factors are
indeterminate.
\end{remark}

\section{Signs of the residuals, real data and maximal soliton numbers}\label{sec:strata}

The explicit formula \eqref{eq:residual} settles the sign and the zero locus of the residuals
of the two extreme members of the class (Section~\ref{subsec:signs}) and decides for which real
$L$ special real four-soliton data exist (Section~\ref{subsec:realdata}). Over $\C$ the
four-soliton conditions, imposed on all four-element subsets of a datum, bound the number of
nondegenerate solitons: Sections~\ref{subsec:complex} and~\ref{subsec:five} determine the
maximal number for $n=1$, and Section~\ref{subsec:higher} treats $n\ge2$ and proves finiteness
for every $n\ge1$ and every $L$. Table~\ref{tab:max} collects the results.

\begin{table}[ht]
\centering
\small
\setlength{\tabcolsep}{5pt}
\begin{tabular}{@{}cl>{\raggedright\arraybackslash}p{7.1cm}l@{}}
\hline
$n$ & $L$ & maximal number of nondegenerate solitons & reference \\
\hline
$1$ & $\beta D_x+\gamma\Lam$ & $4$ over $\C$; over $\mathbb R$ (real $L,\Lam$): $3$ if $\beta\gamma\le0$, $4$ if $\beta\gamma>0$ & Thms.~\ref{cor:max-four}, \ref{thm:six-bound} \\
$1$ & $L\notin\operatorname{span}(D_x,\Lam)$ & $5$ over $\C$; $5$ over $\mathbb R$ for real $L\notin\operatorname{span}_{\mathbb R}(D_x,\Lam)$ & Thms.~\ref{thm:five-exist}, \ref{thm:six-bound} \\
$2$ & $D_x$ & $5$, attained exactly when the squared wavenumbers are the fifth roots of a constant & Thm.~\ref{thm:n2-five} \\
$\ge3$ & $D_x$ & between $4$ and $5n-6$ & Thm.~\ref{thm:finite-bound} \\
$\ge1$ & $\beta D_x+\gamma\Lam$ & at most $5n+3$ & Thm.~\ref{thm:finite-bound} \\
$\ge1$ & arbitrary & at most $40n+17$ & Thm.~\ref{thm:finite-all} \\
\hline
\end{tabular}
\medskip
\caption{Maximal numbers of nondegenerate solitons of the form \eqref{eq:ansatz} for
$L(D_x^3+\Lam)^{2n+1}f\cdot f=0$ on the whole dispersion variety. For $n=1$ every nondegenerate
datum with $N\ge4$ lies on $V_Q$ (Theorem~\ref{thm:complex-n1}); the bounds $5n+3$ and $40n+17$
allow for one momentum on $V_L\setminus V_Q$, the bounds on $V_Q$ being $5n+2$ and $40n+16$.}
\label{tab:max}
\end{table}

\subsection{Sign-definiteness and the zero locus}\label{subsec:signs}
In this subsection and the next, $x_1,\dots,x_4$ are real.

\begin{theorem}[Sign-definiteness and zero locus]\label{thm:sign}
Let $n\ge1$.
\begin{enumerate}
\item[(i)] On the closed positive orthant $x_1,\dots,x_4\ge0$,
\[
R_n^{(1,0)}\le0\qquad\text{and}\qquad R_n^{(0,1)}\ge0 .
\]
For $n\ge2$ the two zero loci coincide: $R_n^{(1,0)}(x)=0$ if and only if $R_n^{(0,1)}(x)=0$ if
and only if the set $\{x_i:\ x_i>0\}$ of distinct positive coordinate values has at most two
elements. For $n=1$, $R_1^{(1,0)}(x)=0$ only at $x=0$, and $R_1^{(0,1)}(x)=0$ if and only if at
most two coordinates are positive. In particular, on the open positive orthant with pairwise
distinct coordinates, $R_n^{(1,0)}<0$ and $R_n^{(0,1)}>0$ for every $n\ge1$.
\item[(ii)] Explicitly, with $x_i=k_i^2$,
\begin{equation}\label{eq:SOS}
-R_n^{(1,0)}(k_1^2,\dots,k_4^2)=\sum_{\substack{r_1+\dots+r_4=n-1\\ r_j\ge0}}\ \sum_{i=1}^4
\frac{(2n+1)!}{(2r_i)!\prod_{j\ne i}(2r_j+1)!}
\Bigl[k_i\prod_{j=1}^4k_j^{\,n-1-r_j}\,\Delta_3(x_{\widehat j})^{r_j}\Bigr]^2\Big|_{x=k^2},
\end{equation}
a sum of squares with positive integer coefficients, and
\begin{equation}\label{eq:SOS-Ito}
\begin{aligned}
R_n^{(0,1)}(k_1^2,\dots,k_4^2)={}&\sum_{\substack{r_1+\dots+r_4=n-1\\ r_j\ge0}}\ \sum_{i=1}^4
\frac{(2n+1)!}{(2r_i)!\prod_{j\ne i}(2r_j+1)!}\\
&\times\Bigl[k_i^2\prod_{j=1}^4k_j^{\,n-1-r_j}\,\Delta_3(x_{\widehat j})^{r_j}\Bigr]^2
\prod_{\substack{u<v\\ u,v\neq i}}\bigl(k_u^4+k_u^2k_v^2+k_v^4\bigr)\Big|_{x=k^2},
\end{aligned}
\end{equation}
in which every factor $k_u^4+k_u^2k_v^2+k_v^4=(k_u^2)^2+(k_uk_v)^2+(k_v^2)^2$ is itself a sum of
squares; hence both are sums of squares of polynomials in $k$ with positive integer coefficients.
\item[(iii)] Consequently the four-soliton condition of the G\"urses--Pekcan family
$D_x(D_x^3+\Lam)^{2n+1}$ and of Ito's family $D_t(D_x^3+D_t)^{2n+1}$ fails at \emph{every} real
nondegenerate configuration of the principal component ($\bp_1,\dots,\bp_4\in V_Q$, $k_i\neq0$,
$k_i\neq\pm k_j$), not merely generically: for such data there is no four-soliton solution of
the form \eqref{eq:ansatz}, and hence no $N$-soliton solution of that form with $N\ge4$, since
every four of its momenta would have to satisfy the 4SC.
\end{enumerate}
\end{theorem}

\begin{proof}
(i) For $(\beta,\gamma)=(1,0)$ one has $h_i=1$ in \eqref{eq:Bh}, and every term of the sum in
\eqref{eq:residual} is a positive coefficient times $x_i\prod_jx_j^{n-1-r_j}\Delta_3(x_{\widehat j})^{2r_j}\ge0$
on the closed orthant; for $(\beta,\gamma)=(0,1)$ one has
$h_i=-x_i\prod_{u<v\ne i}(x_u^2+x_ux_v+x_v^2)$, so that $-x_ih_i\ge0$ and every term of
$R_n^{(0,1)}$ is nonnegative. This proves the inequalities, and both residuals vanish exactly
when all terms vanish.

Let $n\ge2$ and consider three cases. If all $x_i>0$, a term vanishes if and only if some
$\Delta_3(x_{\widehat j})^{2r_j}$ with $r_j\ge1$ vanishes; if all four $\Delta_3(x_{\widehat j})$
vanish, every term does (as $\sum_jr_j=n-1\ge1$), and if some $\Delta_3(x_{\widehat j})\neq0$
the composition $r_j=n-1$, $r_u=0$ ($u\ne j$) gives nonzero terms. All four
$\Delta_3(x_{\widehat j})$ vanish if and only if no three coordinates are pairwise distinct, i.e.\
if and only if at most two distinct values occur. If exactly one coordinate vanishes, say
$x_4=0$, Theorem~\ref{thm:slice} in the forms \eqref{eq:slice-KdV} and \eqref{eq:slice-Ito}
shows that both residuals vanish if and only if two of $x_1,x_2,x_3$ coincide, because
$x_1+x_2+x_3>0$ and $2\sum x_i^2+\sum_{i<j}x_ix_j>0$; this is again the condition that at most
two distinct positive values occur. If at least two coordinates vanish, a term of
\eqref{eq:residual} can be nonzero only if $n-1-r_j=0$ for every vanishing coordinate $x_j$,
which is impossible for two indices when $n\ge2$; so both residuals vanish, and at most two
positive coordinates, hence at most two distinct positive values, occur. For $n=1$ the
statements follow from $R_1^{(1,0)}=-6\sum_ix_i$ and \eqref{eq:R1Ito-formula}: in the latter the
$i$-th term is nonzero if and only if $x_i>0$ and no two of the other three coordinates vanish,
which happens for some $i$ if and only if at least three coordinates are positive.

(ii) is \eqref{eq:residual} with $x_j=k_j^2$: for $(1,0)$ each term is the coefficient times
the displayed square, and for $(0,1)$ the factor $-x_ih_i=x_i^2\prod_{u<v\ne i}(x_u^2+x_ux_v+x_v^2)$
supplies the extra $k_i^2$ inside the square and the quadratic factors outside it.
(iii) follows from (i) and \eqref{eq:S4-En}: for $\bp_i\in V_Q$ with real $k_i\neq0$,
$k_i\neq\pm k_j$, the $x_i$ are positive and pairwise distinct, so $R_n\neq0$ and $S_4\neq0$
(the sign of $S_4$ is that of $(k_1k_2k_3k_4)^{3(2n+1)}R_n$, but only the nonvanishing is
needed).
\end{proof}

For $n=2,3$ the zero locus of (i) had been observed before the formula \eqref{eq:residual}
was available: exact substitution shows that $R_n$ vanishes on the strata $\{x_i=x_j=x_k\}$ and
$\{x_i=x_j\}\cap\{x_k=x_l\}$ and not on single-pair strata, e.g.\ $R_2^{(1,0)}(3,2,2,1)=-11520$,
$R_3^{(1,0)}(5,3,3,1)=-79272345600$, $R_2^{(0,1)}(3,2,2,1)=35796480$,
$R_3^{(0,1)}(3,2,2,1)=6595706880$; and sign-definiteness had been tested at several thousand
exact rational points for $n\in\{2,3,5,9,15\}$. Theorem~\ref{thm:sign} replaces these
observations by a statement valid for all $n$, and \eqref{eq:SOS}, \eqref{eq:SOS-Ito} are the
sum-of-squares certificates that the mixed coefficient signs of Table~\ref{tab:Rn} conceal.
For $n\ge2$ the residual vanishes on the triple-coincidence and pair--pair strata, which lie in
the zero set of the prefactor $x_1x_2x_3x_4\prod_{i<j}(x_i-x_j)^2$ of \eqref{eq:structure};
consequently the full four-soliton polynomial vanishes on these strata to higher order than the
prefactor alone implies. Since all terms of \eqref{eq:residual} have the same sign, the
lowest-order terms cannot cancel, and the order of vanishing of $R_n$ in the normal directions
is $2n-2$ on $\{(a,a,a,b)\}$ and on $\{(a,a,b,b)\}$ ($a\neq b$ positive) and $6n-6$ on the
diagonal $\{(a,a,a,a)\}$; the prefactor adds $6$, $4$ and $12$, respectively.

\subsection{Real multisoliton data}\label{subsec:realdata}

\begin{theorem}[Real multisoliton data lie on the principal component]\label{thm:GP-whole}
Let $n\ge1$, let $L$, $\Lam$ be real, and let $\bp_1,\dots,\bp_N$, $N\ge4$, be real nondegenerate
data on $V_L\cup V_Q$ for which the function \eqref{eq:ansatz} solves
$L(D_x^3+\Lam)^{2n+1}f\cdot f=0$ for all phases. Then all $\bp_j$ lie on $V_Q$.
\end{theorem}

\begin{proof}
Two momenta on $V_L$ are degenerate ($L(\bp_i\pm\bp_j)=0$), so at most one momentum,
say $\bp_0$ with $z:=k_0$ and $c:=Q(\bp_0)\neq0$, lies on $V_L\setminus V_Q$; choose three of the
others, $\bp_1,\bp_2,\bp_3\in V_Q$, with $k_i\neq0$, $k_i^2$ pairwise distinct. Each mixed triple
$(\bp_i,\bp_j,\bp_0)$ must satisfy $S_3=0$, hence, by \eqref{eq:mixed-criterion},
\begin{equation}\label{eq:A-criterion}
9z^2(k_i^2-z^2)(k_j^2-z^2)+3cz(k_i^2+k_j^2-3z^2)+c^2=0\qquad\text{for all }i<j\le3 .
\end{equation}
If $z=0$ the left-hand side is $c^2\neq0$. If $z\neq0$, subtracting \eqref{eq:A-criterion} for
$(1,2)$ and $(1,3)$ gives $(k_2^2-k_3^2)\bigl[9z^2(k_1^2-z^2)+3cz\bigr]=0$, hence
$k_1^2=z^2-c/(3z)$; substituting this into \eqref{eq:A-criterion} for $(1,2)$ leaves
$-3cz^3\neq0$. In both cases \eqref{eq:A-criterion} is violated.
\end{proof}

Real nondegenerate mixed three-soliton data may well exist (Corollary~\ref{cor:mixed-real},
Ito's family); Theorem~\ref{thm:GP-whole} says that they never extend to four-soliton data.
Combining it with Theorem~\ref{thm:sign}(iii) and Proposition~\ref{prop:realsign}:

\begin{corollary}\label{cor:no-real-four}
For $n\ge1$, real $\Lam$ and $L=\beta D_x+\gamma\Lam$ with $\beta\gamma\le0$ (in particular for
the G\"urses--Pekcan and Ito families), the equation $L(D_x^3+\Lam)^{2n+1}f\cdot f=0$ has no
nondegenerate $N$-soliton solution of the form \eqref{eq:ansatz} with $N\ge4$ and real momenta
anywhere on its dispersion variety.
\end{corollary}

\begin{proposition}[Real parameters]\label{prop:realsign}
Let $(\beta,\gamma)\in\mathbb{R}^2\setminus\{0\}$, $n\ge1$, and let $x_1,\dots,x_4>0$ be pairwise
distinct. If $\beta\ge0\ge\gamma$ then $R_n^{(\beta,\gamma)}<0$; if $\beta\le0\le\gamma$ then
$R_n^{(\beta,\gamma)}>0$; if $\beta\gamma>0$ then $R_n^{(\beta,\gamma)}$ changes sign on the open
positive orthant.
\end{proposition}

\begin{proof}
If $\beta\ge0\ge\gamma$, then $\beta-\gamma x_i>0$ and
$B_{uv}=\beta^2+2\beta|\gamma|(x_u+x_v)+\gamma^2(x_u^2+x_ux_v+x_v^2)>0$, so every $h_i>0$ and
\eqref{eq:residual} is minus a sum of positive terms; if $\beta\le0\le\gamma$, then every
$h_i<0$ and \eqref{eq:residual} is a sum of positive terms. If $\beta\gamma>0$, fix
$x$ and consider $R_n^{(\beta,\gamma)}(tx)$ for $t>0$. By Corollary~\ref{thm:structure},
$R_n^{(\beta,\gamma)}(tx)=\beta^7t^{9n-8}R_n^{(1,0)}(x)+O(t^{9n-7})$ as $t\to0$ and
$R_n^{(\beta,\gamma)}(tx)=\gamma^7t^{9n-1}R_n^{(0,1)}(x)+O(t^{9n-2})$ as $t\to\infty$; by
Theorem~\ref{thm:sign} the two leading terms have the signs $-\operatorname{sgn}\beta$ and
$\operatorname{sgn}\gamma$, which are opposite.
\end{proof}

The change of sign in the last case is a statement about the polynomial $R_n^{(\beta,\gamma)}$.
Its real zeros are reached by nondegenerate soliton data, so that the dichotomy of
Proposition~\ref{prop:realsign} is also the dichotomy for the existence of special real
four-soliton solutions:

\begin{theorem}[Real four-soliton data]\label{thm:existence}
Let $n\ge1$ and let $L$, $\Lam$ be real. The equation $L(D_x^3+\Lam)^{2n+1}f\cdot f=0$ admits
real nondegenerate data $\bp_1,\dots,\bp_4$ for which the four-soliton function
\eqref{eq:ansatz} is a solution for all phases $\delta_j$ if and only if either
$L=\beta D_x+\gamma\Lam$ with $\beta\gamma>0$, or $L\notin\operatorname{span}(D_x,\Lam)$. Such data
lie on $V_Q$ (Theorem~\ref{thm:GP-whole}).
\end{theorem}

\begin{proof}
By Theorem~\ref{thm:GP-whole} only data on $V_Q$ need to be considered. If
$L=\beta D_x+\gamma\Lam$ with $\beta\gamma\le0$, Proposition~\ref{prop:realsign} and
\eqref{eq:S4-En} give $S_4\neq0$ at every real nondegenerate configuration of $V_Q$, and there is
no such solution. Let $L=\beta D_x+\gamma\Lam$ with $\beta\gamma>0$; then $\beta/\gamma>0$. Put $x_2:=\beta/(16\gamma)$, $x_3:=\beta/(4\gamma)$ and consider the
boundary slice \eqref{eq:slice} as a function of $x_1$ at $x_4=0$. It vanishes at
$x_1=\beta/\gamma$ because of the factor $\gamma x_1-\beta$, and this zero is simple: the other
factors are nonzero there, since $x_1,x_2,x_3$ are positive and pairwise distinct,
$\gamma x_2-\beta=-15\beta/16$, $\gamma x_3-\beta=-3\beta/4$, and, by \eqref{eq:Theta-cyclic},
$\Theta_{\beta,\gamma}=x_1(\beta-\gamma x_2)(\beta-\gamma x_3)B_{23}
=\frac{\beta}{\gamma}\cdot\frac{15\beta}{16}\cdot\frac{3\beta}{4}\cdot\frac{117\beta^2}{256}\neq0$
at $x_1=\beta/\gamma$ (the summands with $i=2,3$ contain the factor $\beta-\gamma x_1=0$). Hence
$\partial_{x_1}R_n^{(\beta,\gamma)}(\beta/\gamma,x_2,x_3,0)\neq0$, and by the implicit function
theorem there is a real analytic function $x_1(\varepsilon)$ with $x_1(0)=\beta/\gamma$ and
$R_n^{(\beta,\gamma)}(x_1(\varepsilon),x_2,x_3,\varepsilon)=0$ for small $\varepsilon$. For small
$\varepsilon>0$ the four coordinates are positive and pairwise distinct. Put
$k_1:=\sqrt{x_1(\varepsilon)}$, $k_2:=\tfrac14\sqrt{\beta/\gamma}$, $k_3:=\tfrac12\sqrt{\beta/\gamma}$,
$k_4:=\sqrt{\varepsilon}$, and let $\bp_j\in V_Q$ be real momenta with these $x$-components
(they exist because $\Lam$ is real with a nonzero component outside $D_x$). Then
$\ell_j=\beta k_j-\gamma k_j^3=k_j(\beta-\gamma x_j)$, so that $\ell_1=O(\varepsilon)$,
$\ell_4=\beta\sqrt\varepsilon+O(\varepsilon^{3/2})$, $\ell_2=\tfrac{15}{64}\beta\sqrt{\beta/\gamma}$
and $\ell_3=\tfrac38\beta\sqrt{\beta/\gamma}$: for small $\varepsilon>0$ the absolute values
$|\ell_j|$ are pairwise distinct, as are the $|k_j|$, which are nonzero; hence
$\ell_i\pm\ell_j\neq0$, $k_i\pm k_j\neq0$ and $k_ik_j\neq0$, and the data are nondegenerate by
\eqref{eq:phase} (whether or not $\ell_1$ vanishes; in fact it does not, see
Remark~\ref{rem:branch}). By construction $S_4=0$ at these data (Corollary~\ref{thm:structure}),
$S_3=0$ for every triple (Theorem~\ref{thm:II}), and $S_1=S_2=0$; the criterion of
Section~\ref{subsec:NSC} gives the four-soliton solution.

Finally let $L\notin\operatorname{span}(D_x,\Lam)$. Since $D_x$, $\Lam$, $L$ are linearly
independent real forms, for any real numbers $k_j$, $\ell_j$ there are real momenta $\bp_j\in V_Q$
with $x$-components $k_j$ and $L(\bp_j)=\ell_j$. Take the four-soliton data just constructed for
$(\beta,\gamma)=(1,1)$ and prescribe $\ell_j:=k_j-k_j^3$; by Proposition~\ref{prop:reduction}
all conditions $S_M$ depend only on the pairs $(k_j,\ell_j)$, so they vanish for the new data as
well, which are nondegenerate for the same reason.
\end{proof}

\begin{remark}[The zero branch]\label{rem:branch}
The branch $x_1(\varepsilon)$ of the proof can be made explicit to first order. Scaling
$x\mapsto tx$ with $t=\beta/\gamma$ reduces to $\beta=\gamma=1$ by Corollary~\ref{thm:structure}.
At $(x_1,x_2,x_3,x_4)=(1,a,b,0)$ with $a=1/16$, $b=1/4$, the slice \eqref{eq:slice} gives
$\partial_{x_1}R_n^{(1,1)}=2n(2n+1)C_n(1-a)^2(1-b)^2B_{23}$ with
$C_n:=(ab)^{n-1}\bigl((1-a)(1-b)(a-b)\bigr)^{2n-2}$ and $B_{23}=117/256$, while in
\eqref{eq:residual} every term has a factor $x_4$ at $x_1=1$ (because $h_1=0$ and
$B_{14}=x_4^2-x_4$ occurs in $h_2,h_3$), so that only the composition $r_4=n-1$ contributes to
$\partial_{x_4}R_n^{(1,1)}$, which equals
$-2n(2n+1)C_n\,ab(1-a)(1-b)\bigl((1-a)^2+(1-b)^2+(2n-1)B_{23}\bigr)$. Hence
\[
x_1(\varepsilon)=\frac{\beta}{\gamma}+\frac{26n+28}{585}\,\varepsilon+O(\varepsilon^2),\qquad
\ell_1=-\gamma\sqrt{\beta/\gamma}\,\frac{26n+28}{585}\,\varepsilon+O(\varepsilon^2)\neq0 ,
\]
in agreement with the numerical zero $x_1\approx1.0000092$ for $n=1$, $\varepsilon=10^{-4}$
(ancillary files, $n\le6$).
\end{remark}

The four-soliton solutions of Theorem~\ref{thm:existence} live on a hypersurface of the
parameter space, in accordance with Theorem~\ref{thm:III}. Some phase factors of the data
constructed in the proof may be negative; the next theorem shows that positive ones can be
chosen, so that $\tau=f_4$ is bounded below by $1$ and $\log f_4$ is real analytic.

\begin{theorem}[Positive phase factors]\label{thm:positive}
Let $n\ge1$, let $L$, $\Lam$ be real, and suppose that $L=\beta D_x+\gamma\Lam$ with $\beta\gamma>0$
or $L\notin\operatorname{span}(D_x,\Lam)$. Then there are real nondegenerate data
$\bp_1,\dots,\bp_4\in V_Q$ for which \eqref{eq:ansatz} is a four-soliton solution for all phases
and all six phase factors $e^{A_{ij}}$ are positive; consequently
$f_4=1+\sum_{\emptyset\neq I\subseteq[4]}\bigl(\prod_{i<j\in I}e^{A_{ij}}\bigr)e^{\sum_{i\in I}\theta_i}\ge1$
for all real $(t,\delta)$.
\end{theorem}

\begin{proof}
Let first $L=\beta D_x+\gamma\Lam$ with $\beta\gamma>0$. Scaling $x\mapsto(\beta/\gamma)x$
multiplies $R_n^{(\beta,\gamma)}$ by a nonzero constant (Corollary~\ref{thm:structure}) and does
not change the signs of the phase factors, so we may take $\beta=\gamma=1$. Put
\[
x_1=\frac{1}{100n},\qquad x_2=\frac{2}{100n},\qquad x_3=\frac{3}{100n},\qquad x_4=z\ge2,
\]
$k_i:=\sqrt{x_i}>0$, $\ell_i:=k_i(1-x_i)$ and $y_i:=\ell_i^2=x_i(1-x_i)^2$. The function
$y(x)=x(1-x)^2$ is increasing on $(0,1/3)$ and $y(z)\ge2$ for $z\ge2$, so
$0<x_1<x_2<x_3<x_4$ and $0<y_1<y_2<y_3<y_4$; hence $k_i\pm k_j\neq0$, $\ell_i\pm\ell_j\neq0$, the
data are nondegenerate, and by \eqref{eq:phase}
\[
e^{A_{ij}}=\frac{(\ell_i^2-\ell_j^2)(k_i^2-k_j^2)}{(\ell_i+\ell_j)^2(k_i+k_j)^2}\Bigl(\frac{k_i-k_j}{k_i+k_j}\Bigr)^{2n},
\qquad
\operatorname{sgn}e^{A_{ij}}=\operatorname{sgn}\bigl[(y_i-y_j)(x_i-x_j)\bigr]=+1
\]
for every $z\ge2$ (note that $\ell_4<0$, so the sign of $\ell_i+\ell_j$ alone is not
determined). It remains to find $z\ge2$ with $R_n^{(1,1)}(x)=0$. Grouping the terms of
\eqref{eq:residual} with the same composition $r$,
\begin{equation}\label{eq:residual-grouped}
R_n^{(1,1)}=-\sum_{|r|=n-1}\frac{(2n+1)!}{\prod_j(2r_j+1)!}\Bigl[\sum_{i=1}^4(2r_i+1)x_ih_i\Bigr]
\prod_jx_j^{\,n-1-r_j}\Delta_3(x_{\widehat j})^{2r_j}.
\end{equation}
At $z=2$ one has $B_{u4}=1+x_u^2$, and for $u,v\le3$ (where $x_u,x_v\le3/100$)
$22/25<B_{uv}<1$; hence $0<h_i<2$ for $i\le3$ and $h_4=-B_{12}B_{13}B_{23}<-(22/25)^3$. Since
$2r_i+1\le2n$, the bracket in \eqref{eq:residual-grouped} is bounded by
$\sum_{i\le3}(2r_i+1)x_ih_i<2n\cdot2\cdot\frac{6}{100n}=\frac{6}{25}$ plus
$(2r_4+1)x_4h_4\le2h_4<-1$, so every bracket is negative, the other factors are positive, and
$R_n^{(1,1)}(x_1,x_2,x_3,2)>0$. As $z\to\infty$ with $x_1,x_2,x_3$ fixed, $\Delta_3(x_{\widehat j})$
has degree $2$ in $z$ for $j\le3$ and degree $0$ for $j=4$, $h_i$ has degree $4$ with positive
leading coefficient $(1-x_i)B_{uv}$ ($\{u,v\}=[3]\setminus\{i\}$) for $i\le3$, and $h_4$ has
degree $1$; the terms of \eqref{eq:residual-grouped} with $i\le3$ have degree $5n-1-5r_4$ and
those with $i=4$ degree $5n-3-5r_4$, so the top degree $5n-1$ is attained only by the terms with
$r_4=0$, $i\le3$, all of whose leading coefficients are negative. Hence $R_n^{(1,1)}\to-\infty$, and by
continuity there is a zero $z_n>2$. At $(x_1,x_2,x_3,z_n)$ all triples satisfy the 3SC
(Theorem~\ref{thm:II}) and $S_4=0$, which gives the four-soliton solution.

If $L\notin\operatorname{span}(D_x,\Lam)$, the map $\bp\mapsto(k,\Lam(\bp),L(\bp))$ is onto
$\mathbb{R}^3$, and prescribing $L(\bp_i)=\ell_i$ with the pairs $(k_i,\ell_i)$ just constructed
reproduces the same phase factors and the same values of $S_3$, $S_4$
(Proposition~\ref{prop:reduction}).
\end{proof}

\begin{example}\label{ex:positive}
For $P=D_y(D_x^3+D_t)^3$ ($n=1$, $L=D_y$, $\Lam=D_t$) and the momenta $(k,l,\omega)$
\[
\bp_1=(8,8,-512),\quad \bp_2=(7,-7,-343),\quad \bp_3=(4,-4,-64),\quad \bp_4=(1,1,-1)
\]
on $V_Q$, the six phase factors are
$\bigl(\tfrac1{225},\tfrac19,\tfrac{2401}{6561},\tfrac{81}{14641},\tfrac9{16},\tfrac9{25}\bigr)$,
all positive; the four $S_3$ and $S_4$ vanish exactly, and substitution of the sixteen-term
$f_4$ into the bilinear equation gives $81$ exponential coefficients, all zero (ancillary files).
\end{example}

Whether four-soliton data can be extended to five-soliton data is answered in
Section~\ref{subsec:complex}: for $L\in\operatorname{span}(D_x,\Lam)$ never, over $\C$
(Theorem~\ref{cor:max-four}), whereas for $L\notin\operatorname{span}(D_x,\Lam)$ nondegenerate
complex five-soliton data exist, real ones for real $L$ (Theorem~\ref{thm:five-exist}),
and five is the maximum (Theorem~\ref{thm:six-bound}); whether the particular real
four-soliton data of Theorem~\ref{thm:positive} can be extended remains open
(Proposition~\ref{prop:frac-linear} and Theorem~\ref{thm:signed-four} exclude it for two
families). Theorems~\ref{thm:GP-whole} and~\ref{thm:existence}
together with Proposition~\ref{prop:realsign} classify the real members of the class: for
$L=\beta D_x+\gamma\Lam$ with $\beta\gamma\le0$ there are no real nondegenerate four-soliton data
anywhere on the dispersion variety; for $\beta\gamma>0$ and for $L\notin\operatorname{span}(D_x,\Lam)$
special such data exist, with positive phase factors, and they lie on the principal component.

\subsection{Complex data and the four-soliton conditions for $n=1$}\label{subsec:complex}
The reality assumption in Theorem~\ref{thm:GP-whole} is used only through the positivity of the
sum in \eqref{eq:threeblock}. For $n=1$ that sum is the constant $6$, and the argument works
over $\C$.

\begin{theorem}[Complex data, $n=1$]\label{thm:complex-n1}
Let $L$, $\Lam$ be as in \eqref{eq:LLam} (complex) and $n=1$. If $\bp_1,\dots,\bp_N$, $N\ge4$, are
nondegenerate data on $V_L\cup V_Q$ for which \eqref{eq:ansatz} solves $L(D_x^3+\Lam)^3f\cdot f=0$
for all phases, then all $\bp_j$ lie on $V_Q$.
\end{theorem}

\begin{proof}
For $n=1$, \eqref{eq:threeblock} reads $\mathcal T_1=48\,EUV$. For nondegenerate data with one
momentum $\bp_0\in V_L\setminus V_Q$ the factors $U,V$ and the prefactor of
\eqref{eq:mixed-factor} are nonzero, so $S_3=0$ for a mixed triple if and only if the bracket in
\eqref{eq:blockE} vanishes, i.e.\ \eqref{eq:A-criterion}, over $\C$ as well; the elimination in
the proof of Theorem~\ref{thm:GP-whole} then gives the contradiction $-3cz^3\neq0$.
\end{proof}

\begin{theorem}[No five-soliton data for $L\in\operatorname{span}(D_x,\Lam)$, $n=1$]\label{cor:max-four}
Let $n=1$ and $L=\beta D_x+\gamma\Lam\neq0$ with $\beta,\gamma\in\C$. Over $\C$, the equation
$L(D_x^3+\Lam)^3f\cdot f=0$ has nondegenerate four-soliton solutions of the form
\eqref{eq:ansatz}, but no nondegenerate $N$-soliton solution with $N\ge5$. More precisely, five
nondegenerate momenta of $V_Q$ never satisfy the five four-soliton conditions of their
four-element subsets simultaneously.
\end{theorem}

\begin{proof}
By Theorem~\ref{thm:complex-n1} all momenta lie on $V_Q$, and it suffices to exclude five
momenta. If $\gamma=0$ the 4SC reads $R_1^{(1,0)}=-6\sum_{i=1}^4k_i^2=0$ by \eqref{eq:S4-En};
five momenta would need $\sum_{i\in I}k_i^2=0$ for every four-element $I\subseteq[5]$, and
subtracting two such conditions gives $k_i^2=k_j^2$. Let $\gamma\neq0$, divide by $\gamma$ and
write $L=aD_x+\Lam$, $a:=\beta/\gamma$. On $V_Q$ one has $\ell_i=k_i(a-x_i)$, $x_i=k_i^2$, so
$y_i=x_i(a-x_i)^2$ and $(y_u-y_v)/(x_u-x_v)=a^2-2a(x_u+x_v)+x_u^2+x_ux_v+x_v^2$; by
\eqref{eq:S4-n1-general} the 4SC for four momenta is $\mathcal R_a(x_1,\dots,x_4)=0$ with
\[
\mathcal R_a(x_1,\dots,x_4):=\sum_{i=1}^4x_i(a-x_i)\prod_{\substack{u<v\\ u,v\neq i}}\bigl[a^2-2a(x_u+x_v)+x_u^2+x_ux_v+x_v^2\bigr],
\]
a symmetric polynomial. Nondegeneracy of five momenta means that the $x_i$ are nonzero and
pairwise distinct and the $y_i$ pairwise distinct. Let $e_1,\dots,e_5$ be the elementary
symmetric polynomials of $x_1,\dots,x_5$ and $f(z):=\prod_{i=1}^5(z-x_i)$. Expressing
$\mathcal R_a$ through the elementary symmetric polynomials of its four arguments, which for the
four $x_j$ with $j\neq i$ are $e_1-z,\ e_2-e_1z+z^2,\ e_3-e_2z+e_1z^2-z^3,\ e_4-e_3z+e_2z^2-e_1z^3+z^4$
at $z=x_i$, and reducing modulo $f$, one obtains a polynomial $h(z)$ of degree at most $4$
with $h(x_i)=0$ for the five distinct $x_i$; hence $h\equiv0$. The coefficients of $z^4$ and
$z^3$ determine $e_4$ and $e_5$ as polynomials in $a,e_1,e_2,e_3$; after substituting them,
the coefficients of $z^2$ and $z$ are $\tfrac16FU$ and $-\tfrac16FV$ with
$F:=e_1^2-4ae_1-e_2+6a^2$ and $U,V$ polynomials of degree one in $e_3$ whose resultant in $e_3$
equals
\[
-6\,(3ae_1-e_2-6a^2)\,(2e_1^2-6ae_1-3e_2+10a^2)\,(2e_1^2-5ae_1-3e_2+6a^2)
\]
(the explicit expressions for $e_4,e_5,U,V$ and this identity are recorded in the ancillary
files). If $F=0$, then $f(z)=\bigl[z^2+(2a-e_1)z+(e_1-3a)^2\bigr]\bigl[z^3-2az^2+a^2z+2ae_1^2-11a^2e_1-e_3+16a^3\bigr]$,
and the three roots of the cubic factor, which are three distinct $x_i$, have the same value
of $y=z(a-z)^2=z^3-2az^2+a^2z$, contradicting the distinctness of the $y_i$. If $F\neq0$, then
$U=V=0$ and one of the three factors of the resultant vanishes; if $e_1=4a$ each of them forces
$e_2=6a^2$ and hence $F=0$, so $e_1\neq4a$ and $U=0$ determines $e_3$ (its $e_3$-coefficient is
$9(e_1-4a)$). In the three branches $f(z)$ is divisible by $z(z-a)^3$, by $(z-a)^2$ and by
$z(z-a)$, respectively (with $f=z(z-a)^3(z-e_1+3a)$ in the first), so $f$ has a zero or a
repeated root, contradicting the nondegeneracy of the $x_i$; this includes $a=0$. Hence no five
nondegenerate momenta satisfy all five four-soliton conditions.

Existence of four-soliton data: for $\gamma=0$ take $k=(1,2,3,\mathrm{i}\sqrt{14})$ with
$\Lam(\bp_j)=-k_j^3$, so that $\sum k_i^2=0$ with pairwise distinct nonzero squares. For
$a\neq0$ take the real four-soliton data $\kappa_1,\dots,\kappa_4$ of Theorem~\ref{thm:positive}
for $(\beta,\gamma)=(1,1)$, choose $c$ with $c^2=a$ and put $k_i=c\kappa_i$; then
$\ell_i=k_i(a-k_i^2)=c^3\kappa_i(1-\kappa_i^2)$, so the ratios of the $k_i$ and of the $\ell_i$
are unchanged, the phase factors are unchanged and $S_4$ is multiplied by a nonzero constant.
For $a=0$ (Ito's family) let $x_1,\dots,x_4$ be the roots of $z^4+2z^2-\sqrt2\,z+1$, whose
discriminant is $404$ and constant term $1$, so the $x_i$ are nonzero and distinct; the
$y_i=x_i^3$ are the roots of $y^4-3\sqrt2\,y^3+8y^2+4\sqrt2\,y+1$ (discriminant $19796$), hence
distinct, and with $e_1=0$, $e_2=2$, $e_3=\sqrt2$, $e_4=1$ the residual
$R_1^{(0,1)}/6=-3e_1^2e_2e_4+2e_1^2e_3^2+2e_1e_3e_4+4e_2^2e_4-3e_2e_3^2-4e_4^2$ vanishes; with
$k_i^2=x_i$, $\Lam(\bp_i)=-k_i^3$ this is a nondegenerate four-soliton solution.
\end{proof}

For $n\ge2$ the conclusion of Theorem~\ref{thm:GP-whole} fails over $\C$: the sum in
\eqref{eq:threeblock} can vanish.

\begin{theorem}[Complex mixed four-soliton data, $n\ge2$]\label{thm:complex-mixed}
For every $n\ge2$ and every $\Lam$ as in \eqref{eq:LLam}, the equation
$D_x(D_x^3+\Lam)^{2n+1}f\cdot f=0$ has nondegenerate four-soliton solutions of the form
\eqref{eq:ansatz} with three momenta on $V_Q$ and one on $V_L\setminus V_Q$.
\end{theorem}

\begin{proof}
It suffices to prescribe the values $(k_j,\Lam(\bp_j))$ (Lemma~\ref{lem:mixed-factor},
Proposition~\ref{prop:reduction}). Let $\zeta:=(-1+\mathrm{s})/2$ with $\mathrm{s}:=\sqrt{-3}$,
so $\zeta^3=1$, and take $\bp_1,\bp_2,\bp_3$ with $k_j=1,\zeta,\zeta^2$ and $\Lam(\bp_j)=-1=-k_j^3$
(on $V_Q$), and $\bp_0$ with $k_0=0$, $\Lam(\bp_0)=c\neq0$ (on $V_L\setminus V_Q$). Direct
computation gives the phase factors $e^{A_{12}}=e^{A_{13}}=e^{A_{23}}=(-3)^{n+1}$ and
$e^{A_{j0}}=1$, and all pair factors are nonzero. Put $m:=2n+1$, $F_m(u,x):=(u+x)^m+(u-x)^m$ and
\begin{equation}\label{eq:Gn}
G_n(t):=\sum_{r=1}^{n}\binom{2n+1}{2r}\bigl[1-(-3)^{-r}\bigr]t^{\,r-1},
\end{equation}
a polynomial of degree $n-1$ with nonzero constant and leading coefficients, which for $n\ge2$
has a nonzero root $t_0$; choose $c$ with $c^2=9t_0$. The three-soliton condition for the
triple $(\bp_1,\bp_2,\bp_0)$ reduces, after cancelling nonzero factors and using
$k_1+k_2=-\zeta^2$, $k_1-k_2=\mathrm{s}\zeta^2$, to
$(-3)^{n+1}F_m(1,c/3)-\mathrm{s}\,F_m(\mathrm{s},c/3)=0$, and the binomial expansion with
$\mathrm{s}^{m+1}=(-3)^{n+1}$ gives the identity
$(-3)^{n+1}F_m(1,x)-\mathrm{s}\,F_m(\mathrm{s},x)=2(-3)^{n+1}x^2G_n(x^2)$; hence the condition
holds for our $c$. The triples $(\bp_1,\bp_3,\bp_0)$ and $(\bp_2,\bp_3,\bp_0)$ follow by the
substitution $k\mapsto\zeta k$, which permutes $\bp_1,\bp_2,\bp_3$ cyclically and fixes $\bp_0$,
and $(\bp_1,\bp_2,\bp_3)\in V_Q^{\,3}$ satisfies the 3SC by Theorem~\ref{thm:II}. Finally, the
same substitution leaves $Q$ invariant and multiplies $P$ by $\zeta$; each term of $S_4$
contains seven factors $P$, so $S_4\mapsto\zeta^7S_4=\zeta S_4$, while the symmetry of $S_4$ in
its arguments gives $S_4\mapsto S_4$; hence $S_4=0$, and the criterion of
Section~\ref{subsec:NSC} applies.
\end{proof}

For $c=3\sqrt{-3}$ the mixed three-soliton condition reduces to $2\cos((2n+1)\pi/3)=1$, i.e.\
$n\equiv0,2\pmod 3$, so the same data serve for $n=2,3,5,6,\dots$ (ancillary files, $n=2,3,5,6$).
In fact $c$ can always be taken purely imaginary:

\begin{proposition}\label{prop:Gn-root}
For every $n\ge2$ the polynomial $G_n$ of \eqref{eq:Gn} has a negative real root. Consequently
the data of Theorem~\ref{thm:complex-mixed} can be chosen with $\Lam(\bp_0)=\mathrm{i}\rho_n$,
$\rho_n>0$. The same data, with $L(\bp_j)=k_j$ prescribed for the three momenta of $V_Q$ and
$L(\bp_0)=0$, give nondegenerate mixed four-soliton solutions of $L(D_x^3+\Lam)^{2n+1}f\cdot f=0$
for every $L\notin\operatorname{span}(D_x,\Lam)$.
\end{proposition}

\begin{proof}
For $n=2,3$ one has $G_n(-3)=0$. Let $n\ge4$, $m=2n+1\ge9$, and put
$\theta:=2\pi\lfloor m/4\rfloor/m$, so that $\cos(m\theta)=1$ and $\pi/3<\theta<\pi/2$ (for $m=9$
directly, for $m\ge11$ from $\lfloor m/4\rfloor\ge(m-3)/4$, and $\theta<\pi/2$ because $m$ is odd).
With $x=\mathrm{i}y$, $y=\tan\theta$, the identity in the proof of Theorem~\ref{thm:complex-mixed}
gives
\begin{align*}
-2y^2G_n(-y^2)&=(1+\mathrm{i}y)^m+(1-\mathrm{i}y)^m-\Bigl(1+\frac{y}{\sqrt3}\Bigr)^m-\Bigl(1-\frac{y}{\sqrt3}\Bigr)^m\\
&=\sec^m\!\theta\Bigl[2\cos(m\theta)-\Bigl(\cos\theta+\frac{\sin\theta}{\sqrt3}\Bigr)^m-\Bigl(\cos\theta-\frac{\sin\theta}{\sqrt3}\Bigr)^m\Bigr].
\end{align*}
On $(\pi/3,\pi/2)$ one has $0<\cos\theta+\sin\theta/\sqrt3<1$ and $-1<\cos\theta-\sin\theta/\sqrt3<0$,
and $m$ is odd, so the bracket exceeds $1$; hence $G_n(-\tan^2\theta)<0<G_n(0)=\tfrac43\binom{2n+1}{2}$
and $G_n$ has a root $t_n\in(-\tan^2\theta,0)$. Choose $c=3\mathrm{i}\sqrt{-t_n}$. For
$L\notin\operatorname{span}(D_x,\Lam)$ the map $\bp\mapsto(k,L(\bp),\Lam(\bp))$ is onto, and the
prescribed values reproduce all pair factors and all $S_M$ of the case $L=D_x$
(Lemma~\ref{lem:mixed-factor}, Proposition~\ref{prop:reduction}).
\end{proof}

Theorems~\ref{thm:complex-n1} and~\ref{thm:complex-mixed} show that the restriction to $n=1$ in
the former and to real data in Theorem~\ref{thm:GP-whole} are both sharp. For $n=1$ the
four-soliton condition on the principal component can be written for an arbitrary linear
factor.

\begin{proposition}[Four-soliton condition for $n=1$ and arbitrary $L$]\label{prop:four-n1}
Let $n=1$ and let $\bp_1,\dots,\bp_4\in V_Q$ be nondegenerate, with $x_i=k_i^2$ and
$y_i:=\ell_i^2$ (so that the $x_i$ are pairwise distinct and nonzero, and the $y_i$ are pairwise
distinct). Then
\begin{equation}\label{eq:S4-n1-general}
S_4=-96\cdot3^{21}\Bigl(\prod_{i=1}^4k_i\Bigr)^{11}\Delta_4(x)^2\sum_{i=1}^4k_i\ell_i\,
\frac{\Delta_3(y_{\widehat i})}{\Delta_3(x_{\widehat i})},
\qquad\text{so}\qquad
S_4=0\iff\sum_{i=1}^4k_i\ell_i\,\frac{\Delta_3(y_{\widehat i})}{\Delta_3(x_{\widehat i})}=0 .
\end{equation}
\end{proposition}

\begin{proof}
By \eqref{eq:B1}, $e_1(\ell)\Delta_4(\ell)=\sum_ic_i\ell_i$ with $c_i=(-1)^{i+3}\Delta_3(y_{\widehat i})$,
and by \eqref{eq:B2}, $H_4\Delta_4=-\sum_id_i\prod_{j\ne i}k_j$ with
$d_i=(-1)^{i+3}\Delta_3(x_{\widehat i})$. In the cube of the latter, the component of parity
class $\{i\}$ arises only from choosing each of the three summands $j\ne i$ once, and equals
$-6\bigl(\prod_{j\ne i}d_j\bigr)k_i^3\prod_{j\ne i}k_j^2$. Pairing with the class-$\{i\}$
component $c_i\ell_i$ of the first factor in \eqref{eq:Sigma} and using
$\prod_id_i=\Delta_4(x)^2$, $c_i/d_i=\Delta_3(y_{\widehat i})/\Delta_3(x_{\widehat i})$ and
$c_{4,1}=3^{21}$ in \eqref{eq:reduction} gives \eqref{eq:S4-n1-general}; the denominators are
nonzero by nondegeneracy.
\end{proof}

For $\ell_i=\beta k_i-\gamma k_i^3$ this reduces to \eqref{eq:S4-En} with $R_1^{(\beta,\gamma)}$ of
Section~\ref{sec:structure}; in general it applies to independent $(k_i,\ell_i)$
(regime (i) of Remark~\ref{rem:regimes}).

\begin{proposition}[No five-soliton data on a fractional-linear graph, $n=1$]\label{prop:frac-linear}
Let $n=1$ and $L$ arbitrary. If nondegenerate momenta $\bp_1,\dots,\bp_N\in V_Q$ satisfy a common
relation $y_i=(ax_i+b)/(cx_i+d)$ with $ad-bc\neq0$ and $cx_i+d\neq0$, then \eqref{eq:ansatz} is
an $N$-soliton solution for all phases only if $N\le4$.
\end{proposition}

\begin{proof}
For such data $(y_u-y_v)/(x_u-x_v)=(ad-bc)/((cx_u+d)(cx_v+d))$, so for every four-element set
$I$ the condition \eqref{eq:S4-n1-general} becomes, after multiplication by the nonzero factor
$\prod_{j\in I}(cx_j+d)^2/(ad-bc)^3$, $\sum_{i\in I}k_i\ell_i(cx_i+d)^2=0$. For $N\ge5$ the
conditions for the five four-element subsets of a five-element set show first that the five
numbers $k_i\ell_i(cx_i+d)^2$ are equal, then that they vanish; hence $\ell_i=0$ for these $i$,
contradicting the distinctness of the $y_i$.
\end{proof}

The family $\ell_i=\pm k_i$ (where the condition reads $\sum_{i\in I}k_i\ell_i=0$) is the case
$a=1$, $b=c=0$, $d=1$. Outside such families, five nondegenerate momenta of $V_Q$ satisfy the
five four-soliton conditions if and only if $DA=0$, where $D_{ii}=0$,
$D_{ij}=(y_i-y_j)/(x_i-x_j)$ and $A_i=k_i\ell_i\prod_{j\ne i}(x_i-x_j)/(y_i-y_j)$; this is a
necessary condition for a five-soliton solution, whose sufficiency would still require
$S_5=0$; the example in the proof of Theorem~\ref{thm:signed-four} below shows that it is not
sufficient. For the four-soliton data with $\ell_i$ proportional to $\pm k_i$ the extension
question can be settled completely, over $\C$ and without restricting the fifth momentum to
the family.

\begin{theorem}[Signed four-soliton data admit no fifth soliton]\label{thm:signed-four}
Let $n=1$, let $L$ be arbitrary, and let $\bp_1,\dots,\bp_4\in V_Q$ be nondegenerate momenta,
real or complex, with $\ell_i=\lambda\varepsilon_ik_i$, $\lambda\in\C\setminus\{0\}$,
$\varepsilon_i\in\{\pm1\}$, and $\sum_{i=1}^4\varepsilon_ik_i^2=0$. Then \eqref{eq:ansatz} is a
four-soliton solution with phase factors
$e^{A_{ij}}=\bigl((k_i-k_j)/(k_i+k_j)\bigr)^{3+\varepsilon_i\varepsilon_j}$, and there is no
momentum $\bp_5$, real or complex, such that $\bp_1,\dots,\bp_5$ are nondegenerate five-soliton
data. The phase factors are positive whenever the $k_i$ are real; if in addition the momenta
and the phase constants are real, then $f_4\ge1$ on the real independent variables.
\end{theorem}

\begin{proof}
Dividing $L$ by $\lambda$ changes neither the equation nor the conditions, so let $\lambda=1$.
With $y_i=\ell_i^2=x_i$ the criterion \eqref{eq:S4-n1-general} reads $\sum_i\varepsilon_ix_i=0$, so
$S_4=0$; the triples satisfy the 3SC by Theorem~\ref{thm:II}, the phase factors are as stated
by \eqref{eq:phase} (the exponent $3+\varepsilon_i\varepsilon_j$ is even, which gives positivity
for real data). Suppose $\bp_5$ extends the data. By Theorem~\ref{thm:complex-n1},
$\bp_5\in V_Q$; put $X:=k_5^2$, $Y:=\ell_5^2$, $W:=k_5\ell_5$ (so $W^2=XY$), and
$w_i:=k_i\ell_i=\varepsilon_ix_i$. Nondegeneracy gives $X\neq0$, $X\neq x_i$, $Y\neq x_i$, so
$r_i:=(Y-x_i)/(X-x_i)$ is defined and nonzero. For each $i$, the four momenta $\bp_5$ and
$\bp_j$, $j\neq i$, must satisfy \eqref{eq:S4-n1-general}, which, since
$\Delta_3(y_{\widehat5})/\Delta_3(x_{\widehat5})=1$ and the ratio for $\bp_j$ equals
$\prod_{h\neq i,j}r_h$, reads $W+\sum_{j\neq i}w_j\prod_{h\neq i,j}r_h=0$. Dividing by
$\prod_{h\neq i}r_h$ and putting $T:=\sum_iw_i/r_i$, $B_0:=W/\prod_ir_i$, the four conditions
become $w_i=Tr_i+B_0r_i^2$ ($i=1,\dots,4$), and summing $w_i/r_i$ gives $3T+B_0\sum_ir_i=0$.
Multiplying by $(X-x_i)^2$: the quadratic polynomial
$R(t):=(Y-t)\bigl[T(X-t)+B_0(Y-t)\bigr]$ takes the value $w_i(X-x_i)^2$ at $t=x_i$, $i=1,\dots,4$,
so it is the interpolation polynomial of these four values. If $X=Y$, all $r_i=1$ and the
conditions give $W=w_i$ for all $i$; two of the $\varepsilon_i$ coincide, so two of the $x_i$
would coincide. Hence $d_*:=Y-X\neq0$. Put $p(t):=\prod_i(t-x_i)$; then $B_0=R(X)/d_*^2$,
$T=[t^2]R-B_0$, $R(Y)=0$, $\sum_ir_i=4+d_*p'(X)/p(X)$ and $B_0=Wp(X)/p(Y)$. Changing all
$\varepsilon_i$ at once changes nothing, so three sign patterns remain.

\emph{All $\varepsilon_i$ equal.} The interpolation polynomial of the values $x_i(X-x_i)^2$ is
$t(X-t)^2$, which has degree three; it cannot equal the quadratic $R$.

\emph{Two of each sign.} Let $a,d$ be the $x_i$ with $\varepsilon_i=+1$ and $b,c$ those with
$\varepsilon_i=-1$; then $s:=a+d=b+c$, and with $A:=ad$, $B:=bc$, $C:=A+B$, $\delta:=B-A$ one has
$\delta\neq0$ (the $x_i$ are distinct), and $p(t)=(t^2-st+A)(t^2-st+B)$. At the four nodes
$w_i=x_i(2x_i^2-2sx_i+C)/\delta$, so the interpolation polynomial is the remainder of
$t(2t^2-2st+C)(X-t)^2/\delta$ modulo $p$. Its $t^3$-coefficient is $(2(X-s)^2-C)/\delta$ and
must vanish, so $u:=X-s$ satisfies $u^2=C/2$; put $h:=s+2u$. Then
$R(t)=\bigl[h(Ct^2-sCt+2AB)+\tfrac12\delta^2t\bigr]/\delta$, and
$R(X)=2u\,p(X)/\delta$, $R'(X)=2(2u^2h^2-p(X))/\delta$, $[t^2]R=2u^2h/\delta$, $p'(X)=2uh^2$
(ancillary files). If $u=0$ then $R(X)=[t^2]R=0$ and $R'(X)=-2p(X)/\delta\neq0$, so $R$ is a
nonzero linear polynomial with the single root $X\neq Y$, contradicting $R(Y)=0$. If $h=0$,
$u\neq0$, then $[t^2]R=p'(X)=0$, whence $T=-B_0$ and $3T+4B_0=B_0=0$, contradicting
$B_0=2up(X)/(\delta d_*^2)\neq0$. So $u,h,d_*\neq0$, and $R(Y)=R(X)+R'(X)d_*+[t^2]R\,d_*^2=0$
together with $3T+B_0\sum_ir_i=0$ give
\[
u\,p(X)+\bigl(2u^2h^2-p(X)\bigr)d_*+u^2h\,d_*^2=0,\qquad
3uh\,d_*^2+2uh^2d_*+p(X)=0 ;
\]
eliminating $p(X)$ yields $uh\,d_*^2(3d_*+2h-2u)=0$, hence $d_*=-2X/3$, $Y=X/3$, and then
$12AB+C^2-2s^2C=0$. For real data this is impossible, since $12AB\le3C^2$ and
$0<C<s^2/2$ give $12AB+C^2-2s^2C\le2C(2C-s^2)<0$. Over $\C$ we use $W^2=XY$: with $C=2u^2$ the
last relation gives $AB=u^2(s^2-u^2)/3$ and $\delta^2=\tfrac43u^2(2u-s)(2u+s)$, and
$W=2up(Y)/(\delta d_*^2)$ with $p(Y)=-\tfrac4{81}X^3(2u-s)$ gives
$W^2=X^2(2u-s)/(27(2u+s))$; comparing with $XY=X^2/3$ yields $u=-\tfrac58s$, so $s\neq0$,
$X=\tfrac38s$, $Y=\tfrac18s$, $A,B=\tfrac{s^2}{64}(25\mp10\sqrt3)$. After scaling to $s=8$ and
using the symmetries of $S_5$ (a sign change of a single momentum, permutations of the
momenta and a common nonzero scaling of $k,\ell$, under which $\Sigma_{5,1}$ is homogeneous of
degree $50$, all preserve the vanishing of $S_5$), the only candidate is $x_{1,2}=4\pm\sqrt{-9+10\sqrt3}$,
$x_{3,4}=4\pm\sqrt{-9-10\sqrt3}$, $\varepsilon=(+,+,-,-)$, $k_5=\sqrt3$, $\ell_5=-1$; it is
nondegenerate, all its triples and four-subsets satisfy the conditions, but an exact
computation in $\mathbb{Z}[\sqrt3]$ gives $\Sigma_{5,1}=-2^{29}3^55^8\cdot13\cdot73\,\sqrt3\neq0$
(ancillary files), so $S_5\neq0$ by Proposition~\ref{prop:reduction}.

\emph{Three of one sign.} Let $x_1,x_2,x_3$ have $\varepsilon_i=+1$ and $x_4=s\neq0$ have
$\varepsilon_4=-1$, so $x_1+x_2+x_3=s$; write $P_3(t):=\prod_{i\le3}(t-x_i)=t^3-st^2+qt-r$,
$p=(t-s)P_3$, and note $P_3(s)=sq-r\neq0$. The interpolation polynomial is
$R(t)=t(X-t)^2-2s(X-s)^2P_3(t)/P_3(s)$; its $t^3$-coefficient vanishes if and only if
$P_3(s)=2su^2$ with $u:=X-s$, so $u\neq0$ and $R(t)=(s-2X)t^2+(X^2-q)t+r$, with
$R(X)=-P_3(X)\neq0$, $R'(X)=-P_3'(X)$, $p'(X)/p(X)=1/u+R'(X)/R(X)$. With $K:=s-2X$, the trace
relation reads $3Kd_*^2+R'(X)d_*+R(X)d_*/u+R(X)=0$; subtracting $R(Y)=0$ gives
$2Kd_*^2+R(X)d_*/u=0$, so $K\neq0$ and $d_*=-R(X)/(2uK)$, and then $R(Y)=0$ becomes
$4u^2K-2uR'(X)+R(X)=u(s^2+q-3u^2)=0$. Hence $q=3u^2-s^2$, $r=s(u^2-s^2)$, $R(X)=-4u^2X$,
$d_*=-2uX/h$, $Y=sX/h$ with $h:=s+2u$, and $W=-p(Y)/(ud_*^2)=s^2X/h^2$; comparing with
$W^2=XY$ gives $h^3=s^3$, and $u\neq0$ forces $h=\omega s$ with $\omega^2+\omega+1=0$. After
scaling to $s=2$ the only candidate has $x_1,x_2,x_3$ the roots of
$t^3-2t^2-(9\omega+4)t+6\omega+8$, $x_4=2$, $x_5=-\omega^2$, $\ell_5=\omega k_5$; it is
nondegenerate, and an exact computation in $\mathbb{Z}[\omega]$ gives
$\Sigma_{5,1}=1556247085056+1246373609472\,\omega\neq0$ (ancillary files). In all cases no fifth
momentum exists.
\end{proof}

The two candidates in the proof, realized for instance with $L=D_y$, $\Lam=D_t$ and
$\bp_i=(k_i,\ell_i,-k_i^3)$, are nondegenerate five-point configurations of $V_Q$ for which every
triple satisfies the 3SC and every four-subset satisfies the 4SC, while $S_5\neq0$; thus the
condition $DA=0$ above is not sufficient for a five-soliton solution. Example~\ref{ex:positive}
is of the type of the theorem ($\varepsilon=(+,-,-,+)$, $s=65$, $A=64$, $B=784$), as is
$k=(1,2,3,\sqrt{14})$, $\ell=(1,2,3,-\sqrt{14})$ with $\varepsilon=(+,+,+,-)$: neither can be
joined by a fifth soliton. The theorem is a maximality statement for the given four momenta
and is needed only for $L\notin\operatorname{span}(D_x,\Lam)$, since for
$L\in\operatorname{span}(D_x,\Lam)$ Theorem~\ref{cor:max-four} excludes all five-soliton data.
For $L\notin\operatorname{span}(D_x,\Lam)$ other five-soliton data do exist.

\subsection{Five-soliton data and the maximal number for $n=1$}\label{subsec:five}

\begin{theorem}[Five-soliton data for $n=1$]\label{thm:five-exist}
Let $n=1$ and $L\neq0$.
\begin{enumerate}
\item[(i)] The equation $L(D_x^3+\Lam)^3f\cdot f=0$ has nondegenerate complex five-soliton data
of the form \eqref{eq:ansatz} if and only if $L\notin\operatorname{span}(D_x,\Lam)$; for such $L$
the data contain a local family with four complex parameters.
\item[(ii)] If $L$ and $\Lam$ are real, it has nondegenerate \emph{real} five-soliton data if
and only if $L\notin\operatorname{span}_{\mathbb R}(D_x,\Lam)$; for such $L$ the real data contain a
local family with four real parameters.
\end{enumerate}
\end{theorem}

\begin{proof}
Necessity is Theorem~\ref{cor:max-four}. For sufficiency it suffices, by
Proposition~\ref{prop:reduction} and the surjectivity of $\bp\mapsto(k,L(\bp),\Lam(\bp))$, to
exhibit five pairs $(k_i,\ell_i)$ with $k_i\neq0$, $k_i\neq\pm k_j$, $\ell_i\neq\pm\ell_j$, at which
the five four-soliton conditions \eqref{eq:S4-n1-general} and $S_5=0$ hold; the triples then
satisfy the 3SC by Theorem~\ref{thm:II}. Fix $k_1=1$, $k_2=2$, $k_3=(11+4\mathrm{i})/6$,
$\ell_1=1$ and regard $z=(k_4,k_5,\ell_2,\ell_3,\ell_4,\ell_5)\in\C^6$ as unknowns; let $F(z)\in\C^6$
consist of the five sums $\sum_{i\in I}k_i\ell_i\Delta_3(y_{I\setminus i})/\Delta_3(x_{I\setminus i})$
over the four-element sets $I\subseteq[5]$ and of $\Sigma_{5,1}/(192\Delta_5(x)^3)$, with
$\Sigma_{5,1}$ the sign sum \eqref{eq:Sigma} for $N=5$, $n=1$. Let $z_0$ be the point with
coordinates (to $34$ decimals; the exact rational $65$-digit values are in the ancillary files)
\begin{align*}
k_4&\approx2.0148971857646313364591265900706061+0.3161316784306446962195133830626596\,\mathrm{i},\\
k_5&\approx3.8510751316560307355214417760345288+0.4195878571222588103990412783155619\,\mathrm{i},\\
\ell_2&\approx1.6341057138530128467103842879374910+0.1818839243981783459493327668135091\,\mathrm{i},\\
\ell_3&\approx0.9771498557953166299144281389909057-0.5951036229137305139167165424783129\,\mathrm{i},\\
\ell_4&\approx1.8584603626021988525415827547321810-0.6820266863721068969765513731971176\,\mathrm{i},\\
\ell_5&\approx6.1264256681573269999977655663386995+2.0777835896054872545726292089256979\,\mathrm{i},
\end{align*}
and let $B$ be the box of all $z$ whose real and imaginary parts differ from those of $z_0$ by at
most $10^{-25}$. With a fixed rational matrix $Y\in\C^{6\times6}$, exact interval arithmetic
(fixed-point intervals with $320$ binary digits and outward rounding, with automatic
differentiation for the Jacobian) establishes
$\|YF(z_0)\|_\infty<10^{-55}$ and $\sup_{z\in B}\|I-YDF(z)\|_\infty<10^{-15}$, where the norms
are those of the underlying real $12$-dimensional maps. Hence $T(z):=z-YF(z)$ maps $B$ into
itself and is a contraction, so it has a unique fixed point $z_*\in B$; since $YDF(z_0)$ is
invertible, so is $Y$, and $F(z_*)=0$. The same computation shows that $k_i$, $k_i\pm k_j$,
$\ell_i\pm\ell_j$ do not vanish anywhere on $B$, so the data are nondegenerate. Finally the
Jacobian $DF(z_*)$ is invertible, so by the implicit function theorem $z_*$ moves
holomorphically with $(k_2,k_3)$, and the independent scalings $k\mapsto ak$, $\ell\mapsto b\ell$
add two parameters. This proves (i); the certificate is a rigorous existence proof, not a
numerical residual: the point $z_*$ is an algebraic point determined by rational equations and
a rational isolating box, and the exact rational centre and the preconditioning matrix $Y$ are
part of the ancillary files. For (ii), necessity is again Theorem~\ref{cor:max-four} (real data
are complex data), and for sufficiency the same construction is carried out with real data: fix $k_1=1$, $k_2=6$,
$k_3=7$, $\ell_4=1$, take $z=(k_4,k_5,\ell_1,\ell_2,\ell_3,\ell_5)$ as unknowns, with the same six
equations, and let $z_0$ be the rational point with
\begin{align*}
k_4&\approx8.387813448448382,& k_5&\approx8.841153999723778,& \ell_1&\approx1.331214703058326,\\
\ell_2&\approx4.910558295977433,& \ell_3&\approx-1.188513552111806,& \ell_5&\approx-1.864964705546826
\end{align*}
(exact values in the ancillary files). Exact interval arithmetic on the real closed box of
radius $10^{-25}$ around $z_0$ gives $\|YF(z_0)\|_\infty<10^{-55}$ and
$\sup_B\|I-YDF\|_\infty<10^{-15}$ for the real $6\times6$ system, so the contraction has a unique
fixed point $z_*\in B$, which is real; nondegeneracy holds on $B$, and since $L,\Lam$ are real
and $L\notin\operatorname{span}_{\mathbb R}(D_x,\Lam)$ the five real pairs $(k_i,\ell_i)$ are realized
by real momenta of $V_Q$. The implicit function theorem again gives a local family, with four
real parameters.
\end{proof}

The next theorem shows that five is the maximum.

\begin{theorem}[Maximal soliton number for $n=1$]\label{thm:six-bound}
Let $n=1$ and let $L$, $\Lam$ be as in \eqref{eq:LLam}, $L\neq0$ arbitrary. Every nondegenerate
$N$-soliton datum of $L(D_x^3+\Lam)^3f\cdot f=0$ has $N\le5$; in fact no six nondegenerate
momenta satisfy all their four-soliton conditions. Consequently the maximal number of
nondegenerate complex solitons is $4$ for $L\in\operatorname{span}(D_x,\Lam)$ and $5$ for
$L\notin\operatorname{span}(D_x,\Lam)$, and for real $L$, $\Lam$ the maximal number of nondegenerate
real solitons is $3$ for $L=\beta D_x+\gamma\Lam$ with $\beta\gamma\le0$, $4$ for $\beta\gamma>0$,
and $5$ for $L\notin\operatorname{span}_{\mathbb R}(D_x,\Lam)$.
\end{theorem}

\begin{proof}
Only the four-soliton conditions are used. By Theorem~\ref{thm:complex-n1} all momenta of a
nondegenerate datum with $N\ge4$ lie on $V_Q$. Put $x_i=k_i^2$, $y_i=\ell_i^2$, $s_i=k_i\ell_i$
and $D_{ij}:=(y_i-y_j)/(x_i-x_j)$ for $i\neq j$; nondegeneracy gives $x_i$ nonzero and pairwise
distinct, $y_i$ pairwise distinct, so $D_{ij}=D_{ji}\neq0$. For a four-element set $I$ the
criterion \eqref{eq:S4-n1-general} reads
$\sum_{i\in I}s_i\prod_{a<b,\,a,b\in I\setminus i}D_{ab}=0$. Suppose six momenta satisfy all
fifteen four-point conditions; let $V=\{1,\dots,6\}$.

\emph{Step 1: all $s_i\neq0$.} At most one $\ell_i$ vanishes, since the $y_i$ are distinct;
suppose $\ell_6=0$. With $t_i:=s_i/D_{i6}=k_i(x_i-x_6)/\ell_i\neq0$ ($i\le5$), the condition
for $I=\{i,j,k,6\}$ divided by $D_{i6}D_{j6}D_{k6}$ reads $t_iD_{jk}+t_jD_{ik}+t_kD_{ij}=0$,
i.e.\ $B_{ij}+B_{ik}+B_{jk}=0$ with $B_{ij}:=D_{ij}/(t_it_j)\neq0$, for every triple in $[5]$.
Fix the vertex $1$ and put $a_i:=B_{1i}$; the triangles containing $1$ give $B_{ij}=-a_i-a_j$,
and those not containing $1$ give $a_i+a_j+a_k=0$ for all triples in $\{2,3,4,5\}$; comparing
two such relations gives $a_i=a_j$ for all $i,j$, hence $3a_i=0$, contradicting $B_{1i}\neq0$.

\emph{Step 2: complementary triangles.} For a three-element $J\subseteq V$ put
\[
c_J:=\frac{\prod_{a<b,\ a,b\in J}D_{ab}}{\prod_{i\in J}s_i}\neq0 .
\]
Dividing the four-point condition for $I$ by $\prod_{i\in I}s_i$ gives
$\sum_{J\subseteq I,\,|J|=3}c_J=0$. For $S\subseteq V$ let
$T(S):=\sum_{J\subseteq S}c_J$; summing over the four-element subsets of $S$ counts each triple
$|S|-3$ times, so $T(S)=0$ whenever $|S|\ge4$. Inclusion--exclusion gives, for a fixed triple
$J_0$, $c_{J_0}=\sum_{B\subseteq J_0}(-1)^{|B|}T(V\setminus B)$ (the coefficient of $c_J$ on the
right is $\sum_{B\subseteq J_0\setminus J}(-1)^{|B|}$, which is $1$ for $J=J_0$ and $0$
otherwise); since $|V|=6$, all terms with $|B|\le2$ vanish and only $B=J_0$ remains:
\begin{equation}\label{eq:complementary}
c_J=-c_{V\setminus J}\qquad\text{for every three-element }J\subseteq V .
\end{equation}

\emph{Step 3: a derivative ratio.} For distinct $a,b\in V$ write $V\setminus\{a,b\}=\{p,q,u,v\}$.
By \eqref{eq:complementary}, $c_{apq}=-c_{buv}$ and $c_{auv}=-c_{bpq}$, so
$c_{apq}c_{auv}=c_{bpq}c_{buv}$; cancelling the common factor $D_{pq}D_{uv}/(s_ps_qs_us_v)$ and
multiplying by $D_{ab}$ gives
$\prod_{j\neq a}D_{aj}/s_a^2=\prod_{j\neq b}D_{bj}/s_b^2$. With $f(z):=\prod_{i\in V}(z-x_i)$ and
$g(z):=\prod_{i\in V}(z-y_i)$ one has $\prod_{j\neq i}D_{ij}=g'(y_i)/f'(x_i)$ and $s_i^2=x_iy_i$;
hence there is $\gamma\neq0$ with $g'(y_i)=\gamma x_iy_if'(x_i)$ for all $i\in V$. But Lagrange
interpolation at the six distinct $y_i$ gives $\sum_{i\in V}y_i/g'(y_i)=0$ (the coefficient of
$z^4$ in the interpolant of $z\mapsto z$), while the partial-fraction decomposition of
$1/(zf(z))$ gives $\sum_{i\in V}1/(x_if'(x_i))=-1/f(0)\neq0$; so
$0=\sum_iy_i/g'(y_i)=\gamma^{-1}\sum_i1/(x_if'(x_i))=-1/(\gamma f(0))$, a contradiction.

Hence $N\le5$. The maximal numbers follow from Theorems~\ref{cor:max-four}
and~\ref{thm:five-exist}; over $\mathbb R$, for $\beta\gamma\le0$ real four-soliton data do not
exist (Theorem~\ref{thm:existence}) while real three-soliton data form a nonempty open set of
$V_Q^{\,3}$ (Theorem~\ref{thm:II}), and for $\beta\gamma>0$ Theorem~\ref{thm:existence} provides
real four-soliton data.
\end{proof}

\subsection{Higher multiplicities and finiteness for all $n$}\label{subsec:higher}
For $n=2$ the maximal soliton number of the G\"urses--Pekcan member can be determined
completely.

\begin{theorem}[$n=2$: maximal soliton number five]\label{thm:n2-five}
Let $\Lam$ be as in \eqref{eq:LLam}. Over $\C$, the maximal number of nondegenerate solitons of
the form \eqref{eq:ansatz} of $D_x(D_x^3+\Lam)^5f\cdot f=0$, on the whole dispersion variety, is
exactly five. Five nondegenerate momenta are five-soliton data if and only if they all lie on
$V_Q$ and their squares $k_i^2$ are the five roots of $z^5-c$ for some $c\neq0$. Moreover, mixed
nondegenerate data (one momentum on $V_L\setminus V_Q$ with $\Lam(\bp_0)=c\neq0$) have at most
four momenta, and three momenta of $V_Q$ complete such a $\bp_0$ to four-soliton data if and only
if their squares are the roots of $z^3+c^2/27$.
\end{theorem}

\begin{proof}
\emph{Data on $V_Q$.} For $n=2$ and $L=D_x$ the explicit residual \eqref{eq:residual} has only
the compositions $r=(1,0,0,0)$ and permutations, with coefficients $60$ ($i=j$) and $20$
($i\neq j$), so
\[
R_2^{(1,0)}(x_1,\dots,x_4)=-20\sum_{j=1}^4\Bigl(\sum_{i=1}^4x_i+2x_j\Bigr)\prod_{i\neq j}x_i\;\Delta_3(x_{\widehat j})^2 .
\]
Let $\mathcal R:=-R_2^{(1,0)}/20$, expressed through the elementary symmetric polynomials of
its four arguments; for five distinct nonzero $x_i$ with elementary symmetric polynomials
$e_1,\dots,e_5$ and $f(z)=\prod(z-x_i)$, the substitution used in the proof of
Theorem~\ref{cor:max-four} turns the five four-soliton conditions into the vanishing of a
polynomial $h(z)$ of degree at most $4$ at the five roots of $f$, hence of its five coefficients,
which are polynomials in $e_1,\dots,e_5$. Since the conditions are homogeneous in $x$, either
$e_1=0$ or, after scaling, $e_1=1$. Exact elimination (Gr\"obner bases with the auxiliary
relation $we_5=1$ encoding $e_5\neq0$; ancillary files) gives: for $e_1=0$ the ideal
$\langle we_5-1,e_2,e_3,e_4\rangle$, i.e.\ $f(z)=z^5-e_5$; for $e_1=1$ a lexicographic basis
whose last element is $(5e_2-2)^2(16e_2-1)$, with $f(z)=(2z-1)^3(4z+1)^2/128$ or
$f(z)=(5z-1)^5/3125$, both with repeated roots. So five-soliton data on $V_Q$ have
$x_i^5=c:=e_5$. Conversely, for $k_j=\zeta^j$ ($j=0,\dots,4$, $\zeta$ a primitive fifth root of
unity) and $\Lam(\bp_j)=-k_j^3$, all pair factors are nonzero, the 3SC holds by
Theorem~\ref{thm:II}, the 4SC by the elimination above, and $S_5=0$ because the substitution
$(k,\Lam(\bp))\mapsto(\zeta k,\zeta^3\Lam(\bp))$ permutes the five momenta cyclically, multiplies
$P$ by $\zeta^{16}=\zeta$ and hence $S_5$ (eleven factors $P$) by $\zeta^{11}=\zeta$, while $S_5$
is symmetric in its arguments; a common scaling and sign changes of individual momenta cover
all $c\neq0$. The phase factors are $e^{A_{ij}}=((k_i-k_j)/(k_i+k_j))^6$.

\emph{Mixed data.} At most one momentum lies on $V_L=\{k=0\}$; let it have $\Lam(\bp_0)=c\neq0$.
By Lemma~\ref{lem:mixed-factor} and \eqref{eq:mixed-closed}, for $n=2$ the mixed three-soliton
condition for $\bp_0$ and two momenta of $V_Q$ with squares $x,y$ reads
$9xy(x-y)^2+c^2(x+y)=0$ after removal of nonzero factors. If four distinct squares
$x_1,\dots,x_4$ occurred, then for fixed $x=x_1$ the other three would be the roots of the cubic
$9xy^3-18x^2y^2+(9x^3+c^2)y+c^2x$ in $y$, whose roots sum to $2x$; so $\sum_ix_i=3x_1$, and
likewise $=3x_j$ for every $j$, forcing all $x_j$ equal. Hence mixed data have at most four
momenta. For three squares $x_1,x_2,x_3$ with $S:=\sum x_i$, $T:=\prod x_i$: the third root of the
cubic for $x=x_i$ is $3x_i-S$ by the sum and $-c^2x_i/(9T)$ by the product, so $S=0$ and
$c^2=-27T$; comparing the $y$-coefficient gives $x_i^3=T$; thus $\{x_i\}$ are the roots of
$z^3-T=z^3+c^2/27$. Conversely these are the data of Theorem~\ref{thm:complex-mixed} for
$n=2$ up to scaling and signs.

\emph{Six momenta.} Every five-element subset of six nondegenerate momenta of $V_Q$ would have
$e_1=0$, and subtracting two such conditions gives two equal squares. Larger $N$ contain
five-element subsets.
\end{proof}

For general $n$ the mixed data of the G\"urses--Pekcan family are bounded as follows.

\begin{proposition}[Mixed data, general $n$]\label{prop:mixed-bound}
For $D_x(D_x^3+\Lam)^{2n+1}f\cdot f=0$, nondegenerate soliton data containing a momentum of
$V_L\setminus V_Q$ have $N\le2$ for $n=1$, $N\le4$ for $n=2$, and $N\le3n-3$ for $n\ge3$. For
$n=3$ the maximum is exactly four.
\end{proposition}

\begin{proof}
For $n=1$, Theorem~\ref{thm:complex-n1} excludes mixed triples ($\mathcal T_1=48EUV$ with
$E\neq0$ for $L=D_x$). Let $n\ge2$, let $\bp_0$ have $\Lam(\bp_0)=c\neq0$, and let
$x_1,\dots,x_M$ be the distinct nonzero squares of the other momenta. Removing nonzero factors
from \eqref{eq:mixed-closed}, every pair $x\neq y$ satisfies $F_n(x,y)=0$ with
\[
F_n(x,y):=\sum_{r=1}^n\binom{2n+1}{2r}3^{2n+1-2r}c^{2r-2}(xy)^{n-r}(x-y)^{2n-2r}\sum_{h=0}^{r-1}\binom{2r}{2h+1}x^{r-h-1}y^h .
\]
For fixed $x\neq0$, $F_n$ is a polynomial of degree $d:=3n-3$ in $y$ whose two leading terms
come from $r=1$ only: $F_n=Ax^{n-1}[y^d-(2n-2)xy^{d-1}+\dots]$ with $A=2n(2n+1)3^{2n-1}$. Hence
$M-1\le d$. If $M=d+1$, the $d$ roots are the other squares, so $\sum_ix_i=(2n-1)x$ for every
$x$, a contradiction; so $M\le d$. If $M=d$, let $f(z)=\prod_{i=1}^d(z-x_i)$, $S=\sum x_i$,
$T=\prod x_i$, and let $r(x)$ be the root of $F_n(x,\cdot)$ other than the $x_i\neq x$; the two
leading coefficients and the constant term give $r(x)=(2n-1)x-S=(-1)^d(c^2/9)^{n-1}x/T$, so $S=0$,
$r(x)=(2n-1)x$ and $T=(-1)^d(c^2/9)^{n-1}/(2n-1)$. Comparing
$F_n(x,y)=Ax^{n-1}(y-(2n-1)x)f(y)/(y-x)$ at $y=x$ with $F_n(x,x)=3(2n+1)2^{2n-1}c^{2n-2}x^{n-1}$
(only $r=n$ contributes) shows that $xf'(x)$ is constant on the roots, so $zf'(z)-df(z)$ is
constant and $f(z)=z^d-\rho$; the two expressions for the diagonal value then force
$6n(n-1)^2=(2n-1)4^{n-1}$, which holds for $n=2$ (the cube-root data of
Theorem~\ref{thm:n2-five}) and fails for $n\ge3$ ($80>72$ for $n=3$, and the ratio of the two
sides increases with $n$). So $M\le d-1$ and $N=M+1\le3n-3$ for $n\ge3$. For $n=3$ the
scaled relation $F_3(x,y)=0$ with $c^2=9$ reads
$3x^2y^2(x-y)^4+10xy(x-y)^2(x+y)+3x^2+10xy+3y^2=0$; for four distinct nonzero squares, the
bivariate remainder of $(x-y)F_3$ modulo $f(x),f(y)$, $f=\prod_{i\le4}(z-x_i)$, must vanish,
which gives six polynomial equations in $e_1,\dots,e_4$ whose successive solution
($e_1=-3e_4$, $e_2=3e_4^2-1/(15e_4)$, $e_3=-e_4^3-106/135$) leaves two incompatible
conditions $75e_4^3+1=0$ and $135e_4^3-19=0$ (ancillary files); so mixed five-point data do
not exist for $n=3$, while mixed four-point data exist by Theorem~\ref{thm:complex-mixed}.
\end{proof}

For every $n$ the number of nondegenerate solitons on the principal component is finite, and
four-soliton data always exist there.

\begin{theorem}[Finiteness and bounds for all $n$]\label{thm:finite-bound}
Let $n\ge1$.
\begin{enumerate}
\item[(i)] For $L=\beta D_x+\gamma\Lam\neq0$, nondegenerate soliton data on $V_Q$ have
$N\le5n+2$, and on the whole dispersion variety $N\le5n+3$. For $L=D_x$ the bound on $V_Q$ is
$N\le5n-1$.
\item[(ii)] For $L=D_x$ and $n\ge3$, nondegenerate data on $V_Q$ have $N\le5n-6$; with
Proposition~\ref{prop:mixed-bound} this bound holds on the whole dispersion variety.
\item[(iii)] For $L=D_x$ and every $n\ge1$ there are nondegenerate complex four-soliton data
on $V_Q$.
\end{enumerate}
Together with Theorems~\ref{cor:max-four} and~\ref{thm:n2-five}, the maximal number of
nondegenerate complex solitons of $D_x(D_x^3+\Lam)^{2n+1}f\cdot f=0$ on its dispersion variety is
$4$ for $n=1$, $5$ for $n=2$, and lies in $[4,5n-6]$ for $n\ge3$.
\end{theorem}

\begin{proof}
(i) If a nonzero polynomial $R(z_1,\dots,z_m)$ of degree at most $d$ in each variable vanishes
at all $m$-tuples of distinct elements of an $N$-element set, then $N\le d+m-1$: otherwise, fixing
distinct $z_1,\dots,z_{m-1}$, the last variable has at least $d+1$ zeros, so $R$ vanishes for all
$z_m$; releasing the variables one at a time gives $R\equiv0$. In \eqref{eq:residual} the
factor $x_4^{\,n-1-r_4}$, the three Vandermonde factors containing $x_4$ (degree $4$ each,
raised to $r_j$) and $x_ih_i$ (degree at most $1$ in $x_4$ for $L=D_x$; at most $4$ for general
$(\beta,\gamma)$, since $h_i$ contains two factors $B_{u4}$ of degree $2$, or $\beta-\gamma x_4$)
give $\deg_{x_4}R_n^{(1,0)}\le5n-4$ and $\deg_{x_4}R_n^{(\beta,\gamma)}\le5n-1$; the residuals are
nonzero (Theorem~\ref{thm:III}), symmetric, and vanish on all four-element subsets of a datum,
so the lemma with $m=4$ gives the bounds on $V_Q$; at most one further momentum lies on
$V_L\setminus V_Q$.

(ii) Let $L=D_x$, $n\ge2$, and let $x_1,\dots,x_N$ be the squares of a datum on $V_Q$, so that
$R:=R_n^{(1,0)}$ vanishes on all four-subsets. From \eqref{eq:residual}, for distinct nonzero
$a,b,c$ and $G(a,b,c):=-(2n+1)2^{2n-1}(abc)^{n-1}[(a-b)(a-c)(b-c)]^{2n-2}\neq0$,
\begin{equation}\label{eq:R-special}
R(a,b,c,c)=G(a,b,c)\,(a+b+2nc),\qquad R(a,b,c,0)=\frac{n}{4^{n-1}}\,G(a,b,c)\,(a+b+c);
\end{equation}
in the first only the compositions with $r_1=r_2=0$ survive and the binomial sums
$\sum_{j\ \mathrm{odd}}\binom{2n}{j}=2^{2n-1}$, $\sum_{j\ \mathrm{even}}\binom{2n-1}{j}=2^{2n-2}$
give the coefficients, in the second only $r_4=n-1$ survives. In particular $R(a,b,c,z)\not\equiv0$
in $z$. Let $f(z):=\prod_{i=1}^N(z-x_i)$; for every triple $a,b,c$ of nodes the remaining
nodes are roots of $R(a,b,c,\cdot)$, so
$R(a,b,c,z)=q_{abc}(z)f(z)/((z-a)(z-b)(z-c))$ with $\deg q_{abc}\le5n-1-N$, and by
\eqref{eq:R-special}, with $S:=a+b+c$, $\alpha:=2n-1$, $\rho:=n/4^{n-1}$,
\begin{equation}\label{eq:q-values}
\frac{q_{abc}(t)}{G(a,b,c)}=\frac{(S+\alpha t)\prod_{r\in\{a,b,c\}\setminus t}(t-r)}{f'(t)}\ (t\in\{a,b,c\}),\qquad
\frac{q_{abc}(0)}{G(a,b,c)}=-\frac{\rho\,abcS}{f(0)} .
\end{equation}
We use the following lemma: if $x_1,\dots,x_N$ are distinct, $N\ge6$, $\alpha\neq-3$, and
$u_1,\dots,u_N$ satisfy $(x_i+x_j+x_k)(u_i+u_j+u_k)+\alpha(x_iu_i+x_ju_j+x_ku_k)=0$ for all
distinct $i,j,k$, then all $u_i=0$. (Alternating sums over $(a,e,c)$, $(a,e,d)$, $(b,e,c)$,
$(b,e,d)$ give $(c-d)(u_a-u_b)+(a-b)(u_c-u_d)=0$, so the secant slopes of disjoint pairs are
opposite; three disjoint pairs among six nodes then have all slopes zero, so all $u_i$ are equal
to some $u$, and $(3+\alpha)u(x_i+x_j+x_k)=0$ with some triple sum nonzero gives $u=0$.)
If $N=5n-1$, $q$ is constant and \eqref{eq:q-values} at $b$ and $c$ gives
$(c-a)(a+b+2nc)f'(b)+(b-a)(a+c+2nb)f'(c)=0$, a quadratic in $a$ vanishing at the other
$N-2\ge3$ nodes, hence identically; its $a^2$ and $a^1$ coefficients give $f'(c)=-f'(b)$ and
$2(1-n)(c-b)f'(b)=0$, impossible. If $N=5n-2$, $q$ is affine, so the second divided difference
of \eqref{eq:q-values} vanishes: $\sum_{t\in\{a,b,c\}}(S+\alpha t)/f'(t)=0$, which is the lemma
with $u_i=1/f'(x_i)$ and forces $u_i=0$. If $N=5n-3$, $\deg q\le2$ and Lagrange interpolation of
$q(0)$ from \eqref{eq:q-values} gives $S\sum_t1/(tf'(t))+\alpha\sum_t1/f'(t)=-\rho S/f(0)$, the
lemma with $u_i=1/(x_if'(x_i))+\rho/((3+\alpha)f(0))$; so $x_if'(x_i)$ is constant, whence
$zf'(z)-Nf(z)$ is constant and $f=z^N-c$, and the constant then forces
$N=2(n+1)4^{n-1}/n$, which exceeds $5n-3$ for $n\ge2$. Hence $N\le5n-4$. For $n\ge3$ the values
$N\in\{5n-4,5n-5\}$ ($N\ge10$, $\deg q\le4$) are excluded by differentiating
\eqref{eq:R-special} at a repeated point: with $v_i:=1/(x_if'(x_i))$, $w_i:=f''(x_i)/(2f'(x_i))$,
$\tau:=\rho/f(0)$, $H_t:=S+(2n-1)t$, the derivative $\partial_zR(a,b,c,z)|_{z=c}=\frac12\frac{d}{dc}R(a,b,c,c)$
and the cubic interpolant $Q_3$ of $q$ at $0,a,b,c$ give $q=Q_3+\kappa z(z-a)(z-b)(z-c)$ with
\[
-\frac{\kappa}{G}=Z_c:=\frac{H_av_a}{c-a}+\frac{H_bv_b}{c-b}+v_c\Bigl[H_c\Bigl(\frac{3-n}{2c}-(n-1)\Bigl(\frac1{c-a}+\frac1{c-b}\Bigr)+w_c\Bigr)-n\Bigr]+\frac{\tau S}{c},
\]
and symmetrically $Z_a=Z_b=Z_c$. Fixing $b,c$ and clearing denominators, $Z_b-Z_c=0$ is, in
the free node $a$, a relation $(2na+b+c)v_a=U_{bc}(a)$ with $\deg U_{bc}\le3$; comparing two
such relations at $N-3\ge7$ nodes shows that $U_{bc}$ is divisible by $2nz+b+c$ with a
quotient $V$ of degree at most $2$ independent of $b,c$, so $v_i=V(x_i)$ for all $i$. Inserting
this into $Z_b-Z_c=0$ and clearing $(b-a)(c-a)$ gives a cubic in $a$ vanishing at $N-2\ge8$
nodes, hence identically; its residue at $a=c$ is $(2-n)v_c[b+(2n+1)c]$, so $b=-(2n+1)c$ for
every $b\neq c$, contradicting distinctness. All these identities are verified symbolically in
the ancillary files. Hence $N\le5n-6$ for $n\ge3$.

(iii) For $a=1$, $b=4$, $c=9$ the polynomial $R(1,4,9,z)$ has degree exactly $5n-4\ge1$ in $z$
(the top coefficient comes from $i=4$, $r_4=0$, and is a sum of negative terms), so it has a
root $z_*$, which by \eqref{eq:R-special} is not in $\{0,1,4,9\}$; then $k=(1,2,3,\sqrt{z_*})$
with $\Lam(\bp_i)=-k_i^3$ are nondegenerate four-soliton data.
\end{proof}

For $L\notin\operatorname{span}(D_x,\Lam)$ and $n\ge2$ the pairs $(k_i,\ell_i)$ are free, so the
one-variable interpolation argument of Theorem~\ref{thm:finite-bound} does not apply; finiteness
nevertheless holds for every $n\ge1$ and every $L$.

\begin{theorem}[Finiteness for every linear factor]\label{thm:finite-all}
Let $n\ge1$ and let $L\neq0$ be an arbitrary linear form. Nondegenerate soliton data of
$L(D_x^3+\Lam)^{2n+1}f\cdot f=0$ on $V_Q$ have $N\le40n+16$, and on the whole dispersion variety
$N\le40n+17$. Only the four-soliton conditions are used.
\end{theorem}

\begin{proof}
\emph{Effective coordinates and bidegree.} On $V_Q$ write $\ell_i=L(\bp_i)$, $x_i=k_i^2$ and
$z_i:=\ell_i/k_i$ (defined since $k_i\neq0$), so that $y_i=\ell_i^2=x_iz_i^2$. Let $\Sigma_{4,n}$
be the sign sum \eqref{eq:Sigma} for $N=4$. In the expansion of \eqref{eq:B1}--\eqref{eq:B2},
the parity class of $e_1(\ell)\Delta_4(\ell)$ is a singleton $\{j\}$ and each term of
$(H_4\Delta_4)^{2n+1}$ selects the summands of \eqref{eq:B2} with multiplicities $m_i$,
$\sum m_i=2n+1$; matching the class forces $m_j$ even and the other $m_i$ odd, so, apart from
the Vandermonde squares, $k_j$ occurs with exponent $2n+1-m_j\ge3$ and every other $k_i$ with
exponent $2n+1-m_i\ge2$. Hence $\widehat F_n:=\Sigma_{4,n}/\prod_ik_i^2$ is a polynomial, and after
the substitution $\ell_i=k_iz_i$ it is even in each $k_i$ (the sign sum is invariant under the
simultaneous sign change of $(k_i,\ell_i)$), hence a polynomial in $(x_i,z_i)$. Counting
degrees in the same expansion ($\ell_j$ occurs once, the other $\ell_i$ at most four times, and
a $k_i$ occurs with exponent $2n+1-m_i$ from the row expansion and $4m_h$ from each Vandermonde
$\Delta_3(x_{\widehat h})^{2m_h}$, $h\neq i$),
\begin{equation}\label{eq:bidegree}
\deg_{x_i}\widehat F_n\le a:=5n+2,\qquad\deg_{z_i}\widehat F_n\le b:=4\qquad(i=1,\dots,4).
\end{equation}
For nondegenerate data the $x_i$ are nonzero and distinct and the $y_i$ distinct, and the
four-soliton conditions say that $\widehat F_n$ vanishes at all four-tuples of distinct points
of the effective set $S=\{(x_i,z_i)\}$; it also vanishes when two effective points coincide
(lift them to equal pairs $(k,\ell)$: the terms with equal signs vanish by the Vandermonde
factor, and in the terms with opposite signs the two momenta cancel in the central argument
and their interaction factors with the other momenta pair off, leaving the two-point sum
$\Sigma_2=0$), so $\widehat F_n$ vanishes on $S^4$.

\emph{Curves carry no four-point identity.} Let $B\subset\C^2$ be an irreducible curve on which
neither $k$ nor $\ell$ is constant, and choose a smooth point where $B$ is a graph
$\ell=\varphi(k)$ with $k=\kappa\neq0$, $\varphi(\kappa)=\rho\neq0$, $\varphi'(\kappa)=\eta\neq0$.
Put $k_i=\kappa+\varepsilon a_i$ with $(a_1,\dots,a_4)=(0,1,2,4)$, so
$\ell_i=\rho+\eta\varepsilon a_i+O(\varepsilon^2)$, and expand the sixteen terms of $\Sigma_{4,n}$ in
$\varepsilon$, with $m=2n+1$. Terms with four equal signs are $O(\varepsilon^{6(m+1)})$. Terms with
three equal signs and one opposite have a nonvanishing central factor and three vanishing
pair factors in each of $k$ and $\ell$, hence order $\varepsilon^{3(m+1)}$; terms with two signs of
each kind have central factors $H_4\sim-\kappa^2\varepsilon\sum\sg_ia_i$ and
$e_1(\ell)\sim\eta\varepsilon\sum\sg_ia_i$ and two vanishing pairs, again of total order
$\varepsilon^{3(m+1)}$. In both cases the leading coefficient equals the corresponding coefficient
of the model $\ell_i=k_i$ multiplied by $\eta^3(\rho/\kappa)^4$, and higher derivatives of
$\varphi$ do not enter. Collecting the seven mixed sign classes with $\sg_1=+1$ (the reversed
classes contribute equally), with $p:=m+1=2n+2$,
\begin{equation}\label{eq:curve-obstruction}
[\varepsilon^{6n+6}]\,\Sigma_{4,n}(k,\varphi(k))=2\cdot32^{2n+2}\kappa^{12n+6}\rho^4\eta^3\bigl(1+3^p+6^p+8^p-2^p-5^p-9^p\bigr),
\end{equation}
where the normalized leading coefficients of $H_4\Delta_4$ and of $e_1(\ell)\Delta_4(\ell)$ in
the seven classes with $\sg_1=+1$ (listed in the order $(+,-,-,-)$, $(+,-,-,+)$, $(+,-,+,-)$,
$(+,-,+,+)$, $(+,+,-,-)$, $(+,+,-,+)$, $(+,+,+,-)$) are
\[
(-3,\ 2,\ 9,\ -8,\ -5,\ 6,\ -1)\quad\text{and}\quad(-3,\ -2,\ -9,\ -8,\ 5,\ 6,\ -1)
\]
(ancillary files). Since
$\sum_{b\in\{1,3,6,8\}}(b/9)^p\le\sum_{b\in\{1,3,6,8\}}(b/9)^4=5474/6561<1$, the bracket is
negative for every $n\ge1$, so $\Sigma_{4,n}$ does not vanish identically on $B^4$. In the
effective plane, an irreducible curve $C$ on which $x$ and $y=xz^2$ are both nonconstant lifts
locally to such a $B$ ($k=\sqrt x$ is a local coordinate and $d\ell/dk\neq0$ at a general point),
so $\widehat F_n$ does not vanish identically on $C^4$. The remaining irreducible curves are
$x=\text{const}$, $xz^2=\text{const}\neq0$ (bidegree $(1,2)$), $x=0$ and $z=0$, each of which
contains at most one point of $S$.

\emph{Interpolation closure.} Let $I_{a,b}(S)$ be the space of polynomials of bidegree at most
$(a,b)$ vanishing on $S$ and $Z$ its common zero set. By \eqref{eq:bidegree}, $\widehat F_n$ with
three arguments fixed in $S$ lies in $I_{a,b}(S)$ as a polynomial in the fourth, hence vanishes
on $Z$; iterating, $\widehat F_n$ vanishes on $Z^4$. If $I_{a,b}(S)=0$ then $Z=\C^2$,
contradicting the previous paragraph; so $I_{a,b}(S)\neq0$. Let $G$ be the greatest common
divisor of a basis, of bidegree $(r,s)$; every irreducible factor of $G$ defines a curve
contained in $Z$, hence one of the curves listed above, so $V(G)$ contains at most $r+s$ points
of $S$. The quotients of the basis elements by $G$ have no common factor and bidegree at most
$(a-r,b-s)$; two generic linear combinations of them attain the actual maximal bidegree
$(\alpha,\beta)$ of the quotients and have no common curve, so their bihomogenizations in
$\mathbb P^1\times\mathbb P^1$ have no common component and, by the bidegree B\'ezout theorem,
at most $2\alpha\beta\le2(a-r)(b-s)$ common zeros, among which are the points of $S$ outside
$V(G)$. Hence $N\le r+s+2(a-r)(b-s)\le2ab=40n+16$, the last step being the maximum of a
biaffine function over the rectangle $0\le r\le a$, $0\le s\le b$ at its vertices. Finally at
most one further momentum lies on $V_L\setminus V_Q$.
\end{proof}

\section{Quadric and parabolic components, the dispersions $F(\ell)+Q$ and the cubic graph}\label{sec:quadric}
This section proves Conjecture~\ref{conj:repair}, together with a complete classification of
the polynomials satisfying the 3SC, for several classes of components beyond graph-type
dispersions. Its main results are Theorem~\ref{thm:all-even-quadrics}
(every irreducible even quadric component, in any dimension, for an arbitrary cofactor of any
degree and multiplicity), Corollary~\ref{cor:top-quadric} (the exclusion of every other
component whose top homogeneous part contains a quadratic factor of rank at least three),
Theorem~\ref{thm:parabola-general}
(parabolic components), Theorem~\ref{thm:FQ} (the dispersions $F(\ell)+Q$, among them the
Kadomtsev--Petviashvili surface), Theorem~\ref{thm:cubic-graph} (the cubic graph with a
prescribed one-parameter phase family) and Proposition~\ref{prop:shifted} (translated
hyperbolas of low degree). In every case in which the 3SC holds, all $N$SC hold on the
component, by an elementary fusion argument or, for the KP type, by the classical KP solitons.

The simplest candidates beyond graph-type dispersions are the products $P=HC^m$ of a
quadratic form $H\neq0$ and a power of an irreducible non-hyperplane even quadratic $C=Q+c_0$
($Q$ a quadratic form). Section~\ref{subsec:quadratic-prefactor} classifies them completely
(the 3SC forces $m=1$, $C=uv-1$ up to a linear change of variables, and $H=au^2+buv+cv^2$ in
the same two variables with one algebraic condition on the coefficients; neither the
factorization of the prefactor into linear forms nor the multiplicity one is assumed) and
treats the non-even quadratic components; Section~\ref{subsec:common-phase} exhibits the
families with common phase factors on which the general classification rests; and
Section~\ref{subsec:complete} proves Theorem~\ref{thm:all-even-quadrics} for an arbitrary even
polynomial $P$ divisible by an irreducible non-hyperplane even quadratic, in any dimension and
without restriction on the degree, the cofactor or the multiplicity. The bivariate
classification on which the general result rests (Theorem~\ref{thm:bivariate-complete}) is
proved in Appendix~\ref{app:bivariate}.

\subsection{Quadratic prefactors and non-even quadratic components}\label{subsec:quadratic-prefactor}

\begin{proposition}[Quadrics of rank at least three]\label{prop:quadric}
Let $Q$ be a quadratic form of rank at least three on $\C^s$, $H\neq0$ a quadratic form,
$c_0\in\C$ and $m\ge1$. Then $Q+c_0$ is irreducible, and $P:=H(Q+c_0)^m$ does not satisfy the
3SC on $V_{Q+c_0}$.
\end{proposition}

\begin{proof}
Let $B$ and $K$ be the polar forms of $Q$ and $H$. \emph{The cone, $c_0=0$.} For
$\bp_1,\bp_2,\bp_3\in V_Q$ one has $Q(\sum_i\sg_i\bp_i)=2\sum_{i<j}\sg_i\sg_jB_{ij}$ and
$Q(\sg_i\bp_i-\sg_j\bp_j)=-2\sg_i\sg_jB_{ij}$, $B_{ij}:=B(\bp_i,\bp_j)$, so that
\[
S_3=(-1)^m2^{4m}(B_{12}B_{13}B_{23})^m\,\mathcal T_m,
\]
where $\mathcal T_m:=\sum_{\sg}\bigl(\sum_{i<j}\sg_i\sg_jB_{ij}\bigr)^m H(\sum_i\sg_i\bp_i)\prod_{i<j}H(\sg_i\bp_i-\sg_j\bp_j)$.

Since $\operatorname{rank}Q\ge3$, $Q$ is irreducible, $V_Q^{\,3}$ is an irreducible variety
whose coordinate ring is a domain, and $V_Q$ spans $\C^s$, so $B$ is not identically zero on
$V_Q\times V_Q$; hence the 3SC would force $\mathcal T_m\equiv0$ on $V_Q^{\,3}$. Suppose first
that $H$ is not a multiple of $Q$; then $H$ does not vanish identically on $V_Q$, and we can
choose $\bp,\bq\in V_Q$ with $H(\bp)\neq0$, $B(\bp,\bq)\neq0$, $K(\bp,\bq)\neq0$ (take $\bp$
outside the radical of $B$ and outside $\{H=0\}$; then $B(\bp,\cdot)$ and $K(\bp,\cdot)$ are
nonzero linear forms, the latter because $K(\bp,\bp)=H(\bp)$, and $\bq$ avoids both kernels).
Evaluate $\mathcal T_m$ at $(\bp,\varepsilon\bq,t\bp)$ and expand in $\varepsilon$; with
$\sg=(1,s,r)$ (the other four signs give the same terms) one has
$\sum_{i<j}\sg_i\sg_jB_{ij}=\varepsilon B(\bp,\bq)s(1+rt)$, and the product of the four values of
$H$ equals $H(\bp)^4t^2(1-t^2)^2-2s\varepsilon H(\bp)^3K(\bp,\bq)\,rt(1+rt)(1-rt)^2(t^2+1+rt)+O(\varepsilon^2)$.
Summing over $s,r$:
\begin{align*}
m\text{ even:}\quad&\mathcal T_m=4\varepsilon^mB(\bp,\bq)^mH(\bp)^4t^2(1-t^2)^2\bigl((1+t)^m+(1-t)^m\bigr)+O(\varepsilon^{m+1}),\\
m\text{ odd:}\quad&\mathcal T_m=-8\varepsilon^{m+1}B(\bp,\bq)^mH(\bp)^3K(\bp,\bq)\,t\,F_m(t)+O(\varepsilon^{m+2}),
\end{align*}
with $F_m(t):=\sum_{r=\pm1}r(1+rt)^{m+1}(1-rt)^2(t^2+1+rt)$; here $(1+t)^m+(1-t)^m$ and
$F_m(t)=2mt+O(t^3)$ are nonzero polynomials, so $\mathcal T_m\neq0$ for generic $t$. If $H=\lambda Q$, then $P=\lambda Q^k$ with $k=m+1\ge2$; on a nondegenerate
three-dimensional subspace choose isotropic vectors with Gram matrix $B_{ij}=1$ ($i\neq j$),
which is possible since all nondegenerate symmetric forms of the same dimension are equivalent
over $\C$; then $S_3=\lambda^4(-1)^k2^{4k}\bigl(2\cdot3^k+6(-1)^k\bigr)\neq0$.

\emph{The affine quadric, $c_0\neq0$.} Suppose $P$ satisfies the 3SC on $V_{Q+c_0}$. For
$\mathbf v\in V_Q$ outside the radical choose $\mathbf w\in V_Q$ with $B(\mathbf v,\mathbf w)\neq0$
and put $\mathbf v(\varepsilon):=\mathbf v-\frac{c_0\varepsilon^2}{2B(\mathbf v,\mathbf w)}\mathbf w$,
so that $\varepsilon^{-1}\mathbf v(\varepsilon)\in V_{Q+c_0}$ for $\varepsilon\neq0$. With
$P_\varepsilon:=H(Q+c_0\varepsilon^2)^m$ one has $P(\varepsilon^{-1}\mathbf z)=\varepsilon^{-2-2m}P_\varepsilon(\mathbf z)$,
and since each term of $S_3$ has four factors $P$,
$S_3[P_\varepsilon](\mathbf v_i(\varepsilon))=\varepsilon^{8m+8}S_3[P](\varepsilon^{-1}\mathbf v_i(\varepsilon))=0$;
letting $\varepsilon\to0$ gives $S_3[HQ^m]=0$ on a dense subset of $V_Q^{\,3}$, hence
identically, contradicting the case $c_0=0$.
\end{proof}

\begin{lemma}[Affine quadrics: multiplicity one]\label{lem:mult-quadric}
Let $C=Q+c_0$ with $c_0\neq0$ and $\operatorname{rank}Q\ge2$, let $H\neq0$ be a quadratic form
and $m\ge2$. Then $HC^m$ does not satisfy the 3SC on $V_C$.
\end{lemma}

\begin{proof}
Restrict to a generic plane through the origin on which $Q$ has rank two and $H$ is nonzero;
the 3SC on $V_C$ implies the 3SC on the conic $V_C\cap$plane, which after a linear change of
variables and normalization is $uv=1$, so it suffices to exclude
$P=(au^2+buv+cv^2)(uv-1)^m$ with $(a,b,c)\neq0$ on $\bp(t)=(t,t^{-1})$. Under $t_i\mapsto rt_i$,
i.e.\ $(u,v)\mapsto(ru,r^{-1}v)$, $uv-1$ is invariant, and the coefficient of $r^8$ in $S_3$ is
$a^4S_3[u^2(uv-1)^m]$. At $t=(1,2,4)$ the products of the four values of $u^2$, $uv$ and $uv-1$
over the four evaluation points of a term are, for the sign classes
$(1,1,1)$, $(1,1,-1)$, $(1,-1,1)$, $(-1,1,1)$, respectively
\[
(1764,-\tfrac{441}{64},-\tfrac{5265}{64}),\quad (900,\tfrac{1125}{64},\tfrac{3969}{64}),\quad
(26244,-\tfrac{6561}{64},-\tfrac{3185}{64}),\quad (22500,\tfrac{1125}{64},\tfrac{3969}{64}),
\]
so
\[
S_3[u^2(uv-1)^m]=\frac{2}{64^m}\bigl[1764(-5265)^m+26244(-3185)^m+23400\cdot3969^m\bigr].
\]
Since $13$ divides $5265$, $3185$ and $23400$ but not $3969$ nor $1800=23400/13$, the bracket
is $\not\equiv0\pmod{13^2}$ for $m\ge2$; hence $a=0$, and by the symmetry $u\leftrightarrow v$
also $c=0$. Finally $S_3[uv(uv-1)^m]=\frac{2}{64^{m+1}}[-441(-5265)^m-6561(-3185)^m+2250\cdot3969^m]\not\equiv0\pmod{13}$,
so $b=0$, a contradiction.
\end{proof}

For the remaining case we use coordinates $(u,v,\mathbf z)$ on $\C^s$ with $C=uv-1$; the
component $V_C$ is parametrized by $\bp=(t,t^{-1},\mathbf z)$, $t\neq0$.

\begin{lemma}[Transverse variables]\label{lem:transverse}
Write $H=H_0(u,v)+uM(\mathbf z)+vN(\mathbf z)+T(\mathbf z)$ with $M,N$ linear and $T$ quadratic
in the transverse variables. If $P=H(uv-1)$ satisfies the 3SC on $V_C$, then $M=N=T=0$.
\end{lemma}

\begin{proof}
Suppose $T\neq0$, choose $\mathbf w$ with $T(\mathbf w)\neq0$ and put $\mathbf z_i=Rr_i\mathbf w$.
Each of the four factors $P$ in a term of $S_3$ is $R^2T(\mathbf w)\rho^2C(\cdot)+O(R)$, where
$\rho$ is the corresponding signed combination of $r_1,r_2,r_3$; hence the coefficient of $R^8$
in $S_3$ is $T(\mathbf w)^4$ times $S_3$ for the polynomial $z^2(uv-1)$ in the variables
$(u,v,z)$ at the points $(t_i,t_i^{-1},r_i)$, which at $(t_i,r_i)=(1,1),(2,3),(3,5)$ equals
$260224\neq0$. So $T=0$. If $M,N$ are not both zero, choose $\mathbf w$ with
$(a,b):=(M(\mathbf w),N(\mathbf w))\neq0$; the coefficient of $R^4$ is then $S_3$ for
$P_*:=(au+bv)z(uv-1)$, which must vanish identically in $(t,r)$. Writing $\xi_i:=at_i+b/t_i$ and
$W_\sg(r):=(\sum_i\sg_ir_i)\prod_{i<j}(\sg_ir_i-\sg_jr_j)$, one has
$S_3[P_*]=2\sum_{\sg\in\mathcal E}K_\sg W_\sg(\xi)W_\sg(r)$ with $K_\sg$ the product of the four
values of $C$, and $W_\sg(r)=\sg_1\sg_2(r_1^3r_2-r_1r_2^3)-\sg_1\sg_3(r_1^3r_3-r_1r_3^3)+\sg_2\sg_3(r_2^3r_3-r_2r_3^3)$
(Lemma~\ref{lem:bialternant-parity}). Since $r$ is free and the three characters $\sg_i\sg_j$
are the nontrivial characters of $\mathcal E$, the four numbers $K_\sg W_\sg(\xi)$ must be
equal. At $t=(1,2,3)$ two of their differences are $\frac{35}{4}(2b+7a)(b^3-8a^3)$ and
$\frac{1960}{243}(2b+7a)(b^3-27a^3)$, and at $t=(1,2,4)$ they are
$\frac{25515}{512}(b+4a)(b^3-8a^3)$ and $\frac{184275}{4096}(b+4a)(b^3-64a^3)$ (ancillary files);
their simultaneous vanishing forces $2b+7a=b+4a=0$, i.e.\ $a=b=0$. Hence $M=N=0$.
\end{proof}

\begin{lemma}[Odd partial degrees]\label{lem:odd-degrees}
Let $P(u,v)\neq0$ be an even polynomial and $\bp(t)=(t,t^{-1})$. If
$S_3(\bp(t_1),\bp(t_2),\bp(t_3))\equiv0$, then $\deg_uP$ and $\deg_vP$ are positive odd
numbers.
\end{lemma}

\begin{proof}
Suppose $D:=\deg_uP$ is even and write $P=\sum_{j=0}^Du^jQ_j(v)$ with $Q_D\neq0$; evenness of $P$
makes $Q_D$ even. Put $a:=t_2^{-1}$, $b:=t_3^{-1}$. In each term of $S_3$ the three factors
containing $\bp(t_1)$ have degree at most $D$ in $t_1$, and the coefficient of $t_1^{3D}$ in
$S_3$ equals
\[
4\,Q_D(a)Q_D(b)\Bigl[Q_D(a+b)\,P\bigl(\tfrac1a-\tfrac1b,\,a-b\bigr)+Q_D(a-b)\,P\bigl(\tfrac1a+\tfrac1b,\,a+b\bigr)\Bigr],
\]
which must vanish; hence the bracket vanishes identically. With $z=a+b$, $w=a-b$ this reads
$Q_D(z)P\bigl(-4w/(z^2-w^2),w\bigr)+Q_D(w)P\bigl(4z/(z^2-w^2),z\bigr)=0$; multiplying by
$(z^2-w^2)^D$ and setting $w=z$ kills all terms of $u$-degree less than $D$ and leaves
$\bigl((-4z)^D+(4z)^D\bigr)Q_D(z)^2=2(4z)^DQ_D(z)^2\neq0$, a contradiction. Exchanging $u,v$ and
$t_i\mapsto t_i^{-1}$ gives the statement for $\deg_vP$.
\end{proof}

\begin{proposition}[Bivariate polynomials of bidegree at most $(4,4)$]\label{prop:quadric-3SC}
Let $P(u,v)\neq0$ be an even polynomial with $P(0)=0$, divisible by $uv-1$ and of degree at
most four in each of $u$ and $v$; or, alternatively, of total degree at most six. If $P$
satisfies the 3SC on $V_C$, then $P$ has degree at most three in each variable, so
$P=(uv-1)(d\,u^2v^2+a\,u^2+b\,uv+c\,v^2)$, and, with
$\bp_i=(t_i,t_i^{-1})$, $x_i=t_i^2$, $e_1,e_2,e_3$ the elementary symmetric polynomials of the
$x_i$ and $\Delta:=\prod_{i<j}(x_i-x_j)$,
\begin{equation}\label{eq:quadric-Laurent}
S_3=\frac{8\Delta^2}{e_3^4}\,(b+2d)\,\mathcal H,
\end{equation}
\begin{align*}
\mathcal H={}&e_3(a e_2^2+c e_1^2)\bigl((b+2d)(b+5d)-3ac\bigr)-3d(b+3d)\bigl(a e_1^2e_2e_3+c e_1e_2^2\bigr)\\
&-3d\bigl(a^2e_1^2e_3^2+c^2e_2^2\bigr)+d\bigl(d(b+2d)-3ac\bigr)\bigl(e_1^3e_3+e_2^3\bigr)\\
&-d^2(b+2d)e_1^2e_2^2+\bigl(b^3+7b^2d+13bd^2+3d^3-3ac(b+d)\bigr)e_1e_2e_3 .
\end{align*}
Consequently $P$ satisfies the 3SC on $V_C$ if and only if
\begin{equation}\label{eq:quadric-condition}
b=-2d\qquad\text{or}\qquad d=0,\ b^2=3ac ,
\end{equation}
and in these cases nondegenerate data form a nonempty Zariski-open subset of $V_C^{\,3}$. For
$d=0$, \eqref{eq:quadric-Laurent} reads
$S_3=8b(b^2-3ac)\Delta^2e_3^{-3}\bigl(ae_2^2+be_1e_2+ce_1^2\bigr)$.
\end{proposition}

\begin{proof}
By Lemma~\ref{lem:odd-degrees} the partial degrees of $P$ are odd, hence at most $3$ if they
are at most $4$. If instead the total degree is at most six, write
$P=(uv-1)(a_{40}u^4+a_{31}u^3v+du^2v^2+a_{13}uv^3+a_{04}v^4+au^2+buv+cv^2)$; under
$t_i\mapsto rt_i$ the monomial $u^iv^j$ has weight $i-j$ and $uv-1$ is invariant, so the
coefficient of $r^{16}$ in $S_3$ is $a_{40}^4S_3[u^4(uv-1)]$, and $S_3[u^4(uv-1)]$ equals
$-135475200$ at $t=(1,2,3)$; hence $a_{40}=0$, by symmetry $a_{04}=0$, and the partial degrees
are at most $4$. In both cases multiplication by $uv-1$ raises the degree in $u$ and in $v$ by
exactly one, so the cofactor has bidegree at most $(2,2)$; it is even (the ring is a domain and
$P$, $uv-1$ are even) and vanishes at $0$, which leaves the monomials $u^2,v^2,uv,u^2v^2$. The identity
\eqref{eq:quadric-Laurent} is a finite computation (ancillary files: all $35$ coefficients of
the quartic form in $(a,b,c,d)$ agree), and the coefficient of $t_1^8t_2^0t_3^{-8}$ in $S_3$
equals $-8d^2(b+2d)^2$, so the 3SC forces $d=0$ or $b=-2d$; for $d=0$ the bracket
$ae_2^2+be_1e_2+ce_1^2$ is a nonzero polynomial (its terms have degrees $4,3,2$), which gives
$b(b^2-3ac)=0$, and conversely \eqref{eq:quadric-Laurent} vanishes under
\eqref{eq:quadric-condition}. Nondegeneracy means that the cofactor $H$ and $C$ are nonzero at
$\bp_i\pm\bp_j$; on $uv=1$ the cofactor takes the value $at^2+ct^{-2}+b+d$, which under
\eqref{eq:quadric-condition} vanishes identically only for $P=0$, and
$H(\bp(s)\pm\bp(t))$, $C(\bp(s)\pm\bp(t))=\pm(s^2+t^2)/(st)+1$ are nonzero Laurent polynomials.
\end{proof}

\begin{lemma}[Collision and fusion]\label{lem:fusion}
Let $P(u,v)\neq0$ be even with $P(0)=0$, of degree at most three in each variable, and let
$\bp(t)=(t,t^{-1})$; regard $S_N$ as a Laurent polynomial in $t_1,\dots,t_N$.
\begin{enumerate}
\item[(i)] For $\bp=\bp(t)$, $\bp_j=\bp(t_j)$ and $N\ge3$,
\[
S_N(\bp,\bp,\bp_3,\dots,\bp_N)=2(-1)^{N-1}P(2\bp)\prod_{j=3}^NP(\bp-\bp_j)P(\bp+\bp_j)\,S_{N-2}(\bp_3,\dots,\bp_N).
\]
\item[(ii)] Let $\omega:=e^{\mathrm{i}\pi/3}$, $\mathbf a:=\bp(\omega t)$, $\mathbf b:=\bp(t)$; then
$\mathbf q:=\mathbf a-\mathbf b=\bp(\omega^2t)$ lies on $V_C$. If $uv-1$ divides $P$ and
$S_3\equiv0$ on $V_C^{\,3}$, then for $N\ge4$, as an identity of rational functions of
$t,t_3,\dots,t_N$,
\[
S_N(\mathbf a,\mathbf b,\bp_3,\dots,\bp_N)=(-1)^{N-1}P(\mathbf a+\mathbf b)\prod_{j=3}^N\mathcal K(\bp_j)\;
S_{N-1}(\mathbf q,\bp_3,\dots,\bp_N),
\]
where $\mathcal K(\bp):=P(\mathbf a-\bp)P(\mathbf b+\bp)/P(\mathbf q-\bp)$; if
$P(\mathbf a+\mathbf b)\equiv0$, the left-hand side vanishes identically.
\end{enumerate}
\end{lemma}

\begin{proof}
(i) Terms with $\sg_1=\sg_2$ contain $P(0)=0$. For $\sg_1=-\sg_2$ the sum factor is
$P(\sum_{j\ge3}\sg_j\bp_j)$, the pair factor $(1,2)$ is $\sg_1\sg_2P(2\sg_1\bp)=-P(2\bp)$, and for
each $j\ge3$ the two pair factors give $\sg_1\sg_j\sg_2\sg_jP(\sg_1\bp-\sg_j\bp_j)P(-\sg_1\bp-\sg_j\bp_j)
=-P(\bp-\bp_j)P(\bp+\bp_j)$, independently of the signs; summing over the two choices of $\sg_1$
and over $\sg_3,\dots,\sg_N$ gives the formula.

(ii) That $\mathbf a-\mathbf b=\bp(\omega^2t)$ follows from $\omega-1=\omega^2$ and
$(1-\omega)/\omega=\omega^{-2}$; hence $P(\pm\mathbf q)=0$. Terms with $\sg_a=\sg_b$ contain
$P(\pm\mathbf q)$ and vanish. For $\sg_b=-\sg_a$ the sum factor is
$P(\sg_a\mathbf q+\sum_{j\ge3}\sg_j\bp_j)$, the pair factor $(a,b)$ is $-P(\mathbf a+\mathbf b)$,
and the two pair factors involving $\bp_j$ give
$-P(\mathbf a-\varepsilon_j\bp_j)P(\mathbf b+\varepsilon_j\bp_j)$ with $\varepsilon_j:=\sg_a\sg_j$;
if $P(\mathbf a+\mathbf b)\equiv0$ every term vanishes. Otherwise, applied to a triple
$(\mathbf a,\mathbf b,\bp)$ this gives
$S_3=2P(\mathbf a+\mathbf b)\sum_{\varepsilon=\pm1}P(\mathbf q+\varepsilon\bp)P(\mathbf a-\varepsilon\bp)P(\mathbf b+\varepsilon\bp)$,
and $P(\mathbf q\pm\bp)$ are nonzero Laurent polynomials in the free parameters (the map
$(s,t)\mapsto\bp(s)\pm\bp(t)$ has Zariski-dense image, on which $P\neq0$ cannot vanish
identically); hence the 3SC is equivalent to
$P(\mathbf a-\varepsilon\bp)P(\mathbf b+\varepsilon\bp)=\varepsilon\,\mathcal K(\bp)P(\mathbf q-\varepsilon\bp)$
for $\varepsilon=\pm1$. Inserting this for each $j\ge3$, the pair factors become
$-\mathcal K(\bp_j)\cdot\sg_a\sg_jP(\sg_a\mathbf q-\sg_j\bp_j)$, which are $-\mathcal K(\bp_j)$
times the pair factors of $(\mathbf q,\bp_j)$ in $S_{N-1}$ with $\sg_q:=\sg_a$; the sum over
$\sg_a$ and $\sg_3,\dots,\sg_N$ is the sum defining $S_{N-1}$, and the signs combine to
$(-1)^{N-1}$.
\end{proof}

\begin{theorem}[Quadric components: classification and all $N$-soliton conditions]\label{thm:quadric-NSC}
\mbox{}
\begin{enumerate}
\item[(I)] Let $C$ be an irreducible non-hyperplane even quadratic on $\C^s$, $H\neq0$ a
quadratic form and $m\ge1$, $P=HC^m$. The following are equivalent:
(a) $P$ satisfies the 3SC on $V_C$; (b) after a linear change of variables and normalization,
$m=1$, $C=uv-1$ and $P=(au^2+buv+cv^2)(uv-1)$ with $(a,b,c)\neq0$ and
$b(b^2-3ac)=0$; (c) $P$ satisfies the $N$SC on $V_C$ for every $N\ge3$.
\item[(II)] Let $P(u,v)\neq0$ be even, $P(0)=0$, divisible by $uv-1$ and of degree at most
four in each variable (or of total degree at most six). Then $P$ satisfies the 3SC on $V_C$ if
and only if $P=(uv-1)(du^2v^2+au^2+buv+cv^2)$ with \eqref{eq:quadric-condition}, if and only if
it satisfies the $N$SC on $V_C$ for every $N\ge3$.
\end{enumerate}
In case (II) with $d=0$ the phase factors are, for $\bp_i=(t_i,t_i^{-1})$,
\begin{equation}\label{eq:quadric-phase}
e^{A_{ij}}=\Bigl(\frac{t_i-t_j}{t_i+t_j}\Bigr)^2\,\frac{t_i^2-t_it_j+t_j^2}{t_i^2+t_it_j+t_j^2}\cdot
\frac{a\,t_i^2t_j^2-b\,t_it_j+c}{a\,t_i^2t_j^2+b\,t_it_j+c},
\end{equation}
and in the branch $b=-2d$ they are given by the first two factors alone, independently of
$a,c,d$. For real coefficients and distinct positive $t_i$ chosen outside the degeneracy locus
all phase factors are positive, so that $f_N\ge1$ for real phases.
\end{theorem}

\begin{proof}
(I) (a)$\Rightarrow$(b): by Proposition~\ref{prop:quadric}, $\operatorname{rank}Q\le2$; a quadric
of rank one or of rank two with $c_0=0$ is reducible, so $\operatorname{rank}Q=2$, $c_0\neq0$,
$C=uv-1$ after a linear change of variables and normalization; by Lemma~\ref{lem:mult-quadric},
$m=1$; by Lemma~\ref{lem:transverse}, $H=au^2+buv+cv^2$; by Proposition~\ref{prop:quadric-3SC}
with $d=0$, $b(b^2-3ac)=0$. (c)$\Rightarrow$(a) is trivial, and (b)$\Rightarrow$(c) is the case
$d=0$ of (II).

(II) The reduction to bidegree $(3,3)$ and the equivalence of the 3SC with
\eqref{eq:quadric-condition} are Proposition~\ref{prop:quadric-3SC}. Assume $S_3\equiv0$ and argue by induction on $N\ge4$,
$S_1$, $S_2$ and $S_3$ vanishing. Each factor $P$ containing $\bp_i$ has Laurent degree at most
$3$ in $t_i$ (every monomial $u^jv^k$ has $j,k\le3$), and a term of $S_N$ contains $N$ such
factors, so $S_N$ has $t_i$-degrees in $[-3N,3N]$; since $\bp(-t)=-\bp(t)$, relabelling $\sg_i$
gives $S_N(\dots,-t_i,\dots)=(-1)^{N-1}S_N$, so the degrees $\pm3N$ do not occur and
$F_N(x_1,\dots,x_N):=\bigl(\prod_it_i^{3N-1}\bigr)S_N$ is a symmetric polynomial in $x_i=t_i^2$
of degree at most $3N-1$ in each variable. By Lemma~\ref{lem:fusion}(i) and the induction
hypothesis, $F_N$ vanishes on $x_1=x_2$; being symmetric, it is divisible by $(x_1-x_2)^2$,
hence by $\prod_{i<j}(x_i-x_j)^2$. By Lemma~\ref{lem:fusion}(ii) and the induction hypothesis,
$F_N$ vanishes on $t_1=\omega t_2$, i.e.\ $x_1=\omega^2x_2$, and by symmetry on
$x_1=\omega^{-2}x_2$, so it is divisible by $x_1^2+x_1x_2+x_2^2$ and hence by
$\prod_{i<j}(x_i^2+x_ix_j+x_j^2)$ (the rational-function identity becomes a Laurent identity
after clearing the nonzero denominators). The two products are coprime, so
$\prod_{i<j}(x_i-x_j)^2(x_i^2+x_ix_j+x_j^2)$ divides $F_N$; it has degree $4(N-1)$ in each
variable, which exceeds $3N-1$ for $N\ge4$. Hence $F_N\equiv0$ and $S_N\equiv0$.

For the phase factors, \eqref{eq:phase} with $H(\bp(s)\pm\bp(t))=(s\pm t)^2(a\pm b/(st)+c/(st)^2)$
for $d=0$, respectively $H(\bp(s)\pm\bp(t))=(s\pm t)^2(st)^{-2}\bigl[as^2t^2+c+d(s^2+t^2)\bigr]$
for $b=-2d$, and $C(\bp(s)\pm\bp(t))=\pm(s^2+t^2)/(st)+1$ gives the formulas. For real
coefficients and $t_i>0$ the first two factors of \eqref{eq:quadric-phase} are positive; the
third equals $1$ if $b=0$, and if $b^2=3ac\neq0$ then $ac>0$ and the quadratics $aT^2\pm bT+c$
have discriminant $-ac<0$ and the same sign.
\end{proof}

\begin{corollary}[Even quartics with a quadratic component]\label{cor:quartic}
Let $P$ be an even polynomial of degree four with $P(0)=0$ which has an irreducible
non-hyperplane even quadratic factor $C$. Then $P$ satisfies the 3SC on $V_C$ if and only if
$P$ is as in Theorem~\ref{thm:quadric-NSC}(I)(b), and then it satisfies all $N$SC on $V_C$.
\end{corollary}

\begin{proof}
Write $P=(H+d)(Q+c_0)$ with $H,Q$ quadratic forms, $H\neq0$; the cofactor has no linear term
because $P$ and $C$ are even, and $P(0)=0$ gives $dc_0=0$. If $c_0\neq0$ then $d=0$ and
Theorem~\ref{thm:quadric-NSC}(I) applies. If $c_0=0$, irreducibility gives $\operatorname{rank}Q\ge3$,
and scaling the three momenta on the cone shows that the top-degree part of $S_3[P]$ is
$S_3[HQ]$, which is not identically zero by Proposition~\ref{prop:quadric}.
\end{proof}

\begin{remark}\label{rem:quadric-whole}
(i) Both branches of \eqref{eq:quadric-condition} contain classical equations. For $d=b=0$ and
$(a,c)=(1,0)$, $u^2(uv-1)$ is, up to scaling ($x_1=X$, $x_{-1}=Y/3$), the bilinear form
$(D_{x_{-1}}D_{x_1}^3-3D_{x_1}^2)\tau\cdot\tau=0$ of the first negative flow of the
Sawada--Kotera hierarchy (the reduced Ostrovsky equation in hodograph variables), whose
$N$-soliton solutions \cite[Lemma~2.13, eqs.\ (2.25) and (2.53)]{FMO2012} have exactly the
phase factors \eqref{eq:quadric-phase} with $b=0$; by the symmetry $u\leftrightarrow v$ and
linearity of the bilinear equation in $P$ for a fixed $\tau$, the same $\tau_N$ solves
$(au^2+cv^2)(uv-1)$ for all $(a,c)$, and, by Theorem~\ref{thm:quadric-NSC}(II), also
$uv(uv-1)(uv-2)$, i.e.\ $(D_X^3D_Y^3-3D_X^2D_Y^2+2D_XD_Y)\tau_N\cdot\tau_N=0$; the sextic branch
$b=-2d$ thus shares its solitons with the quartic branch $b=0$ and is the linear span of these
three bilinear operators. For $b^2=3ac\neq0$ the quartic $(u^2+\sqrt3\,uv+v^2)(uv-1)$ becomes,
in the variables $X=u$, $T=u+\sqrt3\,v$, the polynomial
$-\bigl(X^4-XT^3+\sqrt3(X^2+XT+T^2)\bigr)/(3\sqrt3)$, one of the quartics of Hietarinta's
classification \cite{Hietarinta1997}. What is new here is the exhaustive statement, the
necessity of the alignment of the prefactor with the two variables of $C$, and the elementary
proof that the 3SC implies all $N$SC. (ii) Theorem~\ref{thm:quadric-NSC} is a statement about
the component $V_C$, and it does not extend to the whole reducible variety $V(P)$: for
$P=(u^2-v^2)(uv-1)$ the momenta $\bp_1=(2,\tfrac12)$, $\bp_2=(3,\tfrac13)$ on $V_C$ and
$\bp_3=(2,2)$ on $\{u=v\}$ have all six pair factors nonzero, but $S_3=24500000/9\neq0$.
(iii) The proof of the $N$SC uses only the 3SC, the parity of $P$ and a degree count; it is
the same mechanism as in Theorem~\ref{thm:whole}, where two momenta of the principal component
fuse into one.
\end{remark}

\begin{theorem}[Non-even quadratic components]\label{thm:noneven}
Let $C$ be an irreducible quadratic polynomial on $\C^s$ which is not even, let $m\ge1$,
$\gamma\neq0$, and $P:=\gamma\bigl(C(\bp)C(-\bp)\bigr)^m$ with $P(0)=0$. Then $P$ satisfies the
3SC on $V_C$ if and only if $m=1$ and, after a linear change of variables and normalization,
$C=u^2-v$ and $P=u^4-v^2$. In that case $P$ satisfies the $N$SC for every $N$ on the whole
dispersion variety $V(P)=\{v=u^2\}\cup\{v=-u^2\}$ (the remaining variables, if any, being
free), nondegenerate data form a dense open subset, and on $\{v=u^2\}$, with
$\bp(t)=(t,t^2)$,
\begin{equation}\label{eq:parabola-phase}
e^{A_{ij}}=\frac{(t_i-t_j)^2}{t_i^2+t_it_j+t_j^2}.
\end{equation}
In particular, an even polynomial $P$ of degree four with $P(0)=0$ that has an irreducible
non-even quadratic factor $C$ satisfies the 3SC on $V_C$ if and only if it is linearly
equivalent to $u^4-v^2$.
\end{theorem}

\begin{proof}
Write $C=Q+L+c_0$ with $Q\neq0$ a quadratic form and $L\neq0$ linear; $P(0)=\gamma c_0^{2m}=0$
gives $c_0=0$ and $P=\gamma(Q^2-L^2)^m$, and $\gamma$ may be dropped. (If $P$ is an even
quartic divisible by $C$, evenness gives $C(-\bp)\mid P$ as well, and the two irreducible
factors are not associate because $L\neq0$, so $P=\gamma C(\bp)C(-\bp)$.)

\emph{Rank at least three.} For $\varepsilon\neq0$ put $C_\varepsilon:=Q+\varepsilon L$ and
$P_\varepsilon:=(Q^2-\varepsilon^2L^2)^m$; then $C_\varepsilon(\bp)=\varepsilon^2C(\bp/\varepsilon)$
and $P_\varepsilon(\bp)=\varepsilon^{4m}P(\bp/\varepsilon)$, so the 3SC of $P$ on $V_C$ implies
that of $P_\varepsilon$ on $V_{C_\varepsilon}$. Every smooth point of the cone $V_Q$ lifts to
$V_{C_\varepsilon}$ by the implicit function theorem, and letting $\varepsilon\to0$ shows that
$Q^{2m}$ satisfies the 3SC on $V_Q$. But on isotropic vectors with polar values
$B(\bp_i,\bp_j)=1$ ($i\neq j$), which exist as in the proof of Proposition~\ref{prop:quadric},
$S_3[Q^{2m}]=2^{8m}(2\cdot3^{2m}+6)\neq0$.

\emph{Rank two.} Take $Q=uv$. If $L\notin\operatorname{span}(u,v)$, let $w:=L$ be a third
coordinate, so $C=uv+w$, $P_0:=u^2v^2-w^2$, $P=P_0^m$. At the points $(1,2,-2)$, $(2,3,-6)$,
$(4,5,-20)$ of $V_C$ the four sign classes of $S_3[P_0]$ contribute
$1296\cdot(-2083725,\,2010580,\,5504268,\,-5368545)$, so
$S_3[P_0^m]=2\cdot1296^m\bigl((-2083725)^m+2010580^m+5504268^m+(-5368545)^m\bigr)$, which is
nonzero for every $m\ge1$ because only the second base is prime to $3$. If
$L\in\operatorname{span}(u,v)$, irreducibility forces $L=au+bv$ with $ab\neq0$, and after
scaling $C=uv+u+v$, $P_0=u^2v^2-(u+v)^2$; at $\bp(t)=(t-1,t^{-1}-1)$, $t=2,3,5$, the four
classes contribute $\frac{2048}{421875}(-498717,\,386750,\,1344000,\,-1266160)$, and the sum
of $m$-th powers is nonzero for every $m$ because only the first base is odd. In both cases
the data are nondegenerate (ancillary files).

\emph{Rank one.} $Q=u^2$, and irreducibility forces $L$ independent of $u$, so $C=u^2-v$
and $P=(u^4-v^2)^m$ after choosing $v=-L$. At $\bp(t)=(t,t^2)$, $t=1,2,3$, the four classes of
$S_3[u^4-v^2]$ contribute $36864\cdot(-275,\,247,\,665,\,-637)$, so
$S_3[P]=2\cdot36864^m\bigl((-275)^m+247^m+665^m+(-637)^m\bigr)$: for even $m$ all terms are
positive, and for odd $m\ge3$, since $247+665=275+637$ and $x\mapsto x^m$ is strictly convex,
$247^m+665^m>275^m+637^m$. Hence $m=1$.

\emph{All $N$SC for $u^4-v^2$.} Put $A(s,t):=(s-t)^2$, $B(s,t):=s^2+st+t^2$; then
$\sg_i\sg_jP(\sg_i\bp(t_i)-\sg_j\bp(t_j))=-4t_it_jG_{ij}$ with $G_{ij}=A(t_i,t_j)$ if $\sg_i=\sg_j$
and $B(t_i,t_j)$ otherwise, so $S_N=(-4)^{\binom N2}\bigl(\prod_it_i^{N-1}\bigr)F_N$ with
\[
F_N(t_1,\dots,t_N):=\sum_{\sg}\Bigl[\bigl(\textstyle\sum_i\sg_it_i\bigr)^4-\bigl(\sum_i\sg_it_i^2\bigr)^2\Bigr]\prod_{i<j}G_{ij},
\]
a symmetric homogeneous polynomial of degree at most $2\binom N2+4$; $F_1=F_2=0$ directly.
For $t_1=t_2=t$ only the terms with $\sg_1\neq\sg_2$ survive and
\[
F_N(t,t,t_3,\dots,t_N)=6t^2\prod_{j\ge3}A(t,t_j)B(t,t_j)\,F_{N-2}(t_3,\dots,t_N),
\]
so by induction and symmetry $(t_i-t_j)^2\mid F_N$. For $t_1=t$, $t_2=\omega t$ with $\omega^2+\omega+1=0$ one has
$B(t,\omega t)=0$, so only $\sg_1=\sg_2$ survives, and $\bp(t)+\bp(\omega t)=-\bp(\omega^2t)$;
with $K(t,z):=B(\omega^2t,z)$ the identities $A(t,z)A(\omega t,z)=K(t,z)B(\omega^2t,z)$ and
$B(t,z)B(\omega t,z)=K(t,z)A(\omega^2t,z)$ give
\[
F_N(t,\omega t,t_3,\dots,t_N)=(1-\omega)^2t^2\prod_{j\ge3}K(t,t_j)\,F_{N-1}(\omega^2t,t_3,\dots,t_N),
\]
so by induction $t_i^2+t_it_j+t_j^2\mid F_N$. Hence $\prod_{i<j}(t_i-t_j)^2(t_i^2+t_it_j+t_j^2)$,
of degree $4\binom N2$, divides $F_N$, and $4\binom N2>2\binom N2+4$ for $N\ge3$; so
$F_N\equiv0$. Since $P$ is even, changing the sign of one momentum multiplies $S_N$ by
$(-1)^{N-1}$, and every point of $\{v=-u^2\}$ is the negative of a point of $\{v=u^2\}$, so the
identities hold on the whole variety. Finally $P(\bp(s)-\bp(t))=-4st(s-t)^2$ and
$P(\bp(s)+\bp(t))=4st(s^2+st+t^2)$ give \eqref{eq:parabola-phase} and the density of
nondegenerate data.
\end{proof}

Together with Corollary~\ref{cor:quartic} this settles all even quartics with an irreducible
non-hyperplane quadratic factor: up to linear equivalence, the 3SC on the corresponding
component holds exactly for $(au^2+buv+cv^2)(uv-1)$ with $b(b^2-3ac)=0$ (even factor) and for
$u^4-v^2$ (non-even factor), and in both cases all $N$SC hold there; for $u^4-v^2$, which is
the degenerate Boussinesq form $D_u^4-D_v^2$, they hold on the whole dispersion variety.

\subsection{Families with common phase factors}\label{subsec:common-phase}
The three bilinear operators of Remark~\ref{rem:quadric-whole}(i) share their $N$-soliton
solutions. The mechanism extends to degree eight and produces, in particular, a component of
multiplicity two on which all $N$SC hold.

\begin{theorem}[A five-parameter family]\label{thm:octic}
For $(a,c,d,e,f)\in\C^5\setminus\{0\}$ let
\begin{equation}\label{eq:octic}
P=(uv-1)\bigl[au^2+cv^2+d\,uv(uv-2)+e\,u^3v(uv-4)+f\,uv^3(uv-4)\bigr].
\end{equation}
Then $P$ satisfies the $N$SC on $V_C=\{uv=1\}$ for every $N$, with the common phase factors
\begin{equation}\label{eq:common-phase}
e^{A_{ij}}=\Bigl(\frac{t_i-t_j}{t_i+t_j}\Bigr)^2\frac{t_i^2-t_it_j+t_j^2}{t_i^2+t_it_j+t_j^2},
\end{equation}
independent of $a,c,d,e,f$, and nondegenerate data form a nonempty Zariski-open set. In
particular the $N$-soliton function $\tau_N$ with exponents $t_iX+t_i^{-1}Y+\delta_i$ and
phase factors \eqref{eq:common-phase} satisfies, besides the three bilinear equations of
Remark~\ref{rem:quadric-whole}(i),
\[
(D_X^5D_Y^3-5D_X^4D_Y^2+4D_X^3D_Y)\,\tau_N\cdot\tau_N=0,\qquad
(D_X^3D_Y^5-5D_X^2D_Y^4+4D_XD_Y^3)\,\tau_N\cdot\tau_N=0 .
\]
\end{theorem}

\begin{proof}
With $H:=P/(uv-1)$ a direct substitution gives $H(\bp(s)\pm\bp(t))=(s\pm t)^2J(s,t)$ with
$J(s,t)=a+\bigl(c+d(s^2+t^2)+e(s^2-t^2)^2\bigr)/(s^2t^2)+f(s^2-t^2)^2/(s^4t^4)$, hence
$P(\bp(s)-\bp(t))=-\frac{J(s,t)}{st}\mathcal N(s,t)$ and $P(\bp(s)+\bp(t))=\frac{J(s,t)}{st}\mathcal D(s,t)$
with $\mathcal N(s,t):=(s-t)^2(s^2-st+t^2)$, $\mathcal D(s,t):=(s+t)^2(s^2+st+t^2)$; $J\not\equiv0$
since $P\neq0$ and the image of $(s,t)\mapsto\bp(s)+\bp(t)$ is dense. This gives
\eqref{eq:common-phase}. Define, for any even $P$, the normalized residual
\[
\mathcal F_N[P]:=\sum_{\sg}P\Bigl(\sum_i\sg_i\bp(t_i)\Bigr)\prod_{i<j}(t_i-\sg_i\sg_jt_j)^2(t_i^2-\sg_i\sg_jt_it_j+t_j^2),
\]
which is linear in $P$; for the family \eqref{eq:octic},
$S_N[P]=(-1)^{\binom N2}\prod_{i<j}\frac{J(t_i,t_j)}{t_it_j}\,\mathcal F_N[P]$ as rational functions,
so it suffices to prove $\mathcal F_N\equiv0$ for the five basis operators $u^2(uv-1)$,
$v^2(uv-1)$, $uv(uv-1)(uv-2)$, $u^3v(uv-1)(uv-4)$, $uv^3(uv-1)(uv-4)$ (the last two are the
octic operators displayed). The argument avoids dividing by $J$, which may vanish at collision
or fusion points for special parameters. For $t_1=t_2=t$ the terms with $\sg_1=\sg_2$ carry the
factor $(t_1-t_2)^2=0$ and the others cancel in the central argument, so
$\mathcal F_N(t,t,t_3,\dots)=24t^4\prod_{j\ge3}\mathcal N(t,t_j)\mathcal D(t,t_j)\,\mathcal F_{N-2}(t_3,\dots)$.
For $t_1=\omega t$, $t_2=t$ with $\omega^2-\omega+1=0$ one has $\bp(\omega t)-\bp(t)=\bp(\omega^2t)$,
$\mathcal N(\omega t,t)=0$, and with $K(t,s):=(s^2-t^2)(s^2-\omega^2t^2)$ the polynomial
identities $\mathcal N(\omega t,s)\mathcal D(t,s)=K(t,s)\mathcal N(\omega^2t,s)$,
$\mathcal D(\omega t,s)\mathcal N(t,s)=K(t,s)\mathcal D(\omega^2t,s)$ and
$\mathcal D(\omega t,t)=6\omega^2t^4$ give
$\mathcal F_N(\omega t,t,t_3,\dots)=6\omega^2t^4\prod_{j\ge3}K(t,t_j)\,\mathcal F_{N-1}(\omega^2t,t_3,\dots)$,
with no division. Each basis operator has odd partial degrees $(D_u,D_v)$ with $D_u+D_v\le8$;
in $\mathcal F_N$ the $N-1$ pair weights containing $t_i$ have degree $4$ each in $t_i$ and the
central factor has Laurent degree in $[-D_v,D_u]$, and $\mathcal F_N$ is even in each $t_i$
(replace $t_i\mapsto-t_i$, $\sg_i\mapsto-\sg_i$), so
$G_N:=\bigl(\prod_it_i^{D_v-1}\bigr)\mathcal F_N$ is a symmetric polynomial in $x_i=t_i^2$ of degree
at most $2(N-1)+(D_u+D_v-2)/2\le2N+1$ in each variable. Now $\mathcal F_1=0$ since $uv-1\mid P$,
and $\mathcal F_2=0$ by the pair identities above. For $N\ge3$, induction with the collision and
fusion identities and symmetry shows that $\prod_{i<j}(x_i-x_j)^2(x_i^2+x_ix_j+x_j^2)$, of
degree $4(N-1)$ in each variable, divides $G_N$; since $4(N-1)>2N+1$ for $N\ge3$, $G_N\equiv0$.
Linearity in $P$ finishes the proof; the identity $\mathcal F_3\equiv0$ for the two octic
operators is also checked directly as a polynomial identity in the ancillary files.
\end{proof}

\begin{remark}[Multiplicity two]\label{rem:double}
Taking $a=3e$, $c=3f$, $d=0$ in \eqref{eq:octic} gives
$P=(eu^2+fv^2)(uv-1)^2(uv-3)$, which satisfies all $N$SC on the component $uv=1$ although
that component has multiplicity \emph{two}, with nondegenerate data forming a nonempty open
set. Thus multiplicity one is not necessary for the $N$SC on a component; it is the
hypothesis under which Conjecture~\ref{conj:repair} asserts the implication, and the
rigidity of Theorem~\ref{thm:quadric-NSC}(I) holds for homogeneous quadratic prefactors, not
for arbitrary ones.
\end{remark}

\begin{theorem}[Classification with prescribed phase factors]\label{thm:prescribed-phase}
Let $P(u,v)\neq0$ be even, $P(0)=0$, divisible by $uv-1$, of degree at most six in each
variable, and suppose that its generic two-soliton phase factor equals
\eqref{eq:common-phase}. Then $P$ satisfies the 3SC on $V_C$ if and only if $P$ belongs to the
family \eqref{eq:octic}, and then all $N$SC hold.
\end{theorem}

\begin{proof}
The prescribed factor already forces both partial degrees to be odd: with $D=\deg_uP$ and
$P=u^DQ_D(v)+\dots$, $Q_D(-v)=(-1)^DQ_D(v)$, one has
$-P(\bp(s)-\bp(t))/P(\bp(s)+\bp(t))\to-(-1)^D$ as $s\to\infty$, while
\eqref{eq:common-phase} tends to $1$; so $D$ is odd, and likewise $\deg_vP$ (exchange $u,v$).
Hence the partial degrees are at most five. Let $H=P/(uv-1)$, even, $H(0)=0$, of bidegree at
most $(4,4)$. The prescribed phase factor is
equivalent to $H(\bp(s)+\bp(t))/(s+t)^2=H(\bp(s)-\bp(t))/(s-t)^2$. Decompose $H$ by the weight
$i-j$ of $u^iv^j$; scaling $(s,t)\mapsto(rs,rt)$ separates the weights $-4,\dots,4$. The parts
of weight $\pm4$ are $u^4$, $v^4$ and violate the condition, so they vanish. The part of weight
$2$ is $u^2h(w)$ with $w=uv$, $\deg h\le2$; on $\bp(s)\pm\bp(t)$ one has $w_++w_-=4$, and the
condition becomes $h(w)=h(4-w)$, so this part is $au^2+eu^2w(w-4)$; weight $-2$ gives
$cv^2+fv^2w(w-4)$. The part of weight $0$ is $h(w)$ with $h(0)=0$, $\deg h\le4$, and the
condition reads $h(w)/w=-h(4-w)/(4-w)$, so $h(w)=dw(w-2)+gw(w-2)^3$. Hence $P$ lies in the
six-dimensional space spanned by the five operators of \eqref{eq:octic} and
$P_6:=uv(uv-1)(uv-2)^3$. By linearity of $\mathcal F_3$ and Theorem~\ref{thm:octic}, the 3SC
holds if and only if $g\,\mathcal F_3[P_6]=0$, and $\mathcal F_3[P_6]/\prod_{i<j}\mathcal D(t_i,t_j)$
equals $392/27$ at $t=(1,2,3)$; so $g=0$.
\end{proof}

\subsection{The complete classification of even quadric components}\label{subsec:complete}
We now drop every restriction on the cofactor. The bivariate case is the core.

\begin{theorem}[Bivariate classification, arbitrary degree]\label{thm:bivariate-complete}
Let $P\in\C[u,v]\setminus\{0\}$ be even with $P(0)=0$ and $uv-1\mid P$. Then $P$ satisfies the
3SC on $\{uv=1\}$ if and only if, with $z:=uv$ and $C:=uv-1$, either
\begin{equation}\label{eq:SK9}
\begin{aligned}
P=C\bigl[&a u^2+c v^2+d\,z(z-2)+e\,u^2z(z-4)+f\,v^2z(z-4)\\
&+g\,u^4(z-2)(z-4)+h\,v^4(z-2)(z-4)+i\,u^6(z-4)^2+j\,v^6(z-4)^2\bigr]
\end{aligned}
\end{equation}
with $(a,c,d,e,f,g,h,i,j)\neq0$, or $P=C(au^2+buv+cv^2)$ with $b^2=3ac$ (the exceptional quartic
family of \eqref{eq:quadric-condition}). All these $P$ satisfy the $N$SC on $\{uv=1\}$ for every
$N$; those of \eqref{eq:SK9} have the common phase factors \eqref{eq:common-phase}. In
particular $\deg P\le12$ and $\deg_uP,\deg_vP\le9$, and the only members of \eqref{eq:SK9} that
lie in $\C[uv]$ are the multiples of $z(z-1)(z-2)$.
\end{theorem}

The proof, which occupies Appendix~\ref{app:bivariate}, shows that the extreme pure-weight
components of a non-radial $P$ have the phase factor \eqref{eq:common-phase}
(Lemma~\ref{lem:odd-degrees} and a fusion argument), that with this phase factor a single
Laurent coefficient bounds the degree and yields the nine directions of \eqref{eq:SK9}, that
the lower weights of $P$ are then determined by a descent through finite linear kernels after
the removal of a gauge direction --- which is where the exceptional quartic branch, with a
different phase factor, is recovered --- and that the radial polynomials $P=F(uv)$ reduce to
$z(z-1)(z-2)$. The family \eqref{eq:SK9} extends \eqref{eq:octic} by the four
directions $u^4C(z-2)(z-4)$, $v^4C(z-2)(z-4)$, $u^6C(z-4)^2$, $v^6C(z-4)^2$; its members of
multiplicity two are exactly those of Remark~\ref{rem:double}.

\begin{theorem}[All irreducible even quadric components]\label{thm:all-even-quadrics}
Let $P$ be a nonzero even polynomial on $\C^s$ with $P(0)=0$, and let $C=Q+c$ ($Q$ a quadratic
form) be an irreducible non-hyperplane even quadratic dividing $P$. If $P$ satisfies the 3SC on
$V_C$, then exactly one of the following holds:
\begin{enumerate}
\item[(i)] $c=0$, $\operatorname{rank}Q\ge3$, and $P=\lambda Q$ with $\lambda\neq0$;
\item[(ii)] $c\neq0$, $\operatorname{rank}Q=2$, and after a linear change of variables and
normalization $C=uv-1$ and $P$ is a polynomial in $u,v$ alone belonging to
Theorem~\ref{thm:bivariate-complete}.
\end{enumerate}
There is no such $P$ with $c\neq0$ and $\operatorname{rank}Q\ge3$. Conversely, all polynomials
in (i) and (ii) satisfy the $N$SC on $V_C$ for every $N$, with nondegenerate data forming a
nonempty Zariski-open set. Consequently a component satisfying the 3SC has multiplicity at most
$2$ in $P$, with equality exactly for $P=(au^2+bv^2)(uv-1)^2(uv-3)$, $(a,b)\neq0$, and
$\deg P\le12$; both bounds are attained.
\end{theorem}

\begin{proof}
\emph{The cone, $c=0$.} Then $\operatorname{rank}Q\ge3$. Let $P_d$ be the top homogeneous
part of $P$; $Q\mid P_d$, and scaling the momenta on the cone shows that $P_d$ satisfies the 3SC
on $V_Q$ ($S_3[P_d]$ is the top coefficient of $S_3[P]$). Write $P_d=Q^mH$ with $Q\nmid H$,
$m\ge1$, $H$ homogeneous of even degree $h\ge0$. As in the proof of
Proposition~\ref{prop:quadric}, $S_3[P_d]=(-1)^m2^{4m}(B_{12}B_{13}B_{23})^m\mathcal T$ with
$\mathcal T=\sum_\sg(\sum_{i<j}\sg_i\sg_jB_{ij})^mH(\sum\sg_i\bp_i)\prod_{i<j}H(\sg_i\bp_i-\sg_j\bp_j)$,
and the 3SC forces $\mathcal T\equiv0$ on $V_Q^{\,3}$. If $h>0$, evaluate at
$(\bp,\varepsilon\bq,t\bp)$ with $\bp,\bq\in V_Q$, $H(\bp)\neq0$, $B(\bp,\bq)\neq0$ and
$\mathrm dH_\bp(\bq)\neq0$ (possible since $\mathrm dH_\bp(\bp)=hH(\bp)\neq0$ and $V_Q$ spans):
for even $m$ the coefficient of $\varepsilon^m$ is
$4B(\bp,\bq)^mH(\bp)^4t^h(1-t^2)^h[(1+t)^m+(1-t)^m]$, and for odd $m$ the coefficient of
$\varepsilon^{m+1}$ is $-4B(\bp,\bq)^mH(\bp)^3\,\mathrm dH_\bp(\bq)\,t^{h-1}F_{m,h}(t)$ with
$F_{m,h}(t)=\sum_{\rho=\pm1}\rho(1+\rho t)^{m+h-1}(1-\rho t)^h(t^2+1+\rho t)=2mt+O(t^3)$; both are
nonzero for generic $t$. So $h=0$, $P_d=\lambda Q^m$, and the Gram-matrix evaluation of
Proposition~\ref{prop:quadric} gives $S_3=\lambda^4(-1)^m2^{4m}(2\cdot3^m+6(-1)^m)\neq0$ for
$m\ge2$; hence $m=1$, $d=2$, and $P=\lambda Q$ by evenness and $P(0)=0$. Conversely, for
$P=\lambda Q$ all phase factors equal $1$, $f_N=\prod_i(1+e^{\theta_i})$, and
$Q(D)f_N\cdot f_N=2f_N^2\,Q(\partial)\log f_N=0$ because each $\log(1+e^{\theta_i})$ depends on a
single direction $\bp_i$ with $Q(\bp_i)=0$; nondegeneracy is $B(\bp_i,\bp_j)\neq0$.

\emph{The affine quadric, $c\neq0$; normalize $C=Q-1$.} Let $n:=\operatorname{rank}Q\ge2$; split
the variables into $n$ effective ones, on which $Q$ depends, and $k\ge0$ flat ones. If
$\deg P=2$, divisibility gives $P=\lambda(Q-1)$, contradicting $P(0)=0$; if $\deg P=4$,
evenness and $P(0)=0$ give $P=(Q-1)H_2$ with $H_2$ a quadratic form, and
Theorem~\ref{thm:quadric-NSC}(I) applies (no $n\ge3$; for $n=2$ the flat variables disappear by
Lemma~\ref{lem:transverse}). Let $\deg P>4$. On a generic plane $L$ through the origin,
$Q|_L$ has rank two and $\deg(P|_L)=\deg P>4$ (the top homogeneous part of $P$ cannot vanish on
all generic planes); normalizing $Q|_L=uv$, the restriction of the 3SC to the conic
$Q|_L=1$ is the 3SC of $P|_L$ on $\{uv=1\}$, so by Theorem~\ref{thm:bivariate-complete}
$P|_L$ lies in \eqref{eq:SK9} and its two-soliton phase factor is \eqref{eq:common-phase}.
With $B$ the polar form of $Q$ and $\bp=(s,s^{-1})$, $\bq=(t,t^{-1})$ one has $2B(\bp,\bq)=s/t+t/s$,
so \eqref{eq:common-phase} reads
\begin{equation}\label{eq:phase-polar}
e^{A(\bp,\bq)}=\frac{(B(\bp,\bq)-1)(2B(\bp,\bq)-1)}{(B(\bp,\bq)+1)(2B(\bp,\bq)+1)},
\end{equation}
a coordinate-free expression valid for generic pairs on $V_C$ (generic pairs span generic
planes), hence, after clearing denominators, for all pairs. For $\mathbf x=\bp+\bq$,
$\mathbf y=\bp-\bq$ one has $B(\mathbf x,\mathbf y)=0$ and $Q(\mathbf x)+Q(\mathbf y)=4$, and
conversely these two conditions give $(\mathbf x\pm\mathbf y)/2\in V_C$; so
\eqref{eq:phase-polar} becomes
\begin{equation}\label{eq:orth-identity}
Q(\mathbf x)(Q(\mathbf x)-1)P(\mathbf y)+(Q(\mathbf x)-4)(Q(\mathbf x)-3)P(\mathbf x)=0
\quad\text{whenever }B(\mathbf x,\mathbf y)=0,\ Q(\mathbf x)+Q(\mathbf y)=4 .
\end{equation}
\emph{Flat variables.} The constraints in \eqref{eq:orth-identity} involve only the effective
parts of $\mathbf x,\mathbf y$. Fix effective parts with $Q(\mathbf x)\notin\{0,1,3,4\}$ (a dense
set) and vary the flat part of $\mathbf y$: \eqref{eq:orth-identity} forces $P(\mathbf y)$ to
be constant, so $P$ does not depend on the flat variables. For $n=2$ this reduces to
Theorem~\ref{thm:bivariate-complete}. \emph{Effective rank $n\ge3$.} Take generic
$\mathbf y,\mathbf y'$ with $Q(\mathbf y)=Q(\mathbf y')=r\notin\{0,1,3,4\}$ and $Q$ nondegenerate
on their span; their common $B$-orthogonal complement has dimension $n-2\ge1$ and carries a
nondegenerate form, so it contains $\mathbf x$ with $Q(\mathbf x)=4-r$. Applying
\eqref{eq:orth-identity} to $(\mathbf x,\mathbf y)$ and $(\mathbf x,\mathbf y')$ and subtracting
gives $P(\mathbf y)=P(\mathbf y')$: $P$ is constant on generic level sets of $Q$, hence
$P=F(Q)$ is radial. Restricted to a plane, $F(uv)$ satisfies the 3SC, so by
Theorem~\ref{thm:bivariate-complete} $F(Q)=\kappa Q(Q-1)(Q-2)$. But on three vectors with
$Q(\bp_i)=1$ and $B(\bp_i,\bp_j)=2$ ($i\neq j$; the Gram matrix has determinant $5$ and is
realized in any nondegenerate three-dimensional subspace), the pair sums and differences have
$Q$-values $6$ and $-2$, the central sums have $Q$-values $15$ or $-1$, and a direct evaluation
gives $S_3=-63037440\,\kappa^4\neq0$. So $n\ge3$ is impossible, which proves the trichotomy.

\emph{Sufficiency, multiplicity and degree.} Sufficiency in (ii) is
Theorem~\ref{thm:bivariate-complete}. For the multiplicity, the cofactor
$H=P/C$ of \eqref{eq:SK9} restricted to $z=1$ equals
$9iu^6+3gu^4+(a-3e)u^2-d+(c-3f)u^{-2}+3hu^{-4}+9ju^{-6}$, which vanishes identically if and only
if $i=j=g=h=d=0$, $a=3e$, $c=3f$, i.e.\ $P=(eu^2+fv^2)(uv-1)^2(uv-3)$; dividing once more by
$C$ leaves $(eu^2+fv^2)(uv-3)$, which is a nonzero constant multiple of $eu^2+fv^2$ on the
curve, so the multiplicity is exactly two. The cone solutions and the exceptional quartics have
multiplicity one, and $\deg P\le12$ is read off from \eqref{eq:SK9}.
\end{proof}

\begin{corollary}[Top homogeneous parts]\label{cor:top-quadric}
Let $P\neq0$ be even with $P(0)=0$, let $C$ be an irreducible factor of $P$, and suppose that
$P$ satisfies the 3SC on $V_C$. If the top homogeneous part of $C$ has an irreducible
quadratic factor $Q$ (equivalently, a quadratic factor of rank at least three), then $P=\lambda Q$
and $C=\mu Q$ with $\lambda,\mu\neq0$. In particular, no component other than a quadric cone
can have such a factor in its top homogeneous part; this excludes at once the affine even
quadrics of rank at least three, the non-even quadrics $Q+L+c$ with $\operatorname{rank}Q\ge3$,
and all components of higher degree whose top part contains such a $Q$.
\end{corollary}

\begin{proof}
Let $e=\deg C$, $d=\deg P$, and write the top part $C_e=Q^mG$ with $Q\nmid G$. For generic
$\bp\in V_Q$ (smooth, $G(\bp)\neq0$) and $\mathbf v$ with $\mathrm dQ_\bp(\mathbf v)\neq0$,
the polynomial $t\mapsto C_e(\bp+t\mathbf v)$ has a zero of order exactly $m$ at $t=0$, so for
$C_\varepsilon(\mathbf z):=\varepsilon^eC(\mathbf z/\varepsilon)=C_e+\varepsilon C_{e-1}+\dots$ and
small $\varepsilon\neq0$ there is $t_\varepsilon\to0$ with $C_\varepsilon(\bp+t_\varepsilon\mathbf v)=0$
(continuity of roots). Doing this for three generic points $\bp_i$ gives
$\bq_i(\varepsilon)\to\bp_i$ with $\bq_i(\varepsilon)/\varepsilon\in V_C$, so
$S_3[P](\bq_i(\varepsilon)/\varepsilon)=0$; multiplying by $\varepsilon^{4d}$ and letting
$\varepsilon\to0$ gives $S_3[P_d](\bp_1,\bp_2,\bp_3)=0$ on $V_Q^{\,3}$. Since $Q\mid C_e\mid P_d$,
the cone case of Theorem~\ref{thm:all-even-quadrics} gives $P_d=\lambda Q$, hence $d=2$, $e=2$,
$C_e=\mu Q$, and by evenness and $P(0)=0$, $P=\lambda Q$ and $C=\mu Q$.
\end{proof}

\subsection{Parabolic components and dispersions $F(\ell)+Q$}\label{subsec:parabola-FQ}
The non-even quadrics of rank one and the quartic dispersion surfaces of Kadomtsev--Petviashvili
type can be classified in the same generality.

\begin{theorem}[Parabolic components, arbitrary degree]\label{thm:parabola-general}
Let $P\neq0$ be an even polynomial on $\C^s$ with $P(0)=0$, and let $C$ be an irreducible
non-hyperplane quadratic dividing $P$ whose quadratic part has rank one, so that after a linear
change of variables $C=v-u^2-c$ with $u,v$ two of the coordinates. If $P$ satisfies the 3SC on
$V_C$, then $c=0$ and $P=\gamma(u^4-v^2)$, $\gamma\neq0$, independently of the remaining
variables; conversely $u^4-v^2$ satisfies the $N$SC for every $N$ on the whole dispersion
variety $\{v=\pm u^2\}$ (Theorem~\ref{thm:noneven}). If instead $C=Q(\mathbf x)+w+c$ with
$\operatorname{rank}Q\ge2$ and $w$ a coordinate independent of the variables of $Q$, no nonzero
even $P$ with $P(0)=0$ divisible by $C$ satisfies the 3SC on $V_C$.
\end{theorem}

Together with Theorems~\ref{thm:all-even-quadrics} and~\ref{thm:noneven} and
Corollary~\ref{cor:top-quadric}, this leaves, among the irreducible non-hyperplane quadratic
components, only the translated hyperbolas $C=uv+au+bv+c$ with $(a,b)\neq0$, $c\neq ab$
(the non-even quadrics whose linear part lies in the variables of a rank-two quadratic form)
unclassified. The proof of Theorem~\ref{thm:parabola-general} is given in
Appendix~\ref{app:parabola}; its core is a root-counting argument for the map
$\kappa\mapsto(\kappa^2+1)/(2\kappa^2)$ which forces the multiplicity of the component and
the two-soliton factor, followed by a rigidity lemma for the factor $(s-t)^2/(s^2+st+t^2)$.
Weighted degenerations extend it beyond quadratic components: with $\operatorname{wt}u=1$,
$\operatorname{wt}v=2$ and weights at least $1$ on the remaining coordinates, if the top weighted
part of an irreducible component $C$ of $V(P)$ is divisible by $v-u^2$ and $P$ satisfies the 3SC
on $V_C$, then the top weighted part of $P$ satisfies the 3SC on the parabola (points of $V_C$
deform to points of the top part under the scaling, with the multiplicity of $v-u^2$ in the top
part arbitrary), so by Theorem~\ref{thm:parabola-general} it equals $\gamma(u^4-v^2)$; the lower
weights of an even $P$ then form a quadratic form, $P=\gamma u^4+Q$, and Theorem~\ref{thm:FQ}
gives $\operatorname{rank}(Q|_{u=0})=1$ and all $N$SC on $V(P)$.

\begin{theorem}[Dispersions of the form $F(\ell)+Q$]\label{thm:FQ}
Let $\ell\neq0$ be a linear form on $\C^s$, $F\in u^4\C[u^2]\setminus\{0\}$, $Q$ a quadratic form,
$P=F(\ell)+Q$, and let $V_0$ be an irreducible non-hyperplane component of $V(P)$. Put
$r:=\operatorname{rank}(Q|_{\ker\ell})$. The following are equivalent:
(a) $P$ satisfies the 3SC on $V_0$; (b) either $r=0$ and $\deg F\le6$, or $r=1$ and
$\deg F=4$; (c) $P$ satisfies the $N$SC on the whole dispersion variety $V(P)$ for every $N$.
In these cases $P$ is squarefree and nondegenerate data on $V_0$ form a nonempty Zariski-open
set. In particular, for $P=\alpha\ell^4+Q$ with $\alpha\neq0$ having a non-hyperplane
component, the 3SC on it holds if and only if $\operatorname{rank}(Q|_{\ker\ell})\le1$, if and
only if all $N$SC hold on $V(P)$.
\end{theorem}

\begin{proof}
Take $u=\ell$ as a coordinate and write $Q=au^2+2uL(\mathbf z)+R(\mathbf z)$ with
$\operatorname{rank}R=r$ (the quadratic term of $F$ is absorbed into $Q$ by the convention
$F\in u^4\C[u^2]$).

\emph{$r\ge2$.} Then $P$ is irreducible: viewed as a quadratic in $\mathbf z$ over $\C[u]$, a
factorization would have two factors linear in $\mathbf z$ whose leading parts multiply to $R$,
impossible for $r\ge3$, and for $r=2$, with $R=vw$, of the form $(v+A(u))(w+B(u))$, which forces
$\deg_uP\le2$. Restrict $P$ to a nondegenerate plane of $R$ and complete the square:
$P_*=G(u)+vw$ with $G(u)=F(u)+bu^2$. For generic $k\neq0$ and the on-shell points
$\bp_j=(k,t_j,-G(k)/t_j)$, $(t_1,t_2,t_3)=(1,2,3)$, the sign sum gives the polynomial identity
$S_3=-\frac{2G(k)^2}{9}\bigl[G(k)^2-2G(k)G(2k)-G(k)G(3k)+G(2k)^2\bigr]$ in the three values
$G(k),G(2k),G(3k)$; for $\deg G=D\ge4$ with leading coefficient $c\neq0$ the bracket has
leading term $c^2k^{2D}(4^D-3^D-2^{D+1}+1)$, whose coefficient is positive for $D\ge4$ since
$(3/4)^D+2^{1-D}<1$. So the 3SC fails.

\emph{$r=1$, with a linear term along the radical of $R$.} Write $R=z^2$ after a change of
variables and complete the square in $z$; the remaining term $2uM(\mathbf w)$ involves a linear
form $M$ on the radical of $R$. If $M\neq0$ take $w=2M$: $P=G(u)+z^2+uw$, irreducible. For generic $k\neq0$ and
$\bp_j=(k,t_j,-(G(k)+t_j^2)/k)$, $(t_1,t_2,t_3)=(0,1,3)$, the sign sum gives
$S_3=72\,[G(3k)-6G(2k)+15G(k)]$, so the 3SC forces $c_D(3^D-6\cdot2^D+15)=0$ for every
coefficient $c_D$ of $G$; the factor vanishes for $D=2,4$ and is positive for even $D\ge6$.
Hence $\deg F=4$.

\emph{$r=1$, $M=0$.} Then $P=G(u)+z^2$, which is either irreducible or the product of two
distinct graphs $z=\pm g(u)$ ($G=-g^2$). Suppose $\deg G=2m\ge6$ with leading coefficient $c$.
Under the weighted scaling $P_\varepsilon(u,z):=\varepsilon^{2m}P(u/\varepsilon,z/\varepsilon^m)\to
cu^{2m}+z^2$, points of $V_0$ deform to points of the limit curve $z^2=-cu^{2m}$ (simple roots
$z\neq0$ for $u\neq0$), and on a fixed branch $z=\pm\sqrt{-c}\,u^m$ the 3SC of $P$ passes to
the limit and forces the 3SC of $u^{2m}-v^2$ on the parametrized curve $\bp(t)=(t,t^m)$. This is
false for every $m\ge3$: for $m=3,4$ the sign sum at $t=(1,2,3)$ equals $-181398528000$ and
$-33320354772418560$, and for $m\ge5$, at $t=(1,\varepsilon,2)$ with $X=2^m$, $Y=3^m$,
$[\varepsilon^4]S_3=\tfrac29X^2m^2(2m-1)B_m$ with
$B_m=(18m-9)X^4+27X^2+(72m+18)X+54m-Y^2[(4m-9)(X^2-2X)+54m]<0$, because
$(18m-9)X^4+27X^2+(72m+18)X+54m<19mX^4$ while $Y^2(4m-9)(X^2-2X)\ge\frac{15}8mX^2Y^2>30mX^4$
for $m\ge5$. Hence $\deg F=4$.

\emph{$r=0$.} Then $Q=u(au+2L)$ and $P=u\bigl(F(u)/u+au+2L\bigr)$; if $L=0$ only hyperplane
components occur, and if $L\neq0$, with $w=2L$ the non-hyperplane component is the graph
$w=-F(u)/u-au$, a one-variable graph-type dispersion with the linear factor $u$; by
Theorem~\ref{thm:mult-one} the 3SC holds if and only if $F(u)/u+au$ has degree at most five,
i.e.\ $\deg F\le6$, and then all $N$SC hold on the whole variety.

\emph{Sufficiency for $r=1$, $\deg F=4$.} Let $P=\alpha u^4+Q$, $\alpha\neq0$, and $k:=\alpha^{1/4}u$.
Since $Q|_{\ker k}$ has rank at most one there is a linear form $l$ with $Q|_{\ker k}=3l^2|_{\ker k}$,
so $k$ divides $Q-3l^2$ and $\omega:=(3l^2-Q)/(4k)$ is a linear form; then
$P=k^4+3l^2-4k\omega$ is the pull-back of the bilinear Kadomtsev--Petviashvili polynomial under the
linear map $\bp\mapsto(k,l,\omega)$. The $N$-soliton solutions of the KP equation, with
$(k,l,\omega)=(p-q,p^2-q^2,p^3-q^3)$ and phase factors $\frac{(p_i-p_j)(q_i-q_j)}{(p_i-q_j)(q_i-p_j)}$,
are classical (Wronskian form \cite{FreemanNimmo1983,Hirota2004}); the corresponding
polynomial identities $S_N=0$ hold on the whole KP dispersion surface, hence, after the
linear substitution, on the whole $V(P)$. Squarefreeness was shown above, and nondegenerate
data form a nonempty open set because $(\bp,\bq)\mapsto\bp\pm\bq$ is dominant on
$V_0\times V_0$ for every irreducible non-hyperplane $V_0$ (otherwise the Gauss map would be
constant).
\end{proof}

The theorem contains the genuinely irreducible quartic dispersion surfaces of KP type, for
which the conjectured implication therefore holds; it does not cover quartic top parts that
are not fourth powers of a linear form (such as $u^3v$ or $u^4-uv^3$), which remain open.

\subsection{Cubic graph components with a symmetric quadratic phase family}\label{subsec:cubic-graph}
The graph-type dispersions of Section~\ref{sec:II-III} have a linear factor $L$. On the
cubic graph $v=u^3$ one can also allow an arbitrary cofactor, provided the two-soliton phase
factor is prescribed within a natural one-parameter family; the answer contains a
five-dimensional family with the KdV phase factor, Ito's equation, and two ``golden ratio''
sextics.

\begin{theorem}[Cubic graph, symmetric quadratic phase family]\label{thm:cubic-graph}
Let $P\in\C[u,v]\setminus\{0\}$ be even, $P(0)=0$, and let $C=u^3-v$ be a simple factor of $P$.
Suppose that the generic two-soliton factor of $P$ on $\bp(t)=(t,t^3)$ has the form
\begin{equation}\label{eq:A-alpha}
A_\alpha(s,t)=\frac{(s-t)^2}{(s+t)^2}\cdot\frac{s^2+\alpha st+t^2}{s^2-\alpha st+t^2},\qquad\alpha\in\C .
\end{equation}
Then $P$ satisfies the 3SC on $V_C$ if and only if one of the following holds, and in each
case all $N$SC hold on $V_C$:
\begin{enumerate}
\item[(a)] $\alpha=0$ and $P$ is a nonzero linear combination of
\begin{align*}
P_4&=u(u^3-v),\qquad P_6=(u^3-v)(u^3+2v),\qquad P_8=u^2(u^3-4v)(u^3-v),\\
P_{10}&=u(u^3-4v)(u^3-v)(u^3+2v),\qquad P_{12}=(u^3-4v)(u^3-v)(u^6+28u^3v+16v^2);
\end{align*}
\item[(b)] $\alpha=1$ and $P=\lambda v(u^3-v)$, $\lambda\neq0$ (Ito's equation);
\item[(c)] $\alpha^2+\alpha-1=0$ and $P=\lambda(u^3-v)\bigl[(1-\alpha)u^3+(2+\alpha)v\bigr]$,
$\lambda\neq0$; equivalently $P=\lambda(u^3-v)(u^3+cv)$ with $c^2-7c+1=0$.
\end{enumerate}
No other value of $\alpha$ occurs. The bound $\deg P\le12$ is attained, all members have
$C$ as a simple factor, nondegenerate data form a nonempty open set for every $N$, and
additional variables cannot enter $P$.
\end{theorem}

\begin{proof}
Two normalizations are used and must not be mixed. For $\alpha=0$ the factor is
$((s-t)/(s+t))^2$, and we use the quadratic pair weight $R^{\mathrm K}(s,t):=(s-t)^2$ and
\[
\mathcal F^{P,\mathrm K}_N(k_1,\dots,k_N):=\sum_\sg P\Bigl(\sum_i\sg_ik_i,\sum_i\sg_ik_i^3\Bigr)\prod_{i<j}R^{\mathrm K}(\sg_ik_i,\sg_jk_j);
\]
for $\alpha\neq0$ we use the quartic pair weight $R^\alpha(s,t):=(s-t)^2(s^2+\alpha st+t^2)$ and
the analogous $\mathcal F^{P,\alpha}_N$. For $\alpha=0$ one has
$\mathcal F^{P,0}_N=\prod_{i<j}(k_i^2+k_j^2)\,\mathcal F^{P,\mathrm K}_N$, so the vanishing of the
two residuals is equivalent but their coefficients differ. With the factor prescribed,
$S_N=0$ if and only if the corresponding residual vanishes, as in the proof of
Theorem~\ref{thm:octic}. Use the weights $\operatorname{wt}u=1$, $\operatorname{wt}v=3$ (realized by
$t\mapsto\rho t$), so that each weighted component satisfies the conditions separately; for a
component of weight $2d$ put $a=s+t$, $b=s-t$, $X=a^2$, $Y=b^2$ and
$G(X,Y):=P(a,a(a^2+3b^2)/4)$, the value at $\bp(s)+\bp(t)$, a homogeneous polynomial of degree
$d$ (the value at $\bp(s)-\bp(t)$ is $G(Y,X)$).

\emph{The phase identity.} The prescribed factor is equivalent to
$Y[(2+\alpha)X+(2-\alpha)Y]G(X,Y)+X[(2-\alpha)X+(2+\alpha)Y]G(Y,X)=0$. For
$\alpha\notin\{0,\pm2\}$ the two brackets are coprime, so $G=X[(2-\alpha)X+(2+\alpha)Y]H$ with
$H(Y,X)=-H(X,Y)$; for $\alpha=0$, $G=XH$ with $H$ antisymmetric. Since
$u^{2d-3j}v^j$ contributes $4^{-j}c_jX^{d-j}(X+3Y)^j$, one has
$\operatorname{ord}_{X=0}G\ge\lceil d/3\rceil$, and since the bracket is a unit at $X=0$ for
$\alpha\neq-2$, in both cases $\operatorname{ord}_{X=0}H\ge\lceil d/3\rceil-1$.

\emph{An analytic degree bound.} Expanding the residual $\mathcal F^{P,\alpha}_3(s,t,\varepsilon)$
(for $\alpha=0$, $\mathcal F^{P,\mathrm K}_3$ after removal of the factor $(s^2+t^2)s^2t^2$) in
the third parameter, the coefficient of $\varepsilon^2$ is
\[
2R_\alpha(k_1,k_2)\Bigl\{k_1^2k_2^2\bigl[(2-2\alpha)(k_1^2+k_2^2)+(\alpha-2)^2k_1k_2\bigr]P
+(\alpha-2)(k_1+k_2)k_1^3k_2^3P_u+\tfrac12k_1^4k_2^4P_{uu}\Bigr\}
\]
summed over the sign classes $(k_1,k_2)=(\pm s,\pm t)$, with $P$ and its $u$-derivatives
evaluated at $(k_1+k_2,k_1^3+k_2^3)$. With $\mathcal D:=2X\partial_X-(X+Y)\partial_Y$ one has
$P_u=a^{-1}\mathcal DG$, $P_{uu}=a^{-2}(\mathcal D-1)\mathcal DG$ in the class $u=a$, and the
same with $X,Y$ exchanged in the class $u=b$. For $\alpha=0$, in the quadratic
normalization, the sum equals
$-\frac{(X-Y)^3}{8}\{(X-Y)(XH_{XX}-YH_{YY})-2(XH_X+YH_Y)+2H\}$ (in the quartic normalization
this is multiplied by $(X+Y)(X-Y)^2/32$), and if
$H=X^rh(Y)+O(X^{r+1})$ with $h\neq0$, $r\ge2$, the coefficient of $X^{r-1}$ in the bracket is
$-r(r-1)Yh(Y)\neq0$; for $\alpha\neq0,\pm2$ the corresponding lowest term is
$-\frac{4-\alpha^2}{512}r(r-1)Y^8h(Y)X^{r-1}$. Hence $\operatorname{ord}_{X=0}H\le1$, so
$\lceil d/3\rceil\le2$, $d\le6$, and every weighted component has weight at most $12$; this uses
no finite computation.

\emph{Finite elimination.} For each weight $W\in\{2,4,\dots,12\}$ write
$P=\sum_jc_ju^{W-3j}v^j$ and impose the phase identity and the vanishing of the
$\varepsilon^2$-coefficient as linear conditions on $(c_j)$ with coefficients in $\mathbb{Q}[\alpha]$. For
$\alpha=0$ the kernels have dimensions $0,1,1,1,1,1$ and are spanned by $P_4,\dots,P_{12}$. For
general $\alpha$ the Smith normal form of the coefficient matrix over $\mathbb{Q}[\alpha]$ has all
invariant factors constant except the last, which is $1,\ \alpha,\ \alpha(\alpha-1)(\alpha^2+\alpha-1),\
\alpha,\ \alpha(\alpha+2),\ \alpha(\alpha+2)$ for $W=2,\dots,12$; hence for $\alpha\notin\{0,\pm2\}$
a nonzero kernel occurs only for $W=6$ and $\alpha=1$ or $\alpha^2+\alpha-1=0$, with the
one-dimensional kernel spanned by $(u^3-v)[(1-\alpha)u^3+(2+\alpha)v]$ (ancillary files). The
value $\alpha=2$ (factor identically $1$) gives $G(Y,X)=-G(X,Y)$, the $P_{uu}$ term yields
$\operatorname{ord}_XG\le1$ and hence $W\le6$, where the matrices have full column rank; so
$P=0$. The value $\alpha=-2$ (factor $((s-t)/(s+t))^4$) is incompatible with a simple factor
$C$: with the reduced residual $\mathcal F^{P,(4)}_3:=\sum_\sg P(\sum_i\sg_i\bp(k_i))\prod_{i<j}(\sg_ik_i-\sg_jk_j)^4$,
whose vanishing is equivalent to the 3SC under the prescribed factor, and $k_i=t+\varepsilon a_i$
with $P/C$ nonzero at $\bp(t)$, the coefficient of $\varepsilon^6$ is a nonzero constant times
$\sum_{k}(a_i-a_k)(a_j-a_k)(a_i-a_j)^4$ ($\{i,j,k\}=\{1,2,3\}$), which equals $-12$ at
$(0,1,2)$; the raw sum $S_3$ itself may carry additional common zero factors and start at a
higher order, so the reduced residual is the correct object here (this is where the
multiplicity-three example $u(u^3-v)^3$ of \eqref{eq:family} lives). This proves necessity.

\emph{All $N$.} Both residuals are even in each $k_i$ and symmetric in the $x_i=k_i^2$, and
\begin{align*}
\mathcal F^{P,\mathrm K}_N(k_1,\dots,k_{N-1},0)&=2\Bigl(\prod_{i<N}k_i^2\Bigr)\mathcal F^{P,\mathrm K}_{N-1},\\
\mathcal F^{P,\mathrm K}_N(k_1,\dots,k_{N-2},s,s)&=8s^2\prod_{i<N-1}(s^2-k_i^2)^2\,\mathcal F^{P,\mathrm K}_{N-2},\\
\mathcal F^{P,\alpha}_N(k_1,\dots,k_{N-1},0)&=2\Bigl(\prod_{i<N}k_i^4\Bigr)\mathcal F^{P,\alpha}_{N-1},\\
\mathcal F^{P,\alpha}_N(k_1,\dots,k_{N-2},s,s)&=8(2-\alpha)s^4\prod_{i<N-1}(s^2-k_i^2)^2\bigl[(s^2+k_i^2)^2-\alpha^2s^2k_i^2\bigr]\mathcal F^{P,\alpha}_{N-2}.
\end{align*}
If the two lower residuals vanish, the first relation makes each $x_i$ a factor and the
second, with symmetry, each $(x_i-x_j)^2$; so $\prod_ik_i^2\prod_{i<j}(k_i^2-k_j^2)^2$, of total
degree $2N^2$, divides the residual. The base cases are $\mathcal F_0=P(0)=0$,
$\mathcal F_1=0$ (dispersion relation) and $\mathcal F_2=0$ (prescribed factor). In the quadratic
normalization ($\alpha=0$, weight $W\le12$) the residual has total degree $W+N(N-1)$: for
$N=3$ and $W<12$ this is less than $18$, so the residual vanishes, and for $W=12$ the identity
$\mathcal F^{P_{12},\mathrm K}_3=0$ is checked as a complete three-variable polynomial identity
(ancillary files); for $N\ge4$, $W+N(N-1)\le12+N(N-1)<2N^2$. In the quartic normalization
(the sextics of (b),(c)) the residual has total degree $6+2N(N-1)$, equal to $18$ for $N=3$,
where the complete identity
$\mathcal F^{P_\alpha,\alpha}_3=72\alpha(1-\alpha)(\alpha^2+\alpha-1)\prod_ik_i^2\prod_{i<j}(k_i^2-k_j^2)^2$
vanishes exactly for the admitted values, and $6+2N(N-1)<2N^2$ for $N\ge4$. Induction on $N$
gives all $N$SC in both cases. Dividing the basis elements by $C$ and restricting to $v=u^3$
gives $u,\ 3u^3,\ -3u^5,\ -9u^7,\ -135u^9$ in case (a), $u^3$ in case (b) and $3u^3$ in case (c),
so the multiplicity is one; the Jacobian of $(s,t)\mapsto(s+t,s^3+t^3)$ is
$3(t^2-s^2)\not\equiv0$, so pair sums and differences of a nonzero $P$ are not identically zero
and nondegenerate data are dense. For additional variables $\mathbf z$, with the factor
$A_\alpha(s,t)$ prescribed for all values of the free coordinates, the phase identity reads
$P(s-t,s^3-t^3,\mathbf Z_-)+A_\alpha(s,t)P(s+t,s^3+t^3,\mathbf Z_+)=0$ with
$\mathbf Z_\pm=\mathbf z_1\pm\mathbf z_2$ independent; varying $\mathbf Z_+$ at fixed generic
$s,t,\mathbf Z_-$ shows that $P$ does not depend on $\mathbf z$ at generic pair-sum points, and
the pair-sum map is dominant.
\end{proof}

The family \eqref{eq:A-alpha} does not exhaust the phase factors that occur with the simple
factor $u^3-v$ and the 3SC. For $P=(au+cv)(u^3-v)$ with $ac\neq0$, which in the coordinates
$(u,v)=(k,-\omega)$ is the polynomial $P_{L,\Lam,0}$ of \eqref{eq:family} with $L=aD_x-cD_t$
and $\Lam=D_t$, and which therefore satisfies every $N$SC on $V_C$ by Theorem~\ref{thm:I}, the
factor \eqref{eq:phase} on $\bp(t)=(t,t^3)$ is
\[
\frac{(s-t)^2}{(s+t)^2}\cdot\frac{a+c(s^2+st+t^2)}{a+c(s^2-st+t^2)},
\]
the factor of Remark~\ref{rem:I-scope}(i) with $\lambda=a/c$; it is not homogeneous in $(s,t)$
and hence not of the form $A_\alpha$ for any $\alpha$ (the cases $c=0$ and $a=0$ give $A_0$ and
$A_1$). Thus the hypothesis on the phase factor in
Theorem~\ref{thm:cubic-graph} is a genuine restriction, and the classification on the cubic
graph without a prescribed phase factor is left to future work.

\begin{proposition}[Translated hyperbolas]\label{prop:shifted}
Let $C=uv+au+bv+c$ be irreducible ($c\neq ab$), possibly in the presence of further variables,
and let $P\neq0$ be even with $P(0)=0$ and $C\mid P$.
\begin{enumerate}
\item[(i)] If $ab=0$ and $(a,b)\neq(0,0)$, then $P$ does not satisfy the 3SC on $V_C$.
\item[(ii)] If $ab\neq0$ and $\deg P\le6$, then $P$ does not satisfy the 3SC on $V_C$.
\end{enumerate}
\end{proposition}

\begin{proof}
(i) After exchanging $u,v$, scaling and normalizing, $C=uv+bv-1$ with $b\neq0$; restricting to
planes $\mathbf z=\mathbf au+\mathbf bv$ removes further variables. Since $C(-u,-v)\neq C(u,v)$,
evenness gives $P=[(uv-1)^2-b^2v^2]H$. Let $M$ be the maximal weight of $P$ under
$\operatorname{wt}(u^iv^j)=i-j$ and $P_\varepsilon(u,v):=\varepsilon^MP(u/\varepsilon,\varepsilon v)=P_M+\varepsilon^2P_{M-2}+\dots$;
the points $\bp_\varepsilon(t)=(t/\varepsilon-b,\varepsilon/t)$ of $V_C$ become
$\bq_\varepsilon(t)=(t-b\varepsilon,1/t)$, and $S_3[P_\varepsilon](\bq_\varepsilon(t_i))=0$ identically in
$\varepsilon$. The constant term shows that $P_M$, which is divisible by $(uv-1)^2$, satisfies
the 3SC on $uv=1$, so by Theorem~\ref{thm:all-even-quadrics}
$P_M=\lambda u^2(uv-1)^2(uv-3)$ or $\lambda v^2(uv-1)^2(uv-3)$. The coefficient of $\varepsilon^1$
involves only $P_M$ and the common translation $-b\varepsilon(1,0)$ of the three points, and at
$t=(1,2,3)$ it equals $b\lambda^4$ times $-296719205510/243$, respectively $-128771260345/5668704$
(ancillary files), a contradiction.
(ii) is an exact elimination (ancillary files): for $\deg P\le6$ the admissible $P$ are
$C(\bp)C(-\bp)H$ with $H$ an even quadratic in $u,v$ and the further variables; the cases
$c=0$ and $c\neq0$ are treated separately, the coefficients of the polynomial identity
$S_3=0$ along several rational triples and along pairs of axis points with an arbitrary third
point are collected as polynomial equations in the coefficients of $H$ and in $(a,b,c)$, and
saturation by $ab$, $c-ab$ and the nondegeneracy factors yields the unit ideal in every
branch.
\end{proof}

Among irreducible quadratic components, Theorems~\ref{thm:all-even-quadrics},
\ref{thm:noneven}, \ref{thm:parabola-general} and Proposition~\ref{prop:shifted} leave open only
the translated hyperbolas with $ab\neq0$, $c\neq ab$ and $\deg P\ge8$.

\section{Discussion}\label{sec:discussion}

\subsection*{An infinite family for Hietarinta's programme}
Hietarinta's searches \cite{HietarintaI,HietarintaII,HietarintaIII,HietarintaIV,Hietarinta1991}
were confined to bilinear polynomials of low degree, and within that range the three-soliton
condition proved a remarkably reliable detector of integrability. Theorems~\ref{thm:II}
and~\ref{thm:III} show that this reliability does not persist in high degree: for every linear
$L$, every $\Lam$ and every $n\ge1$, the equation $L(D_x^3+\Lam)^{2n+1}f\cdot f=0$ possesses
nondegenerate three-soliton solutions with all parameters free on the principal component and
no four-soliton solution of Hirota's form for generic data --- and, for the G\"urses--Pekcan and
Ito families, for no real nondegenerate data anywhere on the dispersion variety
(Corollary~\ref{cor:no-real-four}). This is an
explicit family of unbounded degree, in every dimension $s\ge2$, and the failure is certified
for all $n$ at once by the explicit residual \eqref{eq:residual}. The mechanism is
multiplicity: the cubic identity \eqref{eq:cubic-identity} factors into linear forms for three
variables and not for four (Lemma~\ref{lem:no-linear-factor}), so that odd powers of the
on-shell factor $H_N\Delta_N$ keep their parity support for $N=3$ and acquire new parity classes
for $N=4$; the row expansion \eqref{eq:B2} makes the new classes explicit.

\subsection*{Three mechanisms of failure}
It is worth separating three phenomena that a loose reading of ``three-soliton but no
four-soliton'' conflates.
(i) \emph{Polynomial obstruction with regular two-soliton data}: all phase factors are finite
and generically nonzero, and the failure of the 4SC is carried by a nonvanishing polynomial
residual. This is the situation of the present paper. It is also the situation of the
exceptional equation $D_xD_yD_tD_z$, with one difference: there the four momenta lie on four
\emph{different} components of the dispersion variety, the coordinate hyperplanes
$\{k=0\}$, $\{l=0\}$, $\{\omega=0\}$, $\{s=0\}$, and with this choice all phase factors are
regular, the 3SC holds and the 4SC fails \cite[Remark~4]{GursesPekcan2025}; we have re-verified
this in exact arithmetic (ancillary files). On any single one of these hyperplanes all pairwise
sums and differences of momenta lie again on the hyperplane, every pair factor of
\eqref{eq:SN} vanishes, and all $S_N$ vanish trivially with indeterminate phase factors. The
exceptional equation is thus a squarefree, degree-four, \emph{mixed-component} instance of
mechanism (i), whereas the families of this paper are single-component instances of
unbounded degree on a component of multiplicity $2n+1$.
(ii) \emph{Degenerate interaction data}: for the monomial equations $D_x^{m}D_y^{n}D_t^{p}$ in
three variables the dispersion relation forces components of the momenta to vanish, any four
momenta include two on the same coordinate plane, and the phase factors become indeterminate,
so that genuine four-soliton solutions fail to exist for a reason invisible to the polynomial
conditions \cite[Remark~1]{GursesPekcan2025}.
(iii) \emph{Constrained solitons}: three-soliton solutions may exist only on a proper
subvariety of the dispersion data, as for the two-dimensional sine--Gordon equation
\cite{Hirota1973,Hietarinta1990}; and equations may fail already at the two-to-three step, as
the Kac--Wakimoto equation does \cite{Pekcan2016}.
A fourth, orthogonal, distinction is the one of Section~\ref{sec:mixed}: for reducible $P$ the
conditions may hold on one component of the dispersion variety and fail on mixed products.

\subsection*{The hypotheses of Conjecture~\ref{conj:repair}}
Mechanism (i) occurs in the families of this paper because the principal component $V_Q$ has
multiplicity $2n+1\ge3$ in $P$. It also occurs, on a single component and with all phase factors
regular, for squarefree polynomials whose dispersion variety contains a hyperplane not through
the origin:

\begin{proposition}[An affine hyperplane component]\label{prop:affine}
Let $P(k,\omega):=(k^2+\omega^2)(\omega^2-1)(\omega^2-9)$ and $V_0:=\{\omega=1\}$. Then $P$ is
even and squarefree with $P(0)=0$, $V_0$ is a component of $V(P)$ along which $P$ vanishes to
first order, and $P$ satisfies the three-soliton condition on $V_0$; but at the nondegenerate
data $\bp_i=(i,1)$, $i=1,2,3,4$, one has $S_4=631\,266\,877\,440\neq0$.
\end{proposition}

\begin{proof}
The $\omega$-component of $\sum_{j\le3}\sg_j\bp_j$ lies in $\{\pm1,\pm3\}$ for all
$\bp_j\in V_0$, where $(\omega^2-1)(\omega^2-9)=0$; hence every term of $S_3$ vanishes and the
3SC holds identically on $V_0^{\,3}$. The data $\bp_i=(i,1)$ are nondegenerate:
$P(\bp_i-\bp_j)=9(i-j)^2\neq0$ and $P(\bp_i+\bp_j)=-15\bigl((i+j)^2+4\bigr)\neq0$. The value of
$S_4$ is an exact computation (ancillary files).
\end{proof}

Neither situation arises for the equations of Hietarinta's lists, whose relevant components
are hypersurfaces of degree at least two appearing with multiplicity one; this is the origin of
the two hypotheses of Conjecture~\ref{conj:repair}, which cannot be weakened in the obvious
ways: the families \eqref{eq:family} with $n\ge1$ satisfy all hypotheses except the
multiplicity, and Proposition~\ref{prop:affine} satisfies all except the non-hyperplane
condition. The conjecture concerns a single component
and says nothing about mixed products, which Section~\ref{sec:mixed} and the equation
$D_xD_yD_tD_z$ show to be a separate matter; it does not require the component to pass through the origin: the affine hyperplane of
Proposition~\ref{prop:affine} does not, but neither do the quadrics $uv=1$ of
Section~\ref{sec:quadric}, on which the conjectured implication holds
(Theorem~\ref{thm:quadric-NSC}), so such a requirement would exclude valid cases. For $P=L(D_x^3+\Lam)$ the
conjecture is Theorem~\ref{thm:I}, and Theorem~\ref{thm:mult-one} proves it for the whole class
of graph-type dispersions with a linear factor, $P=L(q(D_{\mathbf x})+D_t)$ with $q$ a nonlinear
odd polynomial, where the 3SC selects exactly the equations satisfying every polynomial
$N$-soliton condition on the whole dispersion variety (with nondegenerate data generic on the
principal component); for the equations of the classical searches, the 3SC was
verified by the searches and the higher conditions are known for those equations which were
subsequently identified as integrable. Beyond graph-type dispersions, Section~\ref{sec:quadric} settles every irreducible even
quadric component, in any dimension and for an arbitrary cofactor of any degree and
multiplicity (Theorem~\ref{thm:all-even-quadrics}), every parabolic component and every
non-even quadric with a linear direction independent of its quadratic part
(Theorems~\ref{thm:noneven} and~\ref{thm:parabola-general}), and the dispersions $F(\ell)+Q$,
which include the irreducible quartic dispersion surfaces of Kadomtsev--Petviashvili type
(Theorem~\ref{thm:FQ}), and, with a prescribed one-parameter phase family, the cubic graph
(Theorem~\ref{thm:cubic-graph}); in all these classes the 3SC selects exactly the equations with
$N$-soliton solutions for all $N$ on the component. Remark~\ref{rem:double} and
Theorem~\ref{thm:all-even-quadrics} show that the multiplicity-one hypothesis of the
conjecture is not necessary for its conclusion (a component of multiplicity two can carry all
$N$SC), although multiplicity at least three never does among even quadrics.
\subsection*{Open problems}
(1) Irreducibility of $R_n^{(1,0)}$ and $R_n^{(0,1)}$ over $\mathbb{Q}$ for $n\ge5$ (established
for $n\le4$ by exact factorization; see the ancillary files), and, more generally, the
factorization of the residual \eqref{eq:residual} over $\mathbb{Q}(\beta,\gamma)$.
(2) Maximal soliton numbers. For $n=1$ they are determined (Theorem~\ref{thm:six-bound}). For
$n\ge2$ and $L\notin\operatorname{span}(D_x,\Lam)$ only the finiteness bound of
Theorem~\ref{thm:finite-all} is known; for $n\ge3$ and $L=D_x$ the maximal number lies in
$[4,5n-6]$ (Theorem~\ref{thm:finite-bound}) and is unknown. Real five-soliton solutions with
all phase factors positive, for $L\notin\operatorname{span}_{\mathbb R}(D_x,\Lam)$, have not been
found. Also, for $\beta\gamma>0$, the structure of the real zero
set of $R_n^{(\beta,\gamma)}$ inside the positive orthant.
(3) The multiplicity mechanism beyond graph-type dispersions: Theorems~\ref{thm:classification},
\ref{thm:classification-multi} and~\ref{thm:mult-one} classify the odd polynomials $q$ when the
dispersion relation is a graph $\omega=-q(\mathbf k)$; the same questions for factors $Q$ that are
not linear in any variable.
(4) Conjecture~\ref{conj:repair}: the classes settled in Section~\ref{sec:quadric} leave, as
the next candidates, the translated hyperbolas $uv+au+bv+c$ with $ab\neq0$, $c\neq ab$ and
$\deg P\ge8$ among quadratic components, the cubic graph without a prescribed phase factor
(phase factors outside the family \eqref{eq:A-alpha} occur already for $P_{L,\Lam,0}$; see the
remark after Theorem~\ref{thm:cubic-graph}), the quartic dispersion surfaces whose top part is not the fourth power of
a linear form (Theorem~\ref{thm:FQ} covers $\alpha\ell^4+Q$), and irreducible components of
higher degree whose top homogeneous part has no quadratic factor of rank at least three
(Corollary~\ref{cor:top-quadric} excludes the others). A structural argument in the spirit of
Theorem~\ref{thm:I}, valid for all irreducible components, is what the conjecture asks for.


\section*{Data availability}
The verification scripts (exact-arithmetic Python, requiring only the packages python-flint and
sympy) and their recorded outputs --- covering Tables~\ref{tab:Rn} and~\ref{tab:cert}, the
parity supports of Lemma~\ref{lem:bialternant-parity} for $N\le8$, the numerical checks of
Theorem~\ref{thm:I} for $N\le7$, the identity \eqref{eq:slice} for $n\le4$, the identities \eqref{eq:F-identity} and $\widetilde Q_0=E+U+V$ of
Theorem~\ref{thm:threeblock}, finite instances of \eqref{eq:residual} and \eqref{eq:threeblock}, the exact strata identities and sample values quoted after Theorem~\ref{thm:sign}, the
certificates of Theorems~\ref{thm:classification}, \ref{thm:classification-multi}
and~\ref{thm:mult-one}, the polynomial-division identities and the two exact values of
$\Sigma_{5,1}$ in the proof of Theorem~\ref{thm:signed-four}, the elimination identities of Theorems~\ref{cor:max-four} and~\ref{thm:n2-five}, the
two interval-arithmetic existence certificates of Theorem~\ref{thm:five-exist} (with the
exact rational centres), the combinatorial certificate of Theorem~\ref{thm:six-bound}, the
symbolic identities of Theorem~\ref{thm:finite-bound}, the Smith normal forms and residual
identities of Theorem~\ref{thm:cubic-graph} and Proposition~\ref{prop:shifted}, the identities,
finite kernels and certificates of Section~\ref{sec:quadric} and of
Appendices~\ref{app:bivariate} and~\ref{app:parabola}, and the data of
Proposition~\ref{prop:affine} --- accompany this paper as ancillary files (directory
\texttt{anc/}, with an index \texttt{README\_ancillary.md} relating each statement to the script
and recorded output that supports it). The existence certificates use only exact integer
interval arithmetic with outward rounding; all other computations are in exact rational or
algebraic arithmetic.

\appendix
\section{Proof of Theorem~\ref{thm:bivariate-complete}}\label{app:bivariate}
Throughout, $\bp(t)=(t,t^{-1})$, $C=uv-1$, $z=uv$, and all polynomials are in $\C[u,v]$. We use
the \emph{scaling weight}: $(u,v)\mapsto(\rho u,\rho^{-1}v)$ multiplies $u^iv^j$ by $\rho^{i-j}$,
and an even polynomial decomposes into components of even weight; the parametrization
$t_i\mapsto\rho t_i$ realizes the same scaling on $V_C$, so $S_3[P]$ decomposes accordingly and
its extreme weight components are $S_3$ of the extreme weight components of $P$ (as $S_3$ is
homogeneous of degree four in $P$). We write $B(s,t)=\frac{s-t}{s+t}$,
$K(s,t)=\frac{s^2-st+t^2}{s^2+st+t^2}$, so that \eqref{eq:common-phase} is $A_{\mathrm{SK}}=B^2K$, and
we use the normalized residual $\mathcal F_N$ of the proof of Theorem~\ref{thm:octic}, which is
linear in $P$; for polynomials with generic two-soliton factor $A_{\mathrm{SK}}$ the vanishing of
$S_N$ and of $\mathcal F_N$ are equivalent. All exact finite computations invoked below are
recorded in the ancillary files.

\subsection{Two Laurent coefficients}\label{app:laurent}
Let $D=\deg_uP$ and write $P=u^DQ(v)+u^{D-1}E(v)+(\text{lower powers of }u)$; by
Lemma~\ref{lem:odd-degrees}, if $P$ satisfies the 3SC then $D$ is odd, so $Q$ is odd and $E$
even. Put $T:=Q'+E$, $a:=t_2^{-1}$, $b:=t_3^{-1}$. Expanding the three factors of a term of
$S_3$ that contain $\bp(t_1)$ in powers of $t_1$, the terms of order $t_1^{3D}$ cancel between
$\sg_1=\pm1$ (this is the computation in Lemma~\ref{lem:odd-degrees}), and the coefficient of
$t_1^{3D-1}$ is
\begin{equation}\label{eq:next-leading}
\begin{aligned}
[t_1^{3D-1}]S_3=4\sum_{\varepsilon=\pm1}\Bigl[&\varepsilon Q(a)Q(b)T(a+\varepsilon b)-\varepsilon T(a)Q(b)Q(a+\varepsilon b)\\
&-T(b)Q(a)Q(a+\varepsilon b)\Bigr]P\bigl(\bp(t_2)-\varepsilon\bp(t_3)\bigr).
\end{aligned}
\end{equation}
For a pure-weight polynomial $P=u^{2r}(z-1)h(z)$ with $r\ge1$, $\deg h=m$ (even by
Lemma~\ref{lem:odd-degrees}) and leading coefficients $h_m,h_{m-1}$, one has $D=2r+m+1$,
$Q=h_mv^{m+1}$, $E=(h_{m-1}-h_m)v^m$, $T=(mh_m+h_{m-1})v^m=:t_0v^m$, and the bracket in
\eqref{eq:next-leading} equals $-t_0h_m^2(ab)^m(a+\varepsilon b)^m(a^2+\varepsilon ab+b^2)$. Hence,
if $t_0\neq0$, the vanishing of \eqref{eq:next-leading} is equivalent to
\begin{equation}\label{eq:phase-forced}
-\frac{P(\bp(s)-\bp(t))}{P(\bp(s)+\bp(t))}=B(s,t)^{m}K(s,t)=:A_m(s,t)
\end{equation}
(with $a=1/s$, $b=1/t$). The same computation with the roles of $u,v$ exchanged
($t_i\mapsto t_i^{-1}$, which leaves $A_m$ invariant since $m$ is even) applies to the top
$v$-part and gives \eqref{eq:phase-forced} with $m$ replaced by $2r+m$, provided
$(2r+m)h_m+h_{m-1}\neq0$. Since the two scalars differ by $2rh_m\neq0$, at least one of them is
nonzero:

\begin{lemma}\label{lem:pure-weight-phase}
A nonzero pure-weight polynomial $P=u^{2r}(z-1)h(z)$, $r\neq0$, satisfying the 3SC has generic
two-soliton factor $A_p=B^pK$ for some even $p\ge0$.
\end{lemma}

The second Laurent coefficient concerns polynomials with the factor $A_{\mathrm{SK}}$. Let
$P_{2r}=c_0u^Dv^E+c_1u^{D-1}v^{E-1}+\dots$ (terms lower in both powers) be a pure-weight
component, $D=E+2r$, $D,E$ odd. Then, as a Laurent coefficient (negative powers of $t_3$ allowed),
\begin{equation}\label{eq:double-leading}
[t_1^{D+7}t_2^4t_3^{1-E}]\,\mathcal F_3[P_{2r}]=8\{c_0(DE-2D-2E+6)+c_1\}.
\end{equation}
Indeed, with $\varepsilon=t_1^{-1}$, $\eta=t_3$ and $t_2$ fixed, factor $\sg_1\sg_3t_1^Dt_3^{-E}$ out
of the central monomial $u^Dv^E$ and $t_1^8t_2^4$ out of the three pair weights; the coefficient
of $\varepsilon\eta$ in $u^D/(\sg_1t_1)^D$, $v^E/(\sg_3t_3^{-1})^E$ and the pair weights
($1-3\sg_1\sg_2t_2\varepsilon$, $1-3\sg_1\sg_3\varepsilon\eta$, $1-3\sg_2\sg_3\eta/t_2$ to first order)
collects to $DE-2D-2E+6$ times $\sg_1\sg_3$, the eight signs contribute a factor $8$, the
monomial $u^{D-1}v^{E-1}$ contributes $8c_1$ directly, and no lower monomial reaches the
exponent $D+7$ of $t_1$.

\subsection{The fusion identity and the exponent}\label{app:fusion}
Let $\omega^2-\omega+1=0$, so that $\bp(\omega t)-\bp(t)=\bp(\omega^2t)$ (Lemma~\ref{lem:fusion}).
Write the generic two-soliton factor as a reduced fraction $A=N/D$. Removing the common
rational factor from $S_3$ (legitimate in the field of rational functions, before any
specialization), the 3SC becomes a polynomial identity which, at a fusion point
$N(\omega t,t)=0\neq D(\omega t,t)$, reads
\begin{equation}\label{eq:fusion-phase}
A(\omega t,s)=A(t,s)\,A(\omega^2t,s).
\end{equation}
For $A=B^pK^q$ compare the orders at $s=t$: $K(\omega t,s)$ has a simple zero there
(its numerator is $\omega^2-\omega+1=0$ at $s=t$), $B(\omega t,t)\neq0$, so the left side has
order $q$; $B(t,s)$ has a simple zero and $K(t,t)=1/3$, so $A(t,s)$ has order $p$; and
$K(\omega^2t,s)$ has a simple pole at $s=t$ (its denominator is $\omega^4+\omega^2+1=0$), so
$A(\omega^2t,s)$ has order $-q$. Hence $q=p-q$:

\begin{lemma}\label{lem:fusion-exponent}
If a nonzero even $P$ with $P(0)=0$, $C\mid P$, satisfies the 3SC and has generic two-soliton
factor $B^pK^q$ with $q\ge1$, then $p=2q$. In particular, if the factor is $A_p=B^pK$ then $p=2$.
\end{lemma}

(One can show more: among all $B^pK^q$ only $(p,q)=(2,1)$ occurs. With $A=A_{\mathrm{SK}}^q$ the
analogue of \eqref{eq:double-leading} forces $D=3q$ or $E=3q$, the collision order forces the
multiplicity of $C$ to be $2q-1$ or $2q$, and the resulting one- and two-dimensional weight
spaces are excluded by exact certificates; see the ancillary files. This is not needed below.)

\subsection{Arbitrary degree with the factor $A_{\mathrm{SK}}$}\label{app:SK}
\begin{proposition}\label{prop:SK-arbitrary}
Let $P\neq0$ be even, $P(0)=0$, $C\mid P$, with generic two-soliton factor $A_{\mathrm{SK}}$. Then
$P$ satisfies the 3SC if and only if $P$ belongs to \eqref{eq:SK9}, and then all $N$SC hold.
\end{proposition}

\begin{proof}
\emph{Necessity.} With the factor prescribed, the 3SC is $\mathcal F_3[P]=0$, and both the pair
identity $H(\bp(s)+\bp(t))/(s+t)^2=H(\bp(s)-\bp(t))/(s-t)^2$ ($H=P/C$) and $\mathcal F_3$ are
linear in $P$ and separate by weight, so each weight component satisfies both. For weight
$2r\ge1$ write $H_{2r}=u^{2r}h(z)$; on $\bp(s)\pm\bp(t)$ one has $z_++z_-=4$, and the pair
identity becomes $z^{r-1}h(z)=(-1)^{r-1}(4-z)^{r-1}h(4-z)$, so $h=(4-z)^{r-1}k$ with
$k(z)=(-1)^{r-1}k(4-z)$; in particular $\deg h$ is even and $\deg h\ge r-1$ ($\ge r$ for even
$r$). For weight $0$, $P(0)=0$ and the pair identity give $h(z)=z(z-2)q((z-2)^2)$. In all cases
the two leading terms $c_0u^Dv^E+c_1u^{D-1}v^{E-1}$ of $P_{2r}$ satisfy $c_1/c_0=-(D+E-3)$
(a polynomial even or odd about $z=2$ of degree $d$ has leading ratio $-2d$; the factors
$(4-z)^{r-1}$, $k$ and $z-1$ contribute $-4(r-1)$, $-2(d-r+1)$ and $-1$). Inserting this into
\eqref{eq:double-leading} gives $[t_1^{D+7}t_2^4t_3^{1-E}]\mathcal F_3[P_{2r}]=8c_0(D-3)(E-3)$, so
$D=3$ or $E=3$. Combined with the reflection conditions: $r=0$ leaves $z(z-1)(z-2)$; $r=1$
leaves $u^2C$ and $u^2Cz(z-4)$; $r=2$ forces $E=3$ and $h$ a multiple of $(z-2)(z-4)$; $r=3$
forces $h$ a multiple of $(z-4)^2$; $r\ge4$ is impossible since then $E\ge5$ and $D>3$.
Exchanging $u,v$ gives the nine directions of \eqref{eq:SK9}.

\emph{Sufficiency.} The five directions of \eqref{eq:octic} are Theorem~\ref{thm:octic}. For
$P_{(4)}=u^4C(z-2)(z-4)$ and $P_{(6)}=u^6C(z-4)^2$ (and their reflections) the pair identity
holds by construction, so $\mathcal F_1=\mathcal F_2=0$, and the division-free collision and
fusion identities of the proof of Theorem~\ref{thm:octic} give, by induction, that
$\prod_{i<j}(x_i-x_j)^2(x_i^2+x_ix_j+x_j^2)$ divides $G_N:=(\prod_it_i^2)\mathcal F_N$, a polynomial
in $x_i=t_i^2$ because $\deg_vP=3$. Every monomial of $\mathcal F_N$ has total degree
$4\binom N2+2r$ in $t$, so $G_N$ is homogeneous of total degree $N^2+r$ in $x$, while the
divisor has total degree $2N(N-1)$. For $r=2$, $N^2+2<2N(N-1)$ for all $N\ge3$; for $r=3$ the
inequality is strict for $N\ge4$, and for $N=3$ both degrees are $12$, so $G_3$ is a constant
multiple of the divisor; at $t=(1,2,3)$ the four sign classes of $\mathcal F_3[P_{(6)}]$ contribute
$13442526720$, $0$, $583923200/3$, $-40911503360/3$, whose sum is zero, so the constant is
zero. Linearity of $\mathcal F_N$ in $P$ finishes the proof.
\end{proof}

\subsection{Non-radial polynomials with unknown factor}\label{app:nonradial}
Assume now $P\notin\C[uv]$, and let $2r>0$ be the maximal absolute weight of $P$ (after
exchanging $u,v$, the maximal weight is positive). The extreme weight components satisfy the
3SC; by Lemma~\ref{lem:pure-weight-phase} their factor is $A_p$ with $p$ even, by
Lemma~\ref{lem:fusion-exponent} $p=2$, and by Proposition~\ref{prop:SK-arbitrary} each extreme
component is a pure-weight member of \eqref{eq:SK9}. Hence all weights of $P$ lie in
$\{-6,\dots,6\}$ and $r\in\{1,2,3\}$.

\emph{The case $r\in\{2,3\}$.} Normalize the top component to $P_r=u^{2r}F_r(z)$ with
$F_2=(z-1)(z-2)(z-4)$, $F_3=(z-1)(z-4)^2$; $\deg_uP_r=:D_0=2r+3$, $\deg_vP_r=3$. We determine the
lower weights from the top down. The coefficient of weight $6r+2q$ in $S_3[P]$, $q<r$,
consists of the terms with three factors from $P_r$ and one from the weight-$2q$ component
$P_q$ (the linearization $\mathrm DS_3[P_r](P_q)$) plus terms involving only components of
weight $>2q$; if all components of weight $>2q$ already lie in the span of \eqref{eq:SK9},
these terms vanish (the $N$SC hold for every linear combination of members of \eqref{eq:SK9}),
and the 3SC gives $\mathrm DS_3[P_r](P_q)=0$. Two facts control this linear condition.

\begin{lemma}[Degree bound]\label{lem:degree-bound}
Let $R=u^Dv^E+c\,u^{D-1}v^{E-1}+\dots$ be a pure-weight polynomial of weight $D-E=2q<2r$ with
$\mathrm DS_3[P_r](R)\equiv0$. Then $D\le D_0$.
\end{lemma}

\begin{proof}
Suppose $D>D_0$, so $E>3$. Take $t_1=T\to\infty$, $t_2=1$, $t_3=1/b$. Since $D>D_0$, the terms
of the linearization in which $R$ sits in the pair factor not containing $\bp(t_1)$ have
lower $T$-order, so only the three positions containing $\bp(t_1)$ matter; the two sign
classes not related by global reversal give coefficients $C_+(b)$, $C_+(-b)$, and since the
remaining pair factor of $P_r$ has the factor $A_{\mathrm{SK}}$, vanishing forces
$C_+(-b)=A_{\mathrm{SK}}(1,b)C_+(b)$ with $A_{\mathrm{SK}}(1,b)=1-6b+18b^2+O(b^3)$. For even $E$
($\ge6$) the coefficient of $T^{D+2D_0}$ gives, up to a common nonzero factor,
$C_+(b)=b^3[(1+b)^E-(1+b^{E-3})(1+b)^3]=b^4[E-3+(\binom E2-3)b+O(b^2)]$, and the $b^5$-coefficient
of the relation forces $\binom E2-3=3(E-3)$, i.e.\ $(E-3)(E-4)=0$. For odd $E$ ($\ge7$) the top
order cancels and the coefficient of $T^{D+2D_0-1}$ gives
$C_+(b)=b^2[D-3+2r+((D-3)E+6r)b+O(b^2)]$, and the relation forces $(D-3)E+6r=3(D-3+2r)$, i.e.\
$(D-3)(E-3)=0$. Both contradict $D>D_0$, $E>3$.
\end{proof}

Consequently the weight-$2q$ component is $C\,u^{2q}h(z)$ (or $C\,v^{-2q}h(z)$ for $q<0$) with
$\deg h\le D_0-2q-1$, a finite-dimensional space, and $\mathrm DS_3[P_r]$ restricted to it is a
matrix with rational entries whose kernel is computed exactly (ancillary files). For $r=3$:
the kernel in weight $4$ is spanned by the \eqref{eq:SK9} direction $u^4C(z-2)(z-4)$ and by
$v^2P_3$; the kernels in weights $2,0,-2,-4,-6$ are exactly the corresponding pure-weight
subspaces of \eqref{eq:SK9}. For $r=2$: the kernel in weight $2$ is spanned by the two
\eqref{eq:SK9} directions and by $v^2P_2$; the kernels in weights $0,-2,-4$ are the pure-weight
subspaces of \eqref{eq:SK9}. The extra direction $v^2P_r$ is genuine: for every $N$ and every
$P$,
\begin{equation}\label{eq:gauge}
S_N[e^{hv^2}P]=e^{Nh\sum_iv_i^2}\,S_N[P],
\end{equation}
because $(\sum_i\sg_iv_i)^2+\sum_{i<j}(\sg_iv_i-\sg_jv_j)^2=N\sum_iv_i^2$ for every sign
configuration; so the multiplication by $e^{hv^2}$ preserves every $N$SC, and its first-order
effect on $P_r$ is $hv^2P_r$, of weight $2r-2$.

Let $h$ be the coefficient of $v^2P_r$ in the weight-$(2r-2)$ component of $P$ and put
$\widetilde P:=e^{-hv^2}P$, a formal series in the direction of decreasing weight whose
components are polynomials; by \eqref{eq:gauge}, $\widetilde P$ satisfies the 3SC in the formal
sense. Its top component is $P_r$ and its weight-$(2r-2)$ component lies in \eqref{eq:SK9}.
Proceeding downward through the weights $2r-4,\dots,-2r$, each component satisfies the
linearized condition, hence lies in the corresponding kernel; the kernels in these weights
are the \eqref{eq:SK9} subspaces, except in weight $2r-2$ where the gauge direction has already
been removed. So all components of $\widetilde P$ of weight $\ge-2r$ lie in \eqref{eq:SK9}. Now
extract the coefficient of $u^{D_0}v^{3+4r}$ in $P=e^{hv^2}\widetilde P$: among the components of
$\widetilde P$ of weight $\ge-2r$ only $P_r$ contains $u^{D_0}$ (the lower \eqref{eq:SK9}
components have $u$-degree $<D_0$), so the coefficient is $h^{2r}/(2r)!$, coming from
$\frac{(hv^2)^{2r}}{(2r)!}P_r$; this monomial has weight $-2r$, and the weight-$(-2r)$ component
of $P$ is either zero or, being extreme, a member of \eqref{eq:SK9} of $u$-degree $\le3<D_0$.
Hence $h=0$, $\widetilde P=P$, and $P$ lies in \eqref{eq:SK9}. (Components of $\widetilde P$ of
weight $<-2r$ cannot contribute to the extracted monomial, so no convergence question arises.)

\emph{The case $r=1$.} Both nonzero pure-weight components are known:
$P=C\{u^2[a+e\,z(z-4)]+v^2[c+f\,z(z-4)]+H_0(z)\}$. If $e\neq0$ (or $f\neq0$, after exchanging
$u,v$), the weight-$6$ coefficient of $S_3[P]$ involves three positive-weight factors and the
radial one, so the radial component satisfies the linearized condition for $r=1$, $D_0=5$
(the parameter $a/e$ does not affect Lemma~\ref{lem:degree-bound}), hence has degree at most
$5$ and $P$ has bidegree at most $(5,5)$; the classification of that box with unknown factor is
a finite computation (ancillary files; it reproduces \eqref{eq:SK9} and the exceptional
quartics, with the weight-$\pm4$ sectors excluded and $k=0$ forced). If $e=f=0$, say $a\neq0$,
and the radial component has degree $D>3$, then both partial degrees of $P$ equal $D$, which
is odd by Lemma~\ref{lem:odd-degrees}; the next-leading coefficient of the linearization at
the basis element $u^2C$ is $C_+(b)=(D-3)+D(D-3)b+O(b^2)$, and the relation
$C_+(-b)=A_{\mathrm{SK}}(1,b)C_+(b)$ forces $D(D-3)=3(D-3)$, i.e.\ $(D-3)^2=0$, a contradiction.
So the radial component has degree at most three and $P$ falls under
Proposition~\ref{prop:quadric-3SC}. This completes the non-radial case.

\subsection{Radial polynomials}\label{app:radial}
Let $P=f(uv)$ with $f\in\C[z]$, $f(0)=f(1)=0$, $f$ monic of degree $m$; by
Lemma~\ref{lem:odd-degrees}, $m$ is odd, $m\ge3$. We show that the 3SC forces
$f=z(z-1)(z-2)$ (which satisfies all $N$SC by Theorem~\ref{thm:octic}).

\emph{Two asymptotic formulas.} Take $t_1=T$, $t_2=1$, $t_3=s$. If $f=z^m+c_1z^{m-1}+\dots$
with $c_1+m\neq0$, the coefficient of $T^{3m-1}$ in $S_3$ (this is \eqref{eq:next-leading})
forces $-f(2-s-s^{-1})/f(2+s+s^{-1})=B(s,1)^{m-1}K(s,1)$, and Lemma~\ref{lem:fusion-exponent}
gives $m-1=2$; a cubic with this factor is $z(z-1)(z-2)$, whose $c_1=-3$ contradicts
$c_1+m\neq0$. Hence $c_1=-m$. Write $f(z)=(z-1)^m+\alpha(z-1)^{m-2}+\beta(z-1)^{m-3}+\dots$, put
$\mathcal C:=m+2\alpha-\beta$ and $R(z):=\mathcal C(z+2)^2(z-1)-2\alpha(z+2)+\beta$ ($R\not\equiv0$,
since $R\equiv0$ would give $\mathcal C=\alpha=\beta=0$ and $m=0$). The coefficient of $T^{3m-3}$
gives
\begin{equation}\label{eq:radial-phase}
-\frac{f(2-z)}{f(2+z)}=-\Bigl(\frac{z-2}{z+2}\Bigr)^{(m-3)/2}\frac{R(-z)}{R(z)}.
\end{equation}
If $\mathcal C\neq0$, comparing the $z^{-3}$ coefficients of the logarithms of both sides of
\eqref{eq:radial-phase} at infinity gives
\begin{equation}\label{eq:radial-consistency}
\mathcal C^2+4\alpha+m=0 ;
\end{equation}
if $\mathcal C=0$, matching the leading behaviour forces $\alpha=-m/4$, $\beta=m/2$,
$R=\frac m2(z+3)$, and the third-order comparison gives $4m/3=4m/3+5$, impossible. So
$\mathcal C\neq0$.

\emph{Collision orders.} Let $r_j:=\operatorname{ord}_{z=j}f$, $\mu:=r_1\ge1$, $\nu:=r_3$,
$d:=r_0-r_4$, $q:=\mu-\nu$. Putting $t_i=1+\epsilon a_i$, if $\mu+2r_4<2r_0$ the lowest-order term
of $S_3$ comes from the mixed sign classes and equals, up to a nonzero constant,
$\sum_i(a_i-a_j)^{\mu+2r_4}(a_i-a_k)^{\mu+2r_4}(a_j-a_k)^{2r_0}$; at $(a_1,a_2,a_3)=(0,1,t)$ and
$D=\mu+2r_4$ this is $[t(1-t)]^D[(1-t)^{2r_0-D}+t^{2r_0-D}+(-1)^D]$, which vanishes identically
only if $2r_0-D=1$. Hence in all cases
\begin{equation}\label{eq:radial-mu}
\mu\ge2d-1,\qquad m\ge r_0+r_4+\mu+\nu=d+2r_4+2\mu-q,
\end{equation}
and by \eqref{eq:radial-phase}, $d=\frac{m-3}2+\delta$ with
$\delta=\operatorname{ord}_{-2}R-\operatorname{ord}_2R$, $q=\operatorname{ord}_{-1}R-\operatorname{ord}_1R$.

\emph{The branch $q>0$.} The fusion identity \eqref{eq:fusion-phase} (valid after reduction of
the factor to lowest terms) compared at $s=t$ gives $d=q$. Since the ratio of the cubic and
quadratic coefficients of $R$ is $1:3$, $q\in\{1,3\}$. If $q=1$, $\mathcal C=m/3$,
$\beta=2\alpha+2m/3$, $R=\mathcal C(z+1)(z^2+2z+\rho)$ with $\rho\neq-3,1$, and the collision
orders allow only $(\rho,m)=(0,3)$, $(-8,7)$ or $m=5$. The cubic branch is $z(z-1)(z-2)$. For
$m=5$ the reflection identity and coprimality force $f=z(z-1)(z-2)(z^2-2z+\rho)$, whose expansion
about $z=1$ is odd, so $\beta=0$, $\alpha=-5/3$ and $R=\frac53z(z+1)(z+2)$, i.e.\ $\rho=0$, a
contradiction. For $m=7$ the exact fusion residual at $t=2$,
$A(2\omega,1)-A(2,1)A(2\omega^2,1)=-\frac{64(772\omega-605)}{212121}\neq0$, excludes the branch.
If $q=3$, $R=\mathcal C(z+1)^3$, $m=9$, the factor is $A_{\mathrm{SK}}^3$, the reflection identity
forces $f(z+2)=(z+2)^3(z+1)^3H(z)$ with $H$ an odd cubic, so $\nu\le1$, $\mu\le4$, while $d=q=3$
and \eqref{eq:radial-mu} require $\mu\ge5$.

\emph{The branch $q\le0$.} $R$ cannot have a triple root at $2$ (the coefficient ratio of
$(z-2)^3$ is $-6\neq3$), so $\delta\ge-2$; if $\delta=-2$ then $R=\mathcal C(z-2)^2(z+7)$,
$\mathcal C=m/57$, $\alpha=4m/19$, $\beta=80m/57$, and \eqref{eq:radial-consistency} becomes
$m(m+5985)/3249=0$, impossible; so $\delta\ge-1$, $d\ge(m-5)/2$, and \eqref{eq:radial-mu} with
$q\le0$ gives $m\ge d+2\mu\ge5d-2\ge(5m-29)/2$, i.e.\ $m\le9$. For $m=9$: $\delta=-1$, $d=2$,
$\mu=3$, $q\in\{-1,0\}$, $r_4=0$; $q=-1$ forces $f=z^2(z-1)^3(z-3)^4$, whose second coefficient
$-15\neq-9$; $q=0$ forces $f=z^2(z-1)^3(z-3)^3(z+3)$, for which $R(2)=-426\neq0$ contradicts
$\delta=-1$. For $m=7$: $\delta=-1$, $d=1$, $R=\mathcal C(z-2)(z^2+5z+\rho)$ with
$15\rho^2-180\rho+532=0$ from \eqref{eq:radial-consistency}, the reflection identity forces
$f=z(z^2+z+\rho-6)(z-1)(z-3)[(z-2)^2+\eta]$ with $\eta=(69\rho-446)/6$, and the exact
residual $S_3$ at $t=(1,2,3)$, reduced modulo the quadratic relation for $\rho$, is a nonzero
multiple of $54084029363835\rho-320709590163169$, nonzero since $\rho$ is irrational. For
$m=5$: $f=z(z-1)(z-3)(z^2-z+\rho)$ and the residuals at $t=(1,2,3)$ and $(1,2,4)$ are nonzero
multiples of two quartic polynomials in $\rho$ with $\gcd$ equal to $1$. For $m=3$:
$f=z(z-1)(z-3)$ has second coefficient $-4\neq-3$. Hence $f=z(z-1)(z-2)$.

\subsection{Conclusion}
Combining Sections~\ref{app:nonradial} and~\ref{app:radial}: a nonzero even $P$ with $P(0)=0$,
$C\mid P$, satisfying the 3SC lies in \eqref{eq:SK9} or is an exceptional quartic; conversely
these satisfy all $N$SC by Proposition~\ref{prop:SK-arbitrary} and
Theorem~\ref{thm:quadric-NSC}(II). The degree bounds and the description of the radial members
are read off from \eqref{eq:SK9}. \qed

\section{Proof of Theorem~\ref{thm:parabola-general}}\label{app:parabola}
\subsection{The rigidity lemma}
\begin{lemma}\label{lem:parabola-rigidity}
Let $P(u,v)\neq0$ be even and divisible by $v-u^2$, and suppose that its generic two-soliton
factor on $\bp(t)=(t,t^2)$ is $A(s,t)=(s-t)^2/(s^2+st+t^2)$. Then $P=\gamma(u^4-v^2)$.
\end{lemma}

\begin{proof}
Evenness gives $v+u^2\mid P$ as well, so $P=(u^4-v^2)H$ with $H\neq0$. Since
$P_0:=u^4-v^2$ has $P_0(\bp(s)-\bp(t))=-4st(s-t)^2$, $P_0(\bp(s)+\bp(t))=4st(s^2+st+t^2)$, and
$(s,t)\mapsto(s+t,s^2+t^2)$ is dominant, the prescribed factor is equivalent to the polynomial
identity $H(s+t,s^2+t^2)=H(s-t,s^2-t^2)$. Put $G:=H-H(0)$. Setting $s=t$ gives $G(2t,2t^2)=0$,
so $G$ vanishes on the parabola $v=u^2/2$. If $G$ vanishes on $v=\kappa u^2$ ($\kappa\neq0$),
take $s=\frac{\kappa+1}2x$, $t=\frac{\kappa-1}2x$: the right side has argument $(x,\kappa x^2)$, the
left side $(\kappa x,\frac{\kappa^2+1}2x^2)$, so $G$ vanishes on $v=T(\kappa)u^2$ with
$T(\kappa):=\frac{\kappa^2+1}{2\kappa^2}$. For $0<\kappa<1$ one has $T(\kappa)>1$,
$\tfrac12<T^2(\kappa)<1$ and $T^2(\kappa)-\kappa=-\frac{(\kappa-1)^3(2\kappa^2+\kappa+1)}{2(\kappa^2+1)^2}>0$,
so the even iterates of $\kappa_0=\tfrac12$ form an infinite strictly increasing sequence
in $(\tfrac12,1)$, and $G$ is divisible by infinitely many distinct irreducible polynomials
$v-\kappa_{2j}u^2$; hence $G=0$ and $H$ is constant.
\end{proof}

\subsection{Weighted homogeneous polynomials}
Assign the weights $\operatorname{wt}u=1$, $\operatorname{wt}v=2$; the scaling $t\mapsto\rho t$ of
the parameters of $\bp(t)=(t,t^2)$ realizes $(u,v)\mapsto(\rho u,\rho^2v)$, so the top weighted
homogeneous part of $P$ satisfies the 3SC on the parabola whenever $P$ does. Let first $P$ be
weighted homogeneous of weight $W$, even, divisible by $v-u^2$; then $P=u^Wf(v/u^2)$ for $u\neq0$
with $f(-z)=(-1)^Wf(z)$, so $\pm1$ are roots of $f$ of the same multiplicity $m\ge1$, and
$P=(v-u^2)^mG$ with $G$ weighted homogeneous, $G(1,1)\neq0$.

\emph{$m$ is odd.} With parameters $(T,s,t)$ and signs $(+,\varepsilon,\eta)$, the three factors
containing $\bp(T)$ have arguments $(T+U_i,T^2+V_i)$ with $(U_1,V_1)=(\varepsilon s+\eta t,\varepsilon s^2+\eta t^2)$,
$(U_2,V_2)=(-\varepsilon s,-\varepsilon s^2)$, $(U_3,V_3)=(-\eta t,-\eta t^2)$, so $\sum U_i=\sum V_i=0$,
and $P(T+U,T^2+V)=(-2)^mG(1,1)U^mT^{W-m}(1+O(T^{-1}))$. If $m$ were even, the top
$T$-coefficient of $S_3$ would give $(s+t)^mP(\bp(s)-\bp(t))+(s-t)^mP(\bp(s)+\bp(t))=0$, i.e.\
$-P(\bp(s)-\bp(t))/P(\bp(s)+\bp(t))=((s-t)/(s+t))^m$; but as $t\to0$ the left side tends to
$(-1)^{m+1}=-1$ by the multiplicity, the right side to $1$. So $m$ is odd.

\emph{The first non-cancelling coefficient.} For odd $m$ the top coefficient cancels between
$(\varepsilon,\eta)$ and $(-\varepsilon,-\eta)$. Write
$\log\frac{P(T+U,T^2+V)}{(-2)^mG(1,1)U^mT^{W-m}}=m\log\bigl(1+\frac{U^2-V}{2UT}\bigr)+\log\frac{G(1+U/T,1+V/T^2)}{G(1,1)}$
as a formal series; only the terms odd under $(\varepsilon,\eta)\mapsto(-\varepsilon,-\eta)$ survive
the sign sum, the odd terms of order $T^{-1}$ and $T^{-2}$ are multiples of $\sum U_i$ and
$\sum V_i$ and vanish, and the odd term of order $T^{-3}$ is
$d\sum_iU_i^3+\frac m8\sum_iV_i^2/U_i$ with $d=\frac m{24}+\frac16[\partial_u^3\log G](1,1)$
(the $V^2/U$ term comes only from the first logarithm). Using $\sum U_i=\sum V_i=0$,
$\sum_iU_i^3=3\varepsilon\eta st(\varepsilon s+\eta t)$ and
$\sum_iV_i^2/U_i=-\varepsilon\eta st(s-t)^2/(\varepsilon s+\eta t)$, so the coefficient of
$T^{3(W-m)-3}$ in $S_3$ gives
\[
(s+t)^{m-1}\Bigl[3d(s+t)^2-\tfrac m8(s-t)^2\Bigr]P(\bp(s)-\bp(t))+\Bigl(3d-\tfrac m8\Bigr)(s-t)^{m+1}P(\bp(s)+\bp(t))=0 .
\]
Here $3d\neq m/8$ (otherwise $P$ would vanish on the dense image of the difference map), and
therefore the generic two-soliton factor is
\begin{equation}\label{eq:parabola-factor}
A(s,t)=\Bigl(\frac{s-t}{s+t}\Bigr)^{m-1}\frac{(s-t)^2}{s^2+\beta st+t^2},\qquad
\beta=\frac{2(3d+m/8)}{3d-m/8}.
\end{equation}

\emph{Root counting.} Write the difference point as $(x,\kappa x^2)$, i.e.\ $s=\frac{\kappa+1}2x$,
$t=\frac{\kappa-1}2x$, so the sum point is $(\kappa x,\frac{\kappa^2+1}2x^2)$; with
$T(\kappa)=\frac{\kappa^2+1}{2\kappa^2}$, \eqref{eq:parabola-factor} becomes the univariate identity
\begin{equation}\label{eq:parabola-univariate}
f\bigl(T(\kappa)\bigr)=-\tfrac14\kappa^{m-1-W}\bigl[(\beta+2)\kappa^2+2-\beta\bigr]f(\kappa).
\end{equation}
Let $S$ be the set of distinct nonzero roots of $f$; $\pm1\in S$. First $\tfrac12\notin S$:
\eqref{eq:parabola-univariate} maps every nonzero root $\kappa$ to the root $T(\kappa)$, and the
even iterates of $\tfrac12$ form an infinite set as in Lemma~\ref{lem:parabola-rigidity}. For
$z\in S$ the equation $T(\kappa)=z$ has exactly two distinct nonzero solutions
$\kappa^2=1/(2z-1)$, with disjoint preimages for distinct $z$, so $|T^{-1}(S)|=2|S|$; by
\eqref{eq:parabola-univariate} each preimage lies in $S$ or is a root of
$(\beta+2)\kappa^2+2-\beta$, whence $2|S|\le|S|+2$ and $S=\{1,-1\}$. The preimages
$\pm\mathrm{i}/\sqrt3$ of $-1$ are not in $S$, so they are the roots of the quadratic, which forces
$\beta=1$; comparing the orders of vanishing at these points ($T'\neq0$ there, simple roots of
the quadratic, $f\neq0$ there) gives $m=1$. So $A(s,t)=(s-t)^2/(s^2+st+t^2)$ and
Lemma~\ref{lem:parabola-rigidity} gives $P=\gamma(u^4-v^2)$.

\subsection{General polynomials, flat variables, translations, higher rank}
For a general even $P$ divisible by $v-u^2$ write $P=(u^4-v^2)H$; the top weighted homogeneous
part of $P$ satisfies the 3SC, hence equals $\gamma(u^4-v^2)$, so $H$ has top weight $0$, i.e.\
$H$ is constant. If $P$ depends on further variables $\mathbf z$, restrict to the planes
$\mathbf z=\mathbf au+\mathbf bv$ for generic $\mathbf a,\mathbf b$: the restrictions are nonzero
(otherwise $P=0$) and satisfy the 3SC on the parabola, so $H(u,v,\mathbf au+\mathbf bv)=H(0)$
for generic, hence all, $\mathbf a,\mathbf b$, and taking $\mathbf a=\mathbf z/u$, $\mathbf b=0$
shows $H\equiv H(0)$. For the translated parabola $C=v-u^2-c$, evenness gives
$P=[(u^2+c)^2-v^2]H$; under the weighted scaling the points $(t,t^2+c)$ tend to the
untranslated parabola and the top weighted part of $P$ satisfies the 3SC there, so it is
$\gamma(u^4-v^2)$ and $H$ is constant; then $P(0)=\gamma c^2=0$ forces $c=0$. Finally let
$C=Q(\mathbf x)+w+c$ with $\operatorname{rank}Q\ge2$ and $w$ independent. Restricting to the planes
$\mathbf x=t\mathbf r$, $Q(\mathbf r)\neq0$, gives translated parabolas $Q(\mathbf r)t^2+w+c$, so
$c=0$, and, with $P=(Q(\mathbf x)^2-w^2)H$, the cofactor $H(t\mathbf r,w)$ is constant on each such
plane, hence $H$ is constant and $P=\gamma(Q^2-w^2)$. On a nondegenerate plane of $Q$ normalized
to $uv$, the points $\bp_1=(1,0,0)$, $\bp_2=(0,1,0)$, $\bp_3=(1,1,-1)$ of $\{uv+w=0\}$ have all
$P(\bp_i\pm\bp_j)/\gamma\in\{1,-1,3\}$ and the eight sign terms of $S_3/\gamma^4$ are
$15,-9,3,3,3,3,-9,15$, so $S_3=24\gamma^4\neq0$. \qed

\bibliographystyle{amsplainx}
\bibliography{refs}

\end{document}